\documentclass{lmcs}


\usepackage{hyperref}
 
\usepackage[utf8]{inputenc}

\usepackage{url}
\usepackage{longtable}
\usepackage{wrapfig}
\usepackage{rotating}
\usepackage[normalem]{ulem}

\usepackage{amssymb}
\usepackage{capt-of}
\usepackage{scalerel}
\usepackage{xspace}

\usepackage{amsmath} 
 
\usepackage{multirow}
\usepackage{tabularx}

\usepackage{graphicx}
\usepackage{tikz,tikz-qtree}
 \usetikzlibrary{arrows.meta,positioning,fit,backgrounds}
\usetikzlibrary{decorations.pathreplacing}  

\usepackage{booktabs}
\usepackage{xcolor}
  \usepackage{makecell}

\makeatletter
\newcommand{\StrongDis}{\mathbin{\mathpalette\make@circled\vee}}
\newcommand{\make@circled}[2]{%
  \ooalign{$\m@th#1\smallbigcirc{#1}$\cr\hidewidth$\m@th#1#2$\hidewidth\cr}%
}
\newcommand{\smallbigcirc}[1]{%
  \vcenter{\hbox{\scalebox{0.77778}{$\m@th#1\bigcirc$}}}%
}
\makeatother

\newcommand{\details}[1]{} 

\newcommand{\LogicName}[1]{\ensuremath{\text{\textup{\sffamily #1}}}}

 \newcommand{\Nat}{\mathbb{N}}
\newcommand{\Lang}{\mathcal{L}}
\newcommand{\Tower}{\textsf{Tower}}
\def\Bool{{\mathbb{B}}}
 \newcommand{\Fam}{\mathcal{F}}
 
 \newcommand{\Com}{\mathcal{S}}
 \newcommand{\bu}{\textsf{b}}
 \newcommand{\co}{\textsf{c}}
 \newcommand{\tr}{\textsf{t}} 
 
\newcommand{\DefinedAs}{\ensuremath{\,:=\,}}
\newcommand{\Val}[1]{\ensuremath{\text{$\parallel$\hspace{0.02cm}$#1$\hspace{0.02cm}$\parallel$}}}
\newcommand{\tpl}[1]{\langle #1 \rangle}
\newcommand{\AP}{\textit{AP}}
\newcommand{\MP}{\textsf{mp}}

\newcommand{\Au}{\mathcal{A}}
\newcommand{\Nu}{\mathcal{N}}
\newcommand{\Mu}{\mathcal{M}}
\newcommand{\Lab}{\textit{Lab}}
\newcommand{\Paths}{\textit{Paths}}  
\newcommand{\KS}{\textit{K}} 
\newcommand{\Card}{\textsf{card}}
\newcommand{\Enc}{\textsf{enc}}  
\newcommand{\False}{\mathsf{false}}
\newcommand{\True}{\mathsf{true}}

\newcommand{\Instance}{\mathcal{I}}

\newcommand{\MaxCol}{\textsf{MaxCol}}

\newcommand{\Symb}{\$}
\newcommand{\bl}{\textit{bl}}
\newcommand{\sbl}{\textit{sb}}
\newcommand{\bla}{\textit{bl}\kern .1em '}
\newcommand{\sbla}{\textit{sb}\kern .1em '}
\newcommand{\col}{\textit{\#}} 

\newcommand{\dep}{\textit{dep}}
\newcommand{\Until}{\mathsf{U}}

\newcommand{\Release}{\mathsf{R}}
\newcommand{\Next}{\mathsf{X}}
\newcommand{\Future}{\mathsf{F}}
\newcommand{\Always}{\mathsf{G}}
\newcommand{\StrongNeg}{{\sim}}

\newcommand{\EQ}{\mathsf{E}}
\newcommand{\AQ}{\mathsf{A}}
\newcommand{\EQOne}{\mathsf{E}_1}
\newcommand{\AQOne}{\mathsf{A}_1}

  \newcommand{\TeamLTL}{\LogicName{TeamLTL}}
   
      \newcommand{\ITL}{\LogicName{ITL}}
  \newcommand{\InqLTL}{\LogicName{InqLTL}}

  \newcommand{\ICTL}{\LogicName{ICTL}}
  \newcommand{\IATL}{\LogicName{IATL}}

 \newcommand{\LTL}{\LogicName{LTL}}
\newcommand{\CTL}{\LogicName{CTL}}
\newcommand{\CTLStar}{\LogicName{CTL$^{*}$}}

\newcommand{\NWA}{\LogicName{NWA}}
\newcommand{\HAA}{\LogicName{HAA}}

\newcommand{\HQPTL}{\LogicName{HyperQPTL}}
\newcommand{\HLTL}{\LogicName{HyperLTL}}
\newcommand{\QPTL}{\LogicName{QPTL}}
\newcommand{\HCTLStar}{\LogicName{HyperCTL$^{*}$}}

\newcommand{\HPDL}{\LogicName{HyperPDL$-\Delta$}}
\newcommand{\PDL}{\LogicName{PDL}}

\def\PSPACE{{\sc PSPACE}}
\def\EXPSPACE{{\sc EXPSPACE}}

\def\NEXPTIME{{\sc NEXPTIME}}

\begin{document}

\title[Inquisitive Team Semantics of LTL]{Inquisitive Team Semantics of LTL\rsuper*}
\titlecomment{{\lsuper*}This paper is an extended version of the conference paper~\cite{BozzelliLMM26}, completed with full proofs as well as additional results and examples.}

\author[L.~Bozzelli]{Laura Bozzelli\lmcsorcid{0009-0004-8555-8229}}	
\address{Pegaso University, Napoli, Italy}	
\email{laura.bozzelli@unipegaso.it}

\author[T.~Litak]{Tadeusz Litak\lmcsorcid{0000-0003-2240-3161}}	
\address{University of Padua, Italy}	
\email{tadeusz.litak@unipd.it}

\author[M.~Mittelmann]{Munyque Mittelmann\lmcsorcid{0000-0002-4664-8406}}	
\address{CNRS, LIPN, Université Sorbonne Paris Nord, France}	
\email{mittelmann@lipn.univ-paris13.fr}

\author[A.~Murano]{Aniello Murano\lmcsorcid{0000-0003-4876-3448}}	
\address{University of Napoli ``Federico II'', Italy}	
\email{aniello.murano@unina.it}

\begin{abstract}
In recent years, hyperproperties (i.e., properties of sets of traces) have been identified as a crucial concept for capturing information-flow security policies
and knowledge of agents in distributed systems. In the synchronous linear-time setting, a well-known approach for the specification of hyperproperties
is based on the set semantics of standard $\LTL$. The resulting logic ($\TeamLTL$) inherits the powerful split interpretation of disjunction from dependency logic.
In this paper, we introduce a novel team semantics of $\LTL$ inspired by inquisitive logic, whose main features are the
intuitionistic interpretation of implication and the inquisitive semantics of disjunction. We investigate expressiveness, decidability, and complexity issues of the novel logic that we call $\InqLTL$ and its extension
with strong negation. We show that $\InqLTL$ with strong negation is highly undecidable and strictly less expressive than $\TeamLTL$ with strong negation.
On the positive side, we identify a meaningful fragment of
$\InqLTL$ with a decidable model-checking problem that can express relevant classes of hyperproperties.  To the best of our knowledge, this fragment represents the first hyper logic with a decidable model-checking problem that allows unrestricted use of temporal modalities and conditional second-order universal quantification over system refinements via intuitionistic implication.

\end{abstract}
\maketitle

\section{Introduction}\label{sec:Intro}

Hyperproperties~\cite{ClarksonS10} are a specification paradigm that generalizes trace properties to properties of sets of traces 
by allowing the comparison of distinct execution traces of a system. They play a crucial role in capturing
information-flow security requirements such as  noninterference~\cite{goguen1982security,McLean96} and observational
determinism~\cite{ZdancewicM03}, which relate  observations of an external low-security agent
along distinct traces resulting from different values of not directly observable inputs.  
These  requirements are not regular properties and so cannot be expressed in traditional temporal logics like
$\LTL$, $\CTL$, and $\CTLStar$~\cite{Pnueli77,EmersonH86}. Other relevant examples of
hyperproperties include epistemic properties specifying the
 knowledge  of agents in distributed systems \cite{HalpernV86,HalpernO08}, bounded termination of programs, the symmetrical access to critical resources in distributed protocols~\cite{FinkbeinerRS15},
and diagnosability of critical systems~\cite{SampathSLST95,BittnerBCGTV22}.

Two main approaches have been proposed for the formal specification, analysis, and automatic verification (model-checking) of 
hyperproperties in a synchronous setting. The first  
extends standard temporal logics like Linear Temporal Logic ($\LTL$), $\CTLStar$, $\QPTL$~\cite{SistlaVW87}, and
$\PDL$~\cite{FischerL79}
with explicit first-order quantification over traces (and trace variables to refer to multiple traces at the same time)
yielding logics like $\HLTL$~\cite{ClarksonFKMRS14}, $\HCTLStar$~\cite{ClarksonFKMRS14},
$\HQPTL$~\cite{Rabe2016,CoenenFHH19}, and $\HPDL$~\cite{GutsfeldMO20}.  These logics enjoy 
a decidable, although nonelementary, model checking problem and have a synchronous semantics: temporal modalities
advance by a lockstepwise traversal of all the quantified traces. The second approach adopts a set semantics of temporal logics, in particular 
$\LTL$, resulting in the logic  $\TeamLTL$~\cite{KrebsMV018}, where the semantical entities are sets of traces (\emph{teams})
instead of single traces, and temporal operators advance in lockstep on all the traces of the current team. Moreover,
 $\TeamLTL$  inherits the powerful split interpretation of disjunction from dependence logic~\cite{Vaananen07}, which also allows expressing \emph{existential} quantification over subteams of the current team.
 The main advantage of the team approach in relation to the explicit quantification over traces is that it 
preserves the modal nature of temporal logics and enables
a more readable and compact formulation of hyperproperties.  
Moreover, a distinctive expressivity feature of team temporal logics  
is the ability to relate an unbounded number of traces,   
which makes global bounded-time requirements expressible. 
An example of such requirements is bounded termination, which consists in checking whether a system terminates in bounded time, i.e., for the (possibly infinite) set of system traces,
there is a time $t$ such that each trace has terminated at time $t$.
  On the other hand,
very few positive decidability results are known for $\TeamLTL$ and 
its extensions~\cite{KrebsMV018,VirtemaHFK021}. For example, model checking the extension with strong/Boolean negation $\StrongNeg$ is highly
undecidable (allowing encoding of \emph{third-order} arithmetic)~\cite{Luck20}, while   
 for $\TeamLTL$ itself and its extensions with dependence and/or inclusion atoms of dependence logic, the corresponding decidability problem is an intriguing open question. The only known positive results have been achieved by imposing drastic syntactical or semantical restrictions~\cite{KrebsMV018,VirtemaHFK021}. 
Asynchronous variants of $\HLTL$ and  $\TeamLTL$ 
have been recently investigated~\cite{BaumeisterCBFS21,BozzelliPS21,GutsfeldMOV22}. These logics have an undecidable model checking problem, 
so the research focuses on better behaved fragments~\cite{BaumeisterCBFS21,BozzelliPS21,GutsfeldMOV22} or more tractable \emph{lax} semantics~\cite{KontinenSV25}.  \vspace{0.25cm}
  
\noindent\textbf{Inquisitive logic} Propositional inquisitive logic~\cite{CiardelliR11,Ciardelli16,Ciardelli2022} provides an approach for the formal analysis of information exchange 
which allows for a unified treatment of declarative and interrogative statements using \emph{inquisitive disjunction} and a notion of entailment where questions can occur as premises, conclusions,
or arbitrary subformulas.
One can see it as a conservative extension of \emph{classical} logic replacing classical objects of evaluation (e.g., first-order relational structures) with collections (\emph{teams} or \emph{information sets}) of such objects. Intuitively, a team encodes the informational uncertainty 
about the current state of affairs. If we use the De Morgan's laws to remove classical disjunction from the primitives, the formalism becomes a specialization of \emph{intuitionistic} logic over teams, with the reverse inclusion being the intuitionistic partial order.  
Since atomic propositions (but not arbitrary formulas) inquisitively satisfy 
the double negation law\footnote{Let us note already here that this is just one example of how the novel logic  $\InqLTL$  we propose in Section \ref{sec:InquisitiveLTL} has more theorems than intuitionistic $\LTL$ of Balbiani et al.~\cite{BalbianiBDF20}. 
Note that our semantics can be encoded into a strict subclass of the birelational models of~\cite{BalbianiBDF20}.  See Section \ref{sec:EmbeddingBalbiani} for a detailed discussion of the connection between these two frameworks.},  disjunction-free inquisitive propositional formulas are indeed classical; in this context, recall the G\"odel-Gentzen translation~\cite[Ch. 2]{ChagrovZ97:ml} \cite{FerreiraO12:full,litakpr17}  \cite[Ch. 6--7]{SorensenU06:book} \cite[Ch. 2.3]{TroelstraVD}. 
 The intuitionistic character of team semantics~\cite{AbramskyV09} and inquisitive aspects of dependencies~\cite{Ciardelli16dep} have been noted before, alongside extensive studies on inquisitive first-order formalisms ~\cite{Ciardelli2016a,Ciardelli2022,CiardelliG22,ciardelli2009inquisitive,grilletti2019disjunction,Grilletti2021,LitakS25}.
 The situation is somewhat different when it comes to inquisitive modal systems. While there are numerous papers investigating extension of the propositional base with modalities ~\cite{Ciardelli2016a,Ciardelli22,Ciardelli2022,CiardelliO17,CiardelliO22,CiardelliR15,MaricP24,Nygren23}, none of those published so far extends to temporal setting. Furthermore, these references restricted attention to  \emph{declarative} modalities (also unlike the present paper, typically over some kind of neighbourhood semantics), which would rule out connectives such as $\Future$. More information is provided in Remark~\ref{rem:decl} of Section~\ref{sec:InquisitiveLTL}, but a detailed discussion of different inquisitive modal frameworks is postponed to future work.
%

\paragraph{Our contribution} 
In this paper, we advance the research on temporal logics with set semantics by introducing an inquisitive synchronous team semantics of $\LTL$. 
The resulting team logic $\InqLTL$ is interpreted on sets of traces (or \textit{teams}) and enjoys both downward closure and related meta-properties from the inquisitive tradition. 
Its distinguishing feature is that it replaces the split disjunction of $\TeamLTL$   with inquisitive disjunction and intuitionistic implication, thereby capturing the dynamics of information-seeking behaviour.
While $\TeamLTL$ without split disjunction was already investigated in~\cite{KrebsMV018} and with both split and inquisitive disjunctions and strong negation in~\cite{Luck20,VirtemaHFK021}, none of these works consider  \emph{intuitionistic} implication combined with inquisitive disjunction.

  We investigate expressiveness, decidability, and complexity issues of $\InqLTL$  and its extension with Boolean negation, denoted $\InqLTL(\StrongNeg)$. We show that the inquisitive team semantics can be expressed in 
  $\TeamLTL(\StrongNeg)$,   
  and that the logic $\TeamLTL(\StrongNeg)$ turns out to be strictly
more expressive than $\InqLTL(\StrongNeg)$. In particular, while it is known that there are satisfiable $\TeamLTL(\StrongNeg)$ formulas whose models are uncountable teams~\cite{Luck20}, 
we establish that $\InqLTL(\StrongNeg)$ has the countable model property. Moreover, we prove that satisfiability and model checking of $\InqLTL(\StrongNeg)$
are highly undecidable by a reduction from truth of \emph{second-order} arithmetic. 

As a main contribution, we identify a meaningful fragment of $\InqLTL$, which we call \emph{left-positive} $\InqLTL$,
where the nested use of implication in the left side of an implication formula is disallowed. Left-positive $\InqLTL$ can formalize relevant information-flow security requirements and, unlike $\TeamLTL$,
is able to express dependency atoms and
\emph{conditional universal subteam} quantification.
 We show that model checking left-positive $\InqLTL$ is decidable, 
and we precisely pinpoint its complexity: for each $k \geq 0$, model checking formulas of implication depth at most $k$ is $k$-\EXPSPACE-complete (and \PSPACE-complete for $k=0$).
For the upper bounds, we introduce an abstract semantics of $\InqLTL$ where teams of paths are abstracted away by paths of sets of states (\emph{macro-paths}).
We then prove that this abstraction is sound and complete for left-positive $\InqLTL$, and provide an automata-theoretic approach for solving  model checking  under the macro-path semantics.
Matching $k$-\EXPSPACE\ lower bounds are provided by an elegant polynomial-time reduction from a suitable domino-tiling problem. 
  
Overall, rather than introducing an isolated or competing formalism, our study positions $\InqLTL$ as a natural development within the unified framework of synchronous team temporal logics, exploring how inquisitive disjunction and intuitionistic implication interact over sets of traces.

\paragraph{Outline of the paper}  
In Section~\ref{sec:Preliminaries}, we introduce some basic notation and preliminary definitions. Moreover, we recall syntax and semantics of $\TeamLTL$~\cite{KrebsMV018} .
In Section~\ref{sec:InquisitiveLTL}, we introduce the novel logic $\InqLTL$ and the investigated $\InqLTL$ fragments, establish its key structural properties,
and illustrate the framework by meaningful examples. In Section~\ref{sec:ExpressivenessResults}, we address expressiveness issues and in Section~\ref{sec:UndecidabilityResults}
we establish the undecidability of $\InqLTL(\StrongNeg)$. Section~\ref{sec:DecidabilityResults} deals with the decision procedure for model checking the left-positive fragment of
$\InqLTL$  and Section~\ref{sec:LowerBounds} provides lower bounds matching the upper bounds in Section~\ref{sec:DecidabilityResults}. 
 Finally, in Section~\ref{sec:Conclusions}
we draw some concluding remarks and discuss directions for future work.

This paper is an extended and revised version of the conference paper~\cite{BozzelliLMM26}. 
In addition to providing full technical proofs for all expressiveness, decidability, and complexity results (including the detailed matching $k$-\EXPSPACE\ lower-bound construction), the overall presentation has been significantly enhanced with extensive conceptual explanations, intuitive discussions, and detailed figures. Furthermore, the present version introduces several  novel contributions: (i) an embedding of $\InqLTL$ into intuitionistic linear temporal logic over expanding frames~\cite{BalbianiBDF20} (Section~\ref{sec:InquisitiveLTL}); (ii) a comprehensive classification and comparative landscape of synchronous team linear-time temporal logics (Section~\ref{sec:InquisitiveLTL}); (iii) a deeper investigation of the expressive boundaries between $\InqLTL$, $\InqLTL(\sim)$, and $\TeamLTL$, including formal separation conjectures (Section~\ref{sec:ExpressivenessResults}); and (iv) additional illustrative specifications modeling information-flow security requirements and epistemic properties under refinement (Section~\ref{sec:InquisitiveLTL}).
\section{Preliminaries}\label{sec:Preliminaries}

In this section, we fix the basic notation and notions used throughout the paper. Moreover, we recall the syntax and semantics of $\TeamLTL$~\cite{KrebsMV018} (in the synchronous setting).

\noindent Let $\Nat$ be the set of natural numbers.
For all $n,h\in\Nat$ and integer constants \mbox{$c>1$},  $\Tower_c(h,n)$ denotes a tower of exponentials of base $c$, 
height $h$, and argument $n$: $\Tower_c(0,n)=n$ and $\Tower_c(h+1,n)=c^{\Tower_c(h,n)}$.  For each $h\in\Nat$,  
$h$-\EXPSPACE\ is   the class of languages decided by deterministic Turing machines bounded in space
by functions of $n$ in $O(\Tower_c(h,n^d))$ for some integer constants $c> 1$ and $d\geq 1$. Note that
$0$-\EXPSPACE\ coincides with \PSPACE.

Given a (finite or infinite) word $w$ over some alphabet, $|w|$ is the length of
$w$ ($|w|=\infty$ if $w$ is infinite).
For each $0\leq i<|w|$, $w(i)$ is the $(i+1)$-th symbol of $w$ and
$w_{\geq i}$ is the suffix of $w$ from position $i$, that is, the word
$w(i)w(i+1)\ldots$. 
For a set $\Lang$ of infinite words over some alphabet and $i\geq 0$,
$\Lang_{\geq i}$ is the set of suffixes of the words in $\Lang$ from position $i$:
$\Lang_{\geq i}\DefinedAs \{w_{\geq i}\mid w\in \Lang\}$.

We fix a finite set $\AP$ of atomic propositions. 
A \emph{trace} is an infinite word on $2^{\AP}$. 
  
\paragraph{Labelled trees and Kripke Structures} 
A tree $T$ is a prefix closed subset of $\Nat^{*}$.  Elements of $T$ are called nodes, and $\varepsilon$ (the empty word) is the root of $T$. For $x\in T$, a child of $x$ in $T$ is a node in $T$  of the form $x\cdot n$ for some $n\in \Nat$.  A  path of $T$ is a maximal sequence $\pi$ of nodes such that   $\pi(i)$ is a child in $T$ of $\pi(i-1)$ for all $0<i<|\pi|$.   For an alphabet $\Sigma$, a $\Sigma$-labelled   tree is a pair $\tpl{T, V}$ consisting of a  tree $T$ 
 and a labelling $V:T \to \Sigma$  assigning to  each $T$-node  a symbol in $\Sigma$.
 
We describe the dynamic behaviour of systems by standard 
Kripke structures over $\AP$
which are tuples $\KS=\tpl{S,S_0,R,\Lab}$ where $S$ is a nonempty set of states, $S_0\subseteq S$ is a set of initial states,
$R\subseteq S\times S$ is a \emph{left-total} transition relation (i.e., for each $s\in S$, there is $s'\in S$ such that $(s,s')\in R$), and $\Lab:S \rightarrow 2^{\AP}$ is a propositional valuation 
assigning to each state the propositions in $\AP$ which hold at $s$. 
For a state $s$, we write $R[s]$ to mean the set of \emph{successors} of state $s$, i.e., the nonempty set of states $s'$
such that $(s,s')\in R$. A \emph{path} $\pi$ of $\KS$ is an infinite word $\pi=  s_0s_1\ldots $ over $S$ 
such that $(s_i,s_{i+1})\in R$ for each $i\geq 0$. The path $\pi$ is \emph{initial} if it starts at some initial state, that is, $s_0\in S_0$. 
The path $\pi$ induces the trace $\Lab(s_0)\Lab(s_1)\ldots$. 
A trace of $\KS$ is a trace induced by some \emph{initial} path of $\KS$.
We denote by $\Lang(\KS)$
 the set of $\KS$'s traces.

In this paper,  we consider logics whose interpretations (models) are sets $\Lang$ of traces. For such logics, we consider the following decision problems.
\begin{itemize}
  \item \emph{Satisfiability}: checking,  given a formula $\varphi$, whether there is a \emph{nonempty} set of traces satisfying $\varphi$ (the notion of \emph{satisfaction} depends on the specific logic).
  \item \emph{Model checking}: checking,   given a finite Kripke structure $\KS$ and a  formula $\varphi$, whether $\Lang(\KS)$ satisfies $\varphi$.  
\end{itemize} 
As usual for two formulas $\varphi$ and $\varphi'$, we write $\varphi\equiv \varphi'$ to mean that $\varphi$ and $\varphi'$ are equivalent, i.e., they are fulfilled by the same interpretations.

\paragraph{$\TeamLTL$}  
We now  recall  $\TeamLTL$~\cite{KrebsMV018} whose syntax 
is the same as that of standard 
$\LTL$~\cite{Pnueli77} in negation normal form where negation is applied only to atomic propositions.  
Formally, $\TeamLTL$ formulas $\varphi$   (over $\AP$)  are generated by the following grammar: 
\[
\varphi ::=   p  \mid \neg p \mid  \varphi \vee \varphi  \mid  \varphi \wedge \varphi \mid \Next \varphi     \mid \varphi \Until \varphi \mid \varphi \Release \varphi
\]
\noindent where $p\in \AP$  and  $\Next$, $\Until$  and $\Release$ are the \emph{next}, \emph{until},  and
\emph{release} temporal modalities, respectively. The logical constants $\top$ and $\bot$ are defined as usual (e.g., $\bot\DefinedAs p\wedge \neg p$). 
We also use the following abbreviations:
$\Future\varphi \DefinedAs\top \Until \varphi$ (\emph{eventually}) 
and  
$\Always\varphi\DefinedAs\bot \Release \varphi$ (\emph{always}).
 $\TeamLTL$ formulas are interpreted over \emph{arbitrary} (possibly infinite) sets $\Lang$ of traces (also called \emph{teams} in the terminology of $\TeamLTL$).
The satisfaction relation $\Lang\models \varphi$, meaning that the set of traces $\Lang$ satisfies formula $\varphi$, 
is inductively 
defined as follows: 
\[
\begin{array}{ll}
 \Lang \models  p  &  \iff \text{for each $w\in\Lang$,\,} p\in w(0)\\
     \Lang \models  \neg p  &  \iff \text{for each $w\in\Lang$,\,}  p\notin w(0)\\
     \Lang \models  \varphi_1\vee\varphi_2  &  \iff \text{for some teams $\Lang_1,\Lang_2$ with $\Lang=\Lang_1\cup\Lang_2$: }  \Lang_1 \models  \varphi_1 \text{ and }  \Lang_2 \models  \varphi_2\\
     \Lang \models  \varphi_1\wedge\varphi_2  &  \iff \Lang \models  \varphi_1 \text{ and }  \Lang \models  \varphi_2\\
     \Lang \models  \Next\varphi &  \iff \Lang_{\geq 1}\models  \varphi\\  
     \Lang \models  \varphi_1\Until \varphi_2  &
  \iff \text{for some $i\geq 0$}: \Lang_{\geq i}
  \models  \varphi_2
  \text{ and }    \Lang_{\geq k} \models   \varphi_1 \text{ for all }0\leq k<i \\
  \Lang \models  \varphi_1\Release \varphi_2  &
 \iff  \text{for each $i\geq 0$}: \Lang_{\geq i}
  \models  \varphi_2
  \text{ or }   \Lang_{\geq k} \models   \varphi_1 \text{ for some }0\leq k< i 
\end{array}
\]
The semantics of disjunction $\varphi_1 \vee \varphi_2$ in $\TeamLTL$ is the standard \emph{split disjunction} of team semantics~\cite{Vaananen07}, requiring that the given team $\Lang$ can be decomposed into two subteams $\Lang_1$ and $\Lang_2$ covering $\Lang$ (i.e., $\Lang = \Lang_1 \cup \Lang_2$) such that $\Lang_1 \models \varphi_1$ and $\Lang_2 \models \varphi_2$. It is worth noting that while in $\LTL$ the logical constant $\bot$ has no model, in $\TeamLTL$, $\bot$ has as its unique model the empty team.

We consider some semantical properties of formulas $\varphi$ from the team and inquisitive semantics literature:
\begin{itemize}
  \item  \emph{Downward closed}: if $\Lang \models \varphi$ and $\Lang'\subseteq \Lang$ then $\Lang'\models \varphi$.
  \item \emph{Empty team satisfaction}: $\emptyset\models \varphi$.
  \item \emph{Flatness}: $\Lang \models \varphi$ iff $w \models_{\LTL} \varphi$ for each $w\in \Lang$.
    \item \emph{Singleton equivalence}: $w \models_{\LTL} \varphi$ iff $\{w\} \models \varphi$ for each trace $w$.
\end{itemize}
\noindent where $w \models_{\LTL} \varphi$ means that the trace $w$ satisfies $\varphi$ under the standard $\LTL$ semantics~\cite{Pnueli77}. 
One can easily check that $\TeamLTL$ formulas satisfy downward closure, singleton
equivalence, and empty  team satisfaction properties~\cite{KrebsMV018,VirtemaHFK021}. However, $\TeamLTL$ formulas do not satisfy the flatness property in general.   
A standard  example~\cite{VirtemaHFK021} is the formula  $\Future p$ asserting that there is a timestamp $i$ such that $p$  uniformly holds at position $i$ along each trace of the given team $\Lang$.
This formula is not flat (even if the teams are assumed to be finite) and cannot be 
expressed in $\HLTL$~\cite{BozzelliMP15}. Due to the downward closure property and singleton
equivalence, $\TeamLTL$ satisfiability corresponds to standard $\LTL$ satisfiability (which is \PSPACE-complete~\cite{SistlaC85}).
On the other hand, the decidability status of model checking $\TeamLTL$ is an intriguing open question.

We also consider the known extension 
 $\TeamLTL(\StrongNeg)$ of $\TeamLTL$ with the \textit{contradictory negation}, or \emph{strong} negation, which is denoted by $\StrongNeg$ to
distinguish it from $\neg$~\cite{Luck20}. The semantics of $\StrongNeg$ is as follows:
\[
     \Lang \models  \StrongNeg \varphi    \iff \Lang \not\models   \varphi
\]
\noindent Note that for each atomic proposition $p$, $\StrongNeg p$ and $\neg p$ are not equivalent. In particular,
 $\StrongNeg p$ does not satisfy downward closure. Moreover, $\StrongNeg \top$ characterizes the nonempty teams.  
 It is known that the model checking
and satisfiability problems for $\TeamLTL(\StrongNeg)$  are highly undecidable, being equivalent to truth in third-order arithmetic~\cite{Luck20}.

\section{Inquisitive $\LTL$}\label{sec:InquisitiveLTL}

In this section, we introduce \emph{Inquisitive $\LTL$} ($\InqLTL$) which can be seen as the natural $\LTL$ counterpart 
of inquisitive first-order logic~\cite{ciardelli2009inquisitive,Ciardelli2016a,Ciardelli2022,CiardelliG22,grilletti2019disjunction,Grilletti2021,LitakS25}.  
Like $\TeamLTL$, $\InqLTL$ provides an alternative semantics for $\LTL$ where the interpretations are sets of traces (teams). 
The main differences between $\TeamLTL$ and $\InqLTL$ are the intuitionistic  semantics of implication
in $\InqLTL $ and  the replacement of the split disjunction connective $\vee$ of $\TeamLTL$ with  inquisitive disjunction in $\InqLTL$.
In the literature on team temporal logics, inquisitive disjunction is denoted by $\StrongDis$. 

Crucially, the terminology \emph{inquisitive} reflects a foundational shift in how propositions and models are interpreted.
In classical semantics, a proposition simply denotes a truth condition (a set of possible worlds or traces), and a model is a single execution.
By contrast, in inquisitive logic~\cite{Ciardelli2016a,Ciardelli2022}, propositions embody \emph{issues} or \emph{questions}, and teams are not merely combinatorial collections of traces, but represent \emph{information states}.
An information state $\Lang$ supports a formula $\varphi$ when the information available in $\Lang$ is sufficient to \emph{settle} the issue raised by $\varphi$.
In this setting, the inquisitive disjunction $\varphi_1 \StrongDis \varphi_2$ raises an alternative—requiring the current state of information to settle $\varphi_1$ or to settle $\varphi_2$—while the intuitionistic implication $\varphi_1 \rightarrow \varphi_2$ captures dependency between issues, ensuring that any enhancement of information that settles the antecedent $\varphi_1$ also settles the consequent $\varphi_2$. Regarding modal operators, see Remark \ref{rem:decl} for differences between our setup and those proposed so far in the inquisitive literature. 

\paragraph{Syntax and semantics of $\InqLTL$}  Formally, formulas $\varphi$ of $\InqLTL$ (over $\AP$) are generated by the following grammar: 
\[
\varphi ::=  \bot \mid  p    \mid  \varphi \StrongDis \varphi  \mid  \varphi \wedge \varphi \mid  \varphi \rightarrow \varphi \mid \Next \varphi     \mid \varphi \Until \varphi \mid \varphi \Release \varphi
\]
where  $p\in\AP$. 
Negation of $\varphi$ is defined as $\neg \varphi\DefinedAs \varphi \rightarrow \bot$. 
For a set $\Lang$ of traces, the satisfaction relation $\Lang\models \varphi$, meaning that the set of traces $\Lang$ satisfies formula $\varphi$, 
is
inductively defined as follows (the semantics of atomic propositions, conjunction, and temporal operators is the same as $\TeamLTL$):
\[
\begin{array}{ll}
    \Lang \models  \bot   &  \iff  \Lang=\emptyset\\
      \Lang \models  \varphi_1\StrongDis\varphi_2  &   \iff \Lang \models  \varphi_1 \text{ or }  \Lang \models  \varphi_2\\
   \Lang \models  \varphi_1 \rightarrow \varphi_2  &   \iff \text{for all $\Lang'\subseteq \Lang$, } \Lang' \models  \varphi_1 \text{ implies }  \Lang' \models  \varphi_2 
\end{array}
\]
Note that $\varphi_1 \rightarrow \varphi_2$ is checked at all the subsets (subteams) of the given team $\Lang$. Moreover,
$\Lang \models \neg\varphi$ iff for each $\Lang' \subseteq \Lang$ with $\Lang'\neq \emptyset$, $\Lang'\not\models \varphi$.
Intuitively, a team $\Lang$ represents an information state, where smaller subteams $\Lang' \subseteq \Lang$ reflect an increase in information (i.e., ruling out uncertainty about system executions).
Crucially, the \emph{downward closure} property (Proposition~\ref{prop:BasicPropertiesInqLTL}) ensures that information is persistent: once an issue is settled at state $\Lang$, refining the team to $\Lang' \subseteq \Lang$ cannot destroy this established knowledge.
In this light, the intuitionistic implication $\varphi_1 \rightarrow \varphi_2$ requires that the dependency between $\varphi_1$ and $\varphi_2$ is robust under information refinement: whenever any enhancement of knowledge resolves the antecedent $\varphi_1$, it must necessarily establish the consequent $\varphi_2$ as well.
From a system-theoretic standpoint, this guarantees that the specified requirement is preserved under refinement, preventing properties from being 
vacuously satisfied simply because the overall system behavior does not globally validate the precondition $\varphi_1$.

We also consider the extension $\InqLTL(\StrongNeg)$ of $\InqLTL$ with strong negation  $\StrongNeg$. 
\emph{Classical} implication with antecedent $\varphi_1$ and consequent $\varphi_2$ is expressible in $\InqLTL(\StrongNeg)$
as $\StrongNeg \varphi_1 \StrongDis \varphi_2$.  Moreover, note that in $\InqLTL(\StrongNeg)$, the temporal modalities $\Until$ and $\Release$ are interdefinable. 
We now establish some basic properties of $\InqLTL$.

\begin{prop}\label{prop:BasicPropertiesInqLTL}  Like $\TeamLTL$, 
$\InqLTL$ formulas satisfy  downward closure, empty team satisfaction, and  singleton equivalence properties. Moreover,
for all teams $\Lang$ and $\InqLTL$ formulas $\varphi$, the following holds:
\begin{enumerate}
  \item $\Lang\models \neg\varphi$  iff  for each $w\in \Lang$, $w\not\models_{\LTL} \varphi$.
  \item $\Lang\models \neg\neg\varphi$  iff  for each $w\in \Lang$, $w\models_{\LTL} \varphi$.
\end{enumerate}
\end{prop}
\begin{proof}
The empty team and singleton equivalence properties directly follow from the semantics of $\InqLTL$. For downward closure,
the proof is by a straightforward induction on the structure of the given formula $\varphi$. We focus on the case where
$\varphi$ is of the form $\varphi = \varphi_1 \rightarrow \varphi_2$ (the other cases being similar). Let $\Lang\models \varphi_1 \rightarrow \varphi_2$ and $\Lang'\subseteq \Lang$.
We need to prove that $\Lang'\models \varphi_1 \rightarrow \varphi_2$ as well. Since each subset (subteam) of $\Lang'$ is a subteam of $\Lang$ too, the result trivially follows from the intuitionistic
semantics of implication.  For Condition~(1) of Proposition~\ref{prop:BasicPropertiesInqLTL}, it suffices to observe that
by downward closure, we have that $\Lang\models \neg\varphi$  iff 
 for each $w\in \Lang$, $\{w\}\not\models\varphi$. Hence, by the singleton equivalence property, the result follows.
 Finally, Condition~(2) directly follows from Condition~(1).
\end{proof}

Unlike $\InqLTL$, there are $\InqLTL(\StrongNeg)$ formulas that satisfy neither downward closure nor empty team satisfaction. 
An example is the $\InqLTL(\StrongNeg)$
 formula $\StrongNeg\bot$ which characterizes the nonempty teams. Since the empty team is a model of each $\InqLTL$ formula, the
 formula  $\StrongNeg\bot$ cannot be expressed in $\InqLTL$.

By downward closure and singleton equivalence, an $\InqLTL$ formula $\varphi$ is satisfiable iff $\varphi$ has a singleton model iff
$\varphi$ is satisfiable under the standard $\LTL$ semantics. Hence, the set of $\InqLTL$-satisfiable formulas coincides with the set of $\LTL$-satisfiable formulas.
 Thus, by the known \PSPACE-completeness of $\LTL$~\cite{SistlaC85}, we obtain the following result.
  
 \begin{thm}\label{theorem:SatisfiabilityAndValidity}
The satisfiability problem of $\InqLTL$ is \PSPACE-complete. 
\end{thm}



\begin{exa} Let us consider the $\InqLTL$ formula $\varphi$ given by  $\varphi \DefinedAs(\neg\neg \Future p)  \rightarrow \Future p$. Evidently, for each trace $w$,
$\{w\}\models \varphi$. However, there are teams that are not models of $\varphi$: an example is the team $\Lang=\{w_n\}_{n\geq 0}$ where for each $n$, $p$ holds exactly at position $n$ along the trace $w_n$
(i.e., $p\in w_n(i)$ iff $i=n$ for all $i,n\geq 0$).  
\end{exa}

\paragraph{Investigated fragments of $\InqLTL$} 
We consider the so called \emph{positive fragment}  and the \emph{left-positive fragment} of $\InqLTL$. The positive fragment is defined by the following grammar: 
\[
\varphi ::=  \bot \mid  p \mid  \neg p   \mid  \varphi \StrongDis \varphi  \mid  \varphi \wedge \varphi   \mid \Next \varphi     \mid \varphi \Until \varphi \mid \varphi \Release \varphi\vspace{-0.1cm}
\]
Intuitively, the positive fragment of $\InqLTL$ corresponds to the logic obtained from $\TeamLTL$ by replacing split disjunction $\vee$ with inquisitive disjunction $\StrongDis$. The left-positive fragment of $\InqLTL$ subsumes the positive fragment and is defined as follows:  
\[ 
\varphi ::=  \bot \mid  p \mid  \neg \xi   \mid  \varphi \StrongDis \varphi  \mid  \varphi \wedge \varphi \mid  \psi \rightarrow \varphi   \mid \Next \varphi     \mid \varphi \Until \varphi \mid \varphi \Release \varphi 
\]
where $\xi$ is an arbitrary $\InqLTL$ formula and the antecedent $\psi$ in the implication   $\psi \rightarrow \varphi$ is a positive $\InqLTL$ formula.
Thus, in left-positive $\InqLTL$, we allow an unrestricted use of negation $\neg$ and a restricted use of intuitionistic implication where the left operand has to  be a positive $\InqLTL$
formula whenever the right operand is distinct from $\bot$. For each $k\geq 0$, $\InqLTL_k$ denotes the fragment of $\InqLTL$  where the nesting depth of the implication connective---occurrences of negation $\neg$ are \emph{not counted}---is at most $k$.
Note that $\InqLTL_0$ and left-positive $\InqLTL_0$ coincide  and allow an unrestricted use of intuitionistic negation.

\paragraph{Derived operators in $\InqLTL$ and $\InqLTL(\StrongNeg)$} 
We now show that some well-known operators from the team logic literature can be expressed in $\InqLTL(\StrongNeg)$,   
and some can already be expressed within the left-positive fragment of $\InqLTL$. 
In particular, we recall the \emph{universal subteam quantifier} $\AQ$  and the \emph{universal 
singleton quantifier} $\AQOne$ whose semantics are as follows:
\[
\begin{array}{ll}
  \Lang \models \AQ\varphi & \iff  \Lang' \models \varphi \text{ for all } \Lang'\subseteq \Lang\\
  \Lang \models \AQOne\varphi & \iff  \{w\}\models \varphi  \text{ for all }  w\in \Lang 
\end{array}
\]
The quantifier $\AQOne$   can be expressed in left-positive $\InqLTL$ as follows: $\AQOne \varphi \equiv \neg\neg \varphi$.

Concerning the universal subteam quantifier $\mathbf{A}$, while formally definable as $\AQ\varphi\equiv\top\rightarrow\varphi$, it does not provide additional expressive power in  $\InqLTL$ due to downward closure: $\mathcal{L}\models\varphi$ implies $\mathcal{L}'\models\varphi$ for all $\mathcal{L}'\subseteq\mathcal{L}$, collapsing $\mathbf{A}\varphi$ to $\varphi$ for every $\InqLTL$ formula.  The quantifier $\mathbf{A}$ becomes non-trivial only in extensions that break downward closure, such as $\InqLTL(\sim)$. Rather than unconstrained universal subteam quantification, the genuine expressive power of intuitionistic implication lies in conditional universal quantification $\psi\rightarrow\varphi$ over subteams.

By using strong negation, we can formalize in $\InqLTL(\StrongNeg)$ 
the dual $\EQ$ of $\AQ$  (\emph{existential subteam quantifier}) and the dual $\EQOne$ of $\AQOne$ (\emph{existential singleton quantifier}): $\EQ \varphi \equiv \StrongNeg \AQ \StrongNeg \varphi$ and
$\EQOne\varphi \equiv \StrongNeg \AQOne  \StrongNeg \varphi$. Note that:
\[
\begin{array}{ll}
  \Lang \models \EQ\varphi & \iff  \Lang' \models \varphi \text{ for some } \Lang'\subseteq \Lang\\
  \Lang \models \EQOne\varphi & \iff  \{w\}\models \varphi  \text{ for some }  w\in \Lang 
\end{array}
\]
Throughout the paper, we will use the notation $\Card_{\leq 1}$ as a shorthand for the left-positive $\InqLTL$ formula 
\[
\Card_{\leq 1}\DefinedAs \bigwedge_{p\in \AP} \Always (p \StrongDis \neg p)
\]
This formula characterizes the teams of
cardinality at most one: $\Lang\models \Card_{\leq 1}$ iff for all $i\geq 0$ and $w,w'\in \Lang$, $w(i)=w'(i)$.
 
\begin{rem} \label{rem:decl}
As already mentioned in the introduction, there is a significant difference between our setup and existing references on inquisitive modal logic(s). Apart from being the first reference to consider temporal inquisitive systems, we also break with the inquisitive tradition of considering only \emph{truth-conditional} or \emph{flat} modalities; not only $\Until$, but even derived operators such as $\Future$ are not \emph{declarative}. Recall that a declarative operator makes formulas prefixed by it behave in a \emph{truth-conditional} (or \emph{flat} in dependence logic terminology) way. Satisfaction of such formulas at a given $\Lang$ is fully determined by its satisfaction at all $w \in \Lang$ \cite{CiardelliO17,CiardelliO22}. On the other hand, the existing inquisitive modal approaches use neighbourhood-based semantics. Thus, the design principles in these references has not focused on capturing what one may call the modal fragment of inquisitive predicate logic.\footnote{It is to be noted that some of existing references propose  first-order correspondence languages suitable for the neighbourhood-based setting of $\LogicName{InqML}$ and its variants \cite{CiardelliO17,CiardelliO22,MeissnerO22}. Such formalisms are two-sorted and different from the standard inquisitive predicate logic $\LogicName{InqBQ}$.} 
The specific question of designing inquisitive modal formalisms in such a way (of which our system would be the first instance) and comparison with other frameworks is postponed to a separate paper currently in preparation. 
\end{rem}

\subsection{Examples of specifications in $\InqLTL$}\label{sec:SpecificationExamples}
 
 $\InqLTL$ and its left-positive fragment can express  relevant information-flow security properties. 
An example is  \emph{noninterference}~\cite{goguen1982security} which requires that all the traces that globally agree on the 
 low-security inputs also globally agree  on the low-security outputs,
independently of the values of high-security inputs. Noninterference can be expressed in left-positive
$\InqLTL$ as follows,  where $LI$ (resp., $LO$) is the set of propositions describing low-security inputs (resp., low-security outputs):
\[
\displaystyle{[\bigwedge_{p\in LI }\Always (p \StrongDis \neg p)] \rightarrow [\bigwedge_{p\in LO }\Always (p \StrongDis \neg p)]}
\]
Another example is \emph{observational determinism}~\cite{ZdancewicM03}, which states that
traces which have the same initial low inputs are indistinguishable to
a low user. This can be expressed by the left-positive $\InqLTL$ formula:
 \[
[\bigwedge_{p\in LI } (p \StrongDis \neg p)] \rightarrow [\bigwedge_{p\in LO }\Always (p \StrongDis \neg p)]
\] 
More flexible noninterference policies allow  controlled releases of secret information (information \emph{declassification}~\cite{SabelfeldS05}).
For example, a password checker must reveal whether the entered password is correct or not. Let $\varphi$ be an $\InqLTL$ formula 
describing facts about high-security inputs which may be released. Noninterference with declassification \mbox{policy $\varphi$
is expressible as:}
\[
[\varphi\wedge \bigwedge_{p\in LI }\Always (p \StrongDis \neg p)] \rightarrow [\bigwedge_{p\in LO }\Always (p \StrongDis \neg p)] 
\]
Note that unlike left-positive $\InqLTL$, unrestricted $\InqLTL$ allows expressing a hierarchy of preconditions on the current subset of traces.

\paragraph{Refinement verification} 
Unlike $\TeamLTL$ and $\HLTL$~\cite{ClarksonFKMRS14}, $\InqLTL$ allows enforcing properties on all the refinements (subsets of traces) 
of a given Kripke structure  that satisfy specific operational conditions. 
As a primary example, consider the team version of the classical response property:
\[
\Always(q \rightarrow \Future p)
\]
Under inquisitive team semantics, this  formula asserts that for each refinement $\Lang_r$, 
whenever the request $q$ occurs uniformly (i.e., $q$ holds at the current time $i$ on all the traces of $\Lang_r$), then a response $p$ will be given uniformly too
(i.e., for some $j\geq i$, $p$ holds on all the traces of $\Lang_r$). 
We emphasize that this specification, as well as the advanced hyperproperties presented below, crucially belongs to the \emph{left-positive fragment} of $\InqLTL$. 
We conjecture that this property can be expressed neither in $\TeamLTL$ nor in known extensions of $\HLTL$ such as $\HQPTL$~\cite{Rabe2016,CoenenFHH19}, 
as these formalisms lack conditional universal subteam quantification.
Beyond standard response patterns, intuitionistic implication enables $\InqLTL$ to express rich hyperproperties over system refinements that  we conjecture to exceed the expressive power of $\TeamLTL$.
\begin{itemize}
  \item \emph{Robust Uniform Liveness under Operational Restrictions:} in distributed coordination and consensus protocols, a goal $g$ must often be reached synchronously across all executions of an operational phase. While such synchronization may not hold over the entire non-deterministic system, it can be guaranteed for every operational refinement where an environment or scheduler uniformly coordinates requests at all times:
  \[
  \Big[\Always (e \StrongDis \neg e)\Big] \longrightarrow \Future g
  \]
  Since the antecedent $\Always(e \StrongDis \neg e)$ is in the positive fragment of $\InqLTL$, this formula is left-positive. It guarantees that whenever an operational restriction ensures that the decision $e$ is resolved uniformly at each step, the goal $g$ is synchronously achieved at a shared future time.
   In fact, $\TeamLTL$ lacks the intuitionistic implication needed to conditionally constrain subteam behaviors.\vspace{0.1cm}

  \item \emph{Epistemic Robustness and Knowledge Preservation under Refinement:} 
in epistemic reasoning and information-flow security, stating that an agent or a system component ``knows'' a fact $p$ at the current time means that there is no uncertainty regarding $p$, i.e., the truth value of $p$ is uniformly determined across all possible traces of the current information state ($p \StrongDis \neg p$). 
A canonical epistemic hyperproperty is the preservation of certainty (or the \emph{monotonicity of knowledge}):
\[
\Always \Big( (p \StrongDis \neg p) \longrightarrow \Always (p \StrongDis \neg p) \Big)
\]
Semantically, this specification ensures that for every refinement or decision of the agent that resolves the secret or state $p$ at the current instant, such certainty can never subsequently be lost at any future time instant. Crucially, this directly formalizes well-known principles such as \emph{no-forgetting} or \emph{perfect recall} within the team-semantics framework. 
Notice that this specification belongs to left-positive $\InqLTL$, since the antecedent $p \StrongDis \neg p$ is in the positive fragment.  
\end{itemize}

\paragraph{Expressing dependence atoms} $\TeamLTL$ is usually enriched with novel atomic statements describing properties of teams.
 The most studied ones are   \emph{dependence atoms} $\dep(\varphi_1, . . . , \varphi_n, \psi)$, where $\varphi_1,\ldots,\varphi_n,\psi$
are $\LTL$ formulas, stating that for each trace $w$ the truth value $\Val{\psi}_{w}$ 
of $\psi$ is functionally determined by the truth values $\Val{\varphi_1}_{w},\ldots, \Val{\varphi_n}_{w}$ of $\varphi_1,\ldots,\varphi_n$. Formally:
\[
\Lang \models  \dep(\varphi_1, . . . , \varphi_n, \psi) \iff \text{for all $w,w'\in\Lang$},  
      (\displaystyle{\bigwedge_{i=1}^{n}} \Val{\varphi_i}_{w}=\Val{\varphi_i}_{w'})\implies   \Val{\psi}_{w}=\Val{\psi}_{w'}
\]
The atom $\dep(\varphi_1, . . . , \varphi_n, \psi)$ is expressible in $\InqLTL$ as follows:
\[
\displaystyle{[\bigwedge_{i=1}^{n} (\neg \varphi_i  \StrongDis \neg\neg \varphi_i )] \rightarrow [(\neg \psi  \StrongDis \neg\neg \psi )] }
\]
Note that if $\varphi_i$ is propositional, then the formula above can be expressed in left-positive $\InqLTL$ by replacing $\neg\neg \varphi_i$ with $\varphi_i$.

\paragraph{Full $\InqLTL$ versus the Left-Positive Fragment}
The distinction between full $\InqLTL$ and its left-positive fragment lies in the quantification depth over subteams. In the left-positive fragment, antecedents contain no implication $\rightarrow$ (except for negated atomic propositions $\neg p$), meaning they directly constrain the subteam under evaluation without inspecting its further sub-behaviors. Full $\InqLTL$ removes this restriction, enabling properties whose antecedents quantify universally over subteams. A representative example is \emph{robust fairness}, expressed by
\[
(\neg\neg \Always\Future p) \longrightarrow \Always\Future p
\]
where the antecedent $\neg\neg \Always\Future p$ tests that every individual execution in the team satisfies the standard fairness condition, requiring implications within the antecedent that place the specification strictly outside the left-positive fragment.

\subsection{Connection with Intuitionistic Temporal Logics: embedding $\InqLTL$ into $\ITL^e$}\label{sec:EmbeddingBalbiani}

In this section, we illustrate the relationship of $\InqLTL$ with existing intuitionistic temporal frameworks. This connects with the perspective mentioned in the introduction: when inquisitive disjunction and intuitionistic implication are the primitive connectives, inquisitive logic naturally aligns with intuitionistic logic rather than classical logic. 
Specifically, $\InqLTL$ can be embedded into the intuitionistic linear temporal logic over \emph{expanding frames} ($\ITL^e$) introduced by Balbiani et al.~\cite{BalbianiBDF20}.

The syntax of $\ITL^e$ coincides with that of $\InqLTL$, with the inquisitive disjunction $\StrongDis$ in $\InqLTL$ corresponding to the intuitionistic disjunction $\vee$ in $\ITL^e$.
$\ITL^e$ formulas over $\AP$ are interpreted over expanding frames which are tuples $\Mu = \tpl{W,\le,S,V}$, where:
\begin{itemize}
  \item  $(W, \leq)$ is a partially ordered set of information states (or worlds),
  \item $S: W \to W$ is a deterministic transition function which is order-preserving, i.e., such that for all $w, v \in W$, $w \le v$ implies $S(w) \le S(v)$ (\emph{forward confluence}), and
  \item $V: W \rightarrow 2^{\AP}$ is a valuation function that is monotone with respect to $\le$: $w \le v$ implies $V(w) \subseteq V(v)$. 
\end{itemize}
The satisfaction relation $\Mu, w \models_e \varphi$ interprets intuitionistic implication and disjunction as follows:
\[
\begin{array}{ll}
\Mu, w \models_e \varphi_1 \vee \varphi_2 & \iff \Mu, w \models_e \varphi_1 \text{ or } \Mu, w \models_e \varphi_2\\
\Mu, w \models_e \varphi_1 \rightarrow \varphi_2 & \iff \text{for all } v \ge w, \; \Mu, v \models_e \varphi_1 \text{ implies } \Mu, v \models_e \varphi_2
\end{array}
\]
while the temporal modalities are evaluated along the deterministic trajectory generated by $S$: for example $\Mu, w \models_e \varphi_1 \Until \varphi_2$ iff there exists $k \ge 0$ such that $\Mu, S^k(w) \models_e \varphi_2$ and for all $0 \le i < k$, $\Mu, S^i(w) \models_e \varphi_1$.

\paragraph{Embedding $\InqLTL$ into $\ITL^e$}
Given a team $\Lang$, we can canonically associate it with an expanding frame $\Mu_\Lang = (W, \le, S, V)$ defined as follows.  
The set of states $W$ is the set of teams of the form $(\Lang')_{\geq i}$ where $\Lang'\subseteq \Lang$ and  $i\geq 0$.  
The intuitionistic partial order $\le$ corresponds to the reverse of set inclusion:  
\[
 \Lang_1 \le \Lang_2 \iff   \Lang_2 \subseteq \Lang_1
\]
Higher states in the poset thus represent refinements of information (ruling out executions). 
The transition function advances time by shifting traces:
\[
S( \Lang') =  (\Lang')_{\geq 1}
\]
Since $\Lang_2 \subseteq \Lang_1$ immediately implies $(\Lang_2)_{\geq 1} \subseteq (\Lang_1)_{\geq 1}$, the successor function $S$ satisfies the forward confluence condition. 
Finally, the valuation is defined by $p \in V(\Lang')$ iff for all $w \in \Lang'$, $p \in w(0)$, which is trivially monotone with respect to $\le$.

Let $\tau$ be the syntactic mapping from $\InqLTL$ to $\ITL^e$ which replaces each occurrence of the inquisitive disjunction $\StrongDis$ with the disjunction $\vee$ of $\ITL^e$. By a straightforward structural induction, it follows that for every  team $\Lang$ and every $\InqLTL$ formula $\varphi$:
\[
\Lang \models \varphi \iff \Mu_\Lang,  \Lang  \models_e \tau(\varphi)
\]

\paragraph{Decidability and Model Checking Considerations}
We emphasize that this embedding does not directly entail decidability or provide complexity upper bounds for $\InqLTL$. 
Although Balbiani et al.~\cite{BalbianiBDF20} proved that the satisfiability problem for $\ITL^e$ is decidable by establishing an effective finite model property over the general class of expanding frames, their result does not apply to model checking over a fixed, concrete system.
In contrast, in the team-semantics setting, the expanding frame $\Mu_\Lang$ canonically induced by a Kripke structure (or an arbitrary infinite language $\Lang$) is inherently infinite: the associated poset  contains the entire powerset lattice $(2^{\Lang}, \supseteq)$.
General finite model properties for $\ITL^e$ do not provide effective abstractions for model checking over such specific, highly structured, and infinite trace-induced frames. 
Consequently, establishing the decidability of the model checking problem requires dedicated combinatorial techniques, which in this paper we develop specifically for the left-positive fragment of $\InqLTL$. 

\subsection{Complexity and Decidability Landscape}\label{sec:ComplexityLandscape}

To contextualize the computational boundaries of $\InqLTL$ and $\InqLTL(\StrongNeg)$, we compare our results with the existing literature on synchronous team linear-time temporal logics~\cite{KrebsMV018,Luck20,VirtemaHFK021}. 
A fundamental dividing line in synchronous team semantics is drawn by the \emph{downward closure} property (see Section~\ref{sec:Preliminaries} and Proposition~\ref{prop:BasicPropertiesInqLTL}). 
When downward closure holds, the satisfiability problem uniformly collapses; conversely, settling the model-checking problem has proven challenging, remaining an intriguing open question for most fragments.
When downward closure fails—either due to inclusion atoms~\cite{VirtemaHFK021}, contradictory negation $\StrongNeg$, or unconstrained second-order features—both satisfiability and model checking typically cross the decidability boundary, spanning from first-order arithmetic ($\Sigma_1^0$) and analytical degrees ($\Sigma_1^1$) to full second- and third-order arithmetic.

\paragraph{Downward-Closed Fragments}

Table~\ref{tab:downward-closed-fragments} summarizes the complexity landscape of model checking for known and novel downward-closed synchronous fragments. 
In team semantics, all logics—including non-downward-closed formalisms like $\TeamLTL(\StrongNeg)$—satisfy the singleton equivalence property (Section~\ref{sec:Preliminaries}), meaning that truth over a singleton team $\{w\}$ coincides with classical $\LTL$ satisfaction along the individual trace $w$~\cite{KrebsMV018,VirtemaHFK021}. 
Crucially, when a logic additionally enjoys the downward closure property, satisfiability collapses to classical trace satisfiability: a formula has a non-empty team model if and only if it is satisfied by a singleton team. 
Consequently, satisfiability for downward-closed fragments is directly reducible to classical $\LTL$ satisfiability, placing the problem in \PSPACE~\cite{SistlaC85}. 
For rich downward-closed fragments subsuming classical $\LTL$, satisfiability is \PSPACE-complete~\cite{SistlaC85}, while for more restricted sublanguages such as $\TeamLTL - \lor$, the exact complexity remains an open question (though bounded from above by \PSPACE)~\cite{KrebsMV018}. 
For this reason, an explicit column for satisfiability is omitted from Table~\ref{tab:downward-closed-fragments}.

\begin{table}[t]
\centering
\small
\begin{tabular}{@{}lll@{}}
\toprule
\textbf{Logic / Fragment} & \textbf{MC Lower Bound} & \textbf{MC Upper Bound} \\
\midrule
$\TeamLTL - \vee$ & ? & \PSPACE~\cite{KrebsMV018} \\
$\TeamLTL$ & \PSPACE-hard~\cite{SistlaC85} & ? \\
$\TeamLTL(\dep)$ & \NEXPTIME-hard~\cite{KrebsMV018} & ? \\
Left-flat $\TeamLTL(\StrongDis,\AQ_1)$ & \PSPACE-hard~\cite{KrebsMV018} & \EXPSPACE~\cite{VirtemaHFK021} \\
\midrule
\textcolor{red}{$\InqLTL_0$} & \textcolor{red}{\PSPACE-hard [Thm.~\ref{theo:ComplexityMCLeftPositiveInqLTL}]} & \textcolor{red}{\PSPACE\ [Thm.~\ref{theo:ComplexityMCLeftPositiveInqLTL}]} \\
\textcolor{red}{Left-positive $\InqLTL_k$, $k \ge 1$} & \textcolor{red}{$k$-\EXPSPACE-hard [Thm.~\ref{theo:lowerBounds}]} & \textcolor{red}{$k$-\EXPSPACE\ [Thm.~\ref{theo:ComplexityMCLeftPositiveInqLTL}]} \\
\textcolor{red}{$\InqLTL_k$,   $k \ge 1$} & \textcolor{red}{$k$-\EXPSPACE-hard} & \textcolor{red}{?} \\
\bottomrule
\end{tabular}
\caption{Model checking (MC) complexity for downward-closed synchronous team linear-time temporal logics, with our novel results highlighted in red (? denotes an open problem). Satisfiability for all listed fragments is in \PSPACE\ (and \PSPACE-complete whenever standard $\LTL$ is subsumed) via singleton equivalence and downward closure.}
\label{tab:downward-closed-fragments}
\end{table}

The main complexity and decidability results summarized in Table~\ref{tab:downward-closed-fragments}—where our novel contributions are highlighted in red and open problems are marked by ?—can be detailed as follows:
\begin{itemize}
  \item Split-disjunction-free $\TeamLTL$ ($\TeamLTL - \vee$): as shown by Krebs et al.~\cite{KrebsMV018}, in the absence of split disjunction, synchronized macro-states $S_0, S_1, \dots$ across a Kripke structure follow a deterministic and ultimately periodic trace bounded by $2^{|S|}$. Model checking can thus be decided in \PSPACE\ via an on-the-fly B\"{u}chi automaton construction tracking macro-states~\cite{KrebsMV018}.
  \item $\TeamLTL$ and $\TeamLTL(\dep)$: the addition of split disjunction $\vee$ has left the decidability of full $\TeamLTL$ model checking open since 2018~\cite{KrebsMV018}. The presence of temporal modalities interacting with $\vee$ requires partitioning potentially uncountable sets of traces at every time step. The extension $\TeamLTL(\dep)$ of $\TeamLTL$ with dependence atoms (the latter have been recalled in Section~\ref{sec:SpecificationExamples}) retains downward closure; a \NEXPTIME\ lower bound has been proved in~\cite{KrebsMV018} (via reduction from propositional dependence logic validity), while the upper bound remains entirely open.
  \item  Left-flat $\TeamLTL(\StrongDis, \AQ_1)$: Virtema et al.~\cite{VirtemaHFK021} considered $\TeamLTL$ extended with inquisitive disjunction $\StrongDis$ and universal singleton quantification $\AQ_1$. In the original definition, the \emph{left-flat} fragment is formulated using until and weak until, requiring the left operand of every such binary temporal subformula to be flat. Adapted to our release-based syntax, this translates to requiring that the left operand $\psi$ of $\psi \Until \varphi$ and the right operand $\psi$ of $\varphi \Release \psi$ be flat according to the definition given in Section~\ref{sec:Preliminaries}. Note that a formula $\psi$ is flat iff $\psi \equiv \AQ_1\psi$. Model checking for this fragment is known to be decidable in \EXPSPACE\ via a reduction to the $\exists_p^* \forall_\pi$-fragment of \HQPTL~\cite{VirtemaHFK021}.
  \item $\InqLTL_0$: for implication-free $\InqLTL$ formulas (where intuitionistic negation $\neg$ can still be used without restrictions), $\InqLTL_0$ and left-positive $\InqLTL_0$ coincide. As stated in Theorem~\ref{theo:ComplexityMCLeftPositiveInqLTL}, model checking $\InqLTL_0$ is \PSPACE-complete, matching the complexity of classical $\LTL$~\cite{SistlaC85}.
  \item Left-positive $\InqLTL$: in left-positive $\InqLTL$, the split disjunction of $\TeamLTL$ is replaced by inquisitive disjunction $\StrongDis$ alongside intuitionistic implication, while disallowing the left-hand nesting of implication (except for negation $\neg p \equiv p \to \bot$ over atomic propositions). As established in Section~\ref{sec:DecidabilityResults} and Section~\ref{sec:LowerBounds}, model checking is decidable and nonelementary, and its complexity is precisely $k$-\EXPSPACE-complete when the nesting depth of implication is bounded by $k \ge 1$ (Theorem~\ref{theo:ComplexityMCLeftPositiveInqLTL} and Theorem~\ref{theo:lowerBounds}).
  \item $\InqLTL$: when removing the syntactic restriction on the antecedents of implication, full $\InqLTL$ (resp., $\InqLTL_k$) directly inherits the nonelementary (resp., $k$-\EXPSPACE) lower bound from its left-positive fragment. However, the exact decidability status for model checking unrestricted $\InqLTL$ remains an intriguing open question.
\end{itemize}

\paragraph{Non-Downward-Closed Fragments and Undecidability Boundaries}

Table~\ref{tab:non-downward-closed-fragments} summarizes the landscape for non-downward-closed synchronous  team logics, detailing both model checking and satisfiability.
In these formalisms, while the singleton equivalence property still holds, the failure of downward closure prevents satisfiability from collapsing to classical $\LTL$ satisfiability. 

\begin{table}[t]
\centering
\small
\begin{tabular}{@{}lll@{}}
\toprule
\textbf{Logic / Fragment} & \textbf{Model Checking} & \textbf{Satisfiability} \\
\midrule
$\TeamLTL(\subseteq, \StrongDis)$ & $\Sigma_1^0$-hard~\cite{VirtemaHFK021} & $\Sigma_1^0$-hard~\cite{VirtemaHFK021} \\
$\TeamLTL(\subseteq, \StrongDis, \AQ)$ & $\Sigma_1^1$-hard~\cite{VirtemaHFK021} & $\Sigma_1^1$-hard~\cite{VirtemaHFK021} \\
\multirow{2}{*}{$\TeamLTL(\StrongNeg)$} & Complete for & Complete for \\
 & third-order arithmetic~\cite{Luck20} & third-order arithmetic~\cite{Luck20} \\
$k$-coherent $\TeamLTL(\StrongNeg)$ & in \EXPSPACE~\cite{VirtemaHFK021} & in \EXPSPACE~\cite{VirtemaHFK021} \\
\midrule
\multirow{2}{*}{\textcolor{red}{$\InqLTL(\StrongNeg)$}} & \textcolor{red}{Hard for second-order} & \textcolor{red}{Hard for second-order} \\
 & \textcolor{red}{arithmetic [Thm.~\ref{theo:UndecidabilityStrongInqLTL}]} & \textcolor{red}{arithmetic [Thm.~\ref{theo:UndecidabilityStrongInqLTL}]} \\
\bottomrule
\end{tabular}
\caption{Complexity of model checking and satisfiability for non-downward-closed synchronous team logics, with our novel results highlighted in red.}
\label{tab:non-downward-closed-fragments}
\end{table}

\begin{itemize}
  \item $\TeamLTL(\subseteq, \StrongDis)$: following Virtema et al.~\cite{VirtemaHFK021}, this logic extends $\TeamLTL$ with inquisitive/Boolean disjunction $\StrongDis$ and inclusion atoms 
 $\varphi_1,\ldots,\varphi_n \subseteq  \psi_1,\ldots,\psi_n$ with $\varphi_1,\ldots,\varphi_n,\psi_1,\ldots,\psi_n$ being $\LTL$ formulas.  Given a team $\Lang$, the inclusion atom requires that for every trace $w \in \Lang$ there exists a trace $w' \in \Lang$ such that $w \models \varphi_i \iff w' \models \psi_i$ for all $i \in \{1, \dots, n\}$. Inclusion atoms break downward closure, and as shown in~\cite{VirtemaHFK021}, they enable the simulation of lossy 3-counter machines, leading to $\Sigma_1^0$-hardness for both model checking and satisfiability of $\TeamLTL(\subseteq, \StrongDis)$.
  \item $\TeamLTL(\subseteq, \StrongDis,\AQ)$: when universal subteam quantification $\AQ$ is added to $\TeamLTL(\subseteq, \StrongDis)$, Virtema et al.~\cite{VirtemaHFK021} showed that one can enforce the encoding of the recurrence problem for deterministic 2-counter machines, pushing both model checking and satisfiability to analytical undecidability ($\Sigma_1^1$-hardness).
  \item $\TeamLTL(\StrongNeg)$:  L\"{u}ck~\cite{Luck20} demonstrated that the combination of split disjunction $\lor$ and contradictory negation $\StrongNeg$ allows quantifying over arbitrary subteams and sets of subteams, encoding truth in third-order arithmetic.
  \item $k$-coherent $\TeamLTL(\StrongNeg)$: Virtema et al.~\cite{VirtemaHFK021} considered the semantic fragment of $\TeamLTL(\StrongNeg)$ consisting of \emph{$k$-coherent} formulas (for a fixed $k \ge 1$), that is, formulas $\varphi$ such that for every team $\Lang$, $\Lang \models \varphi$ if and only if $\Lang' \models \varphi$ for all $\Lang' \subseteq \Lang$ with $|\Lang'| \le k$. Model checking and satisfiability for this fragment is decidable in \EXPSPACE\ via a reduction to the universal fragment of $\HLTL$~\cite{VirtemaHFK021}.
       \item $\InqLTL(\StrongNeg)$:  in Section~\ref{sec:ExpressivenessResults} and Section~\ref{sec:UndecidabilityResults}, we show that while $\InqLTL(\StrongNeg)$ is highly undecidable (hard for second-order arithmetic), it  retains the Countable Model Property (Proposition~\ref{prop:CountableModelProperty}). Consequently, $\InqLTL(\StrongNeg)$ avoids the divergence into third-order arithmetic and is strictly less expressive than $\TeamLTL(\StrongNeg)$ (Proposition~\ref{prop:FromInqLTLtoTeamLTL}).
\end{itemize}

\section{Expressiveness results on $\InqLTL$ and $\InqLTL(\StrongNeg)$}\label{sec:ExpressivenessResults} 

In this section, we address expressiveness issues about the novel logics $\InqLTL$ and $\InqLTL(\StrongNeg)$. As a main
result, we establish that  $\InqLTL(\StrongNeg)$ is strictly less expressive than $\TeamLTL(\StrongNeg)$.

In Section~\ref{sec:InquisitiveLTL}, we have observed that unlike $\InqLTL$, there are $\InqLTL(\StrongNeg)$ formulas that satisfy neither downward closure nor empty team satisfaction. 
In particular, the formula  $\StrongNeg\bot$ cannot be expressed in $\InqLTL$. Hence, the following holds.

\begin{thm} 
$\InqLTL$ is strictly less expressive than $\InqLTL(\StrongNeg)$.  
\end{thm} 

Next, we show  that each satisfiable 
$\InqLTL(\StrongNeg)$ formula $\varphi$ has a countable model, that is, a countable team satisfying $\varphi$. 
In fact, we prove a stronger result by using a normal form of $\InqLTL(\StrongNeg)$ which is defined as follows: 
\[
\varphi ::=   \bot  \mid \StrongNeg \bot \mid  p \mid    \StrongNeg p   \mid  \varphi \StrongDis \varphi  \mid  \varphi \wedge \varphi \mid    \AQ \varphi \mid  \EQ \varphi   \mid  
  \Next \varphi  \mid \varphi \Until \varphi \mid \varphi \Release \varphi
\]
We observe that this normal form is expressively complete for $\InqLTL(\StrongNeg)$. Indeed,
$\varphi_1 \rightarrow \varphi_2 \equiv \AQ (\StrongNeg \varphi_1\StrongDis \varphi_2)$. Hence, 
since modalities $\AQ$ and $\EQ$ are duals and the temporal modalities $\Until$ and $\Release$ are duals, by pushing strong negation $\StrongNeg$ inward, one can convert an $\InqLTL(\StrongNeg)$ formula $\varphi$ into an equivalent $\InqLTL(\StrongNeg)$ formula in normal form.

\begin{prop}[Countable model property]\label{prop:CountableModelProperty} Let $\varphi$ be an $\InqLTL(\StrongNeg)$ formula.
Then, for each uncountable model $\Lang_u$ of $\varphi$, there is a countable model $\Lang_c$ of $\varphi$ such that 
$\Lang_c\subseteq \Lang_u$ and for each team $\Lang$ such that $\Lang_c\subseteq\Lang\subseteq \Lang_u$, $\Lang$ is still a model of $\varphi$.
\end{prop}
\begin{proof}
Without loss of generality, we can assume that the given $\InqLTL(\StrongNeg)$ formula $\varphi$ is in normal form. Let $\Lang_u$ be an uncountable model of $\varphi$. We prove Proposition~\ref{prop:CountableModelProperty}
  by induction on the structure of $\varphi$. 
  \begin{itemize}
    \item$\varphi\in \{\bot,\StrongNeg\bot\}$ or  $\varphi=p$ or $\varphi=\StrongNeg p$   with $p\in \AP$: these cases are trivial. 
    \item $\varphi = \varphi_1\StrongDis \varphi_2$: hence,  $\Lang_u \models \varphi_1$ or $\Lang_u \models \varphi_2$, and the result directly follows from  the induction hypothesis.
    \item $\varphi = \varphi_1\wedge \varphi_2$: hence,  $\Lang_u \models \varphi_1$ and $\Lang_u \models \varphi_2$. By the induction hypothesis, for each $i=1,2$, there is a countable 
    model $\Lang_{c,i}$ of $\varphi_i$ such that $\Lang_{c,i}\subseteq \Lang_u$ and each team $\Lang$ satisfying $\Lang_{c,i}\subseteq\Lang\subseteq \Lang_u$ is still a model of $\varphi_i$. We define $\Lang_c\DefinedAs \Lang_{c,1}\cup \Lang_{c,2}$, and the result follows. 
    \item $\varphi = \AQ\varphi_1$ or $\varphi = \EQ\varphi_1$: if  $\varphi = \AQ\varphi_1$, then for each $\Lang\subseteq \Lang_u$, $\Lang\models \varphi_1$. Hence, each countable subset $\Lang_c$ of $\Lang_u$ satisfies the thesis of  Proposition~\ref{prop:CountableModelProperty}. If instead $\varphi = \EQ\varphi_1$, then  
 by the induction hypothesis, there exists a countable model $\Lang_c$ of $\varphi_1$ such that $\Lang_c\subseteq \Lang_u$. Moreover, by the semantics of $\EQ$,
    each team $\Lang$ satisfying $\Lang_{c}\subseteq\Lang\subseteq \Lang_u$ is   a model of $\varphi$. 
 \item The root operator of $\varphi$ is a temporal modality. We consider the case where 
 $\varphi$ is of the form $\varphi_1\Release \varphi_2$  (the other cases being similar).
 Since $\Lang_u$ is a model of $\varphi$, either (i) $(\Lang_u)_{\geq i}\models \varphi_2$ for each $i\geq 0$, or (ii) there is $k\geq 0$ such that $(\Lang_u)_{\geq k}\models \varphi_1$ and
 $(\Lang_u)_{\geq i}\models \varphi_2$ for each $0\leq i\leq k$. We focus on the first case (the other case being similar). By the induction hypothesis, for each $i\geq 0$, there is a countable subteam 
 $\Lang_{c,i}$ of $(\Lang_u)_{\geq i}$ such that each team $\Lang$ satisfying $\Lang_{c,i}\subseteq\Lang\subseteq  (\Lang_u)_{\geq i}$ is a model of $\varphi_2$. For each $i\geq 0$, let $\Lang'_{c,i}$ be any subset of
 $\Lang_u$  such that  $(\Lang'_{c,i})_{\geq i}= \Lang_{c,i}$. Since the number of words of length $i$ over the finite alphabet $2^{\AP}$ is finite,  $\Lang'_{c,i}$ is countable.
 We set $\Lang_c\DefinedAs \bigcup_{i\geq 0} \Lang'_{c,i}$.  Since the countable union of countable sets is still countable, by construction the result follows.\qedhere
  \end{itemize}   
\end{proof} 

It is known that satisfiability of $\TeamLTL(\StrongNeg)$
is hard for truth in third-order arithmetics~\cite{Luck20}. The proof   in~\cite{Luck20} entails the existence of satisfiable formulas whose unique models are uncountable.
Here, we give a direct proof.

\begin{prop}\label{prop:UncountableModelPropertyTeamLTLNeg}There exists a satisfiable $\TeamLTL(\StrongNeg)$ formula whose models are all uncountable.
\end{prop}
\begin{proof}
In the proof, we exploit intuitionistic implication $\rightarrow$, inquisitive disjunction $\StrongDis$, and  the subteam quantifiers $\AQ$, $\EQ$, and $\EQ_1$ which can be easily expressed in 
  $\TeamLTL(\StrongNeg)$ as seen in Section~\ref{sec:InquisitiveLTL}.
Let $\AP =\{1,2,\#\}$. A \emph{$1$-trace} is a trace $w$ of the form $\{1\}^{n-1} \{\#,1\} \{1\}^\omega$
for some $n\in\Nat$ (i.e., proposition $1$ holds at each position and $\#$ holds exactly at position $n$): the encoding $enc_1(w)$ of $w$ is the natural number $n$.
A \emph{$2$-trace} is a trace $w$ over $2^{\{2,\#\}}$ such that $2\in w(i)$ for each $i\geq 0$. The trace $w$ encodes the set $enc_2(w)$ of natural numbers $n$ such that
$\#\in w(n)$. Given a $2$-trace $w$ and a set $\Lang$ of $1$-traces, we say that $w$ \emph{encodes} $\Lang$ if   $enc_2(w)$ coincides with the set of natural numbers
encoded by the traces in $\Lang$. 

For each $\ell\in\{1,2\}$, let $\Gamma_\ell$ the set of all $\ell$-traces and $\Lang_{all}\DefinedAs \Gamma_1\cup \Gamma_2$. Since $\Gamma_2$ contains the encodings of all the subsets of natural
numbers, $\Lang_{all}$ is uncountable. We construct a $\TeamLTL(\StrongNeg)$ formula $\varphi_{all}$ whose unique model is $\Lang_{all}$. Hence, the result follows.
For the definition of $\varphi_{all}$, we need two preliminary results, where a team $\Lang$ is \emph{consistent} if it only contains $1$-traces and $2$-traces. The proof of the following
claim is straightforward.

\paragraph{Claim 1} One can construct a $\TeamLTL(\StrongNeg)$ formula $\varphi_{con}$ characterizing the consistent teams.\vspace{0.2cm}

\noindent Next, we show the following.  
 
\paragraph{Claim 2} One can construct a $\TeamLTL(\StrongNeg)$ formula $\varphi_{check}$ such that for each consistent team $\Lang$,
$\Lang\models \varphi_{check}$ iff there is a unique $2$-trace $w$ in $\Lang$, and this unique trace  encodes $\Lang\cap \Gamma_1$.\vspace{0.2cm} 

\noindent Formula $\varphi_{check}$ in Claim~2 is defined as follows.
 \[
 \begin{array}{ll}
 \varphi_{check} \DefinedAs & \varphi_{sing,2} \wedge [(\varphi_{sing,1}\wedge \varphi_{sing,2}) \rightarrow \Future \#] \wedge  \Always[(\StrongNeg \EQ_1(2\wedge \#)) \StrongDis \EQ_1(1\wedge \#)]\vspace{0.2cm}\\
 \varphi_{sing,\ell} \DefinedAs & (\EQ_1\ell) \wedge (\ell \rightarrow \Card_{\leq 1}) \quad   \text{ for each }\ell\in  \{1,2\}
 \end{array}
 \]
Formula $ \varphi_{sing,\ell}$ requires that the current consistent  team contains exactly one $\ell$-trace.
Now, let us consider the formula $\psi_2$ given by  $\psi_2 = \StrongNeg (2 \vee \varphi_{check})$ (recall that $\vee$ is split disjunction).   
Given a consistent team $\Lang$, by Claim~2, the previous formula asserts that there is \emph{no} $2$-trace $w$ in $\Lang$ which encodes 
$\Lang\cap \Gamma_1$. Hence, given a consistent team $\Lang$, the  formula  $1 \vee \psi_2$, asserts that there is a subteam $\Lang_1\subseteq \Lang\cap \Gamma_1$
of $1$-traces such that no $2$-trace in $\Lang$ encodes $\Lang_1$. Therefore, the desired formula $\varphi_{all}$ is defined as follows.   
\[ 
\Always \EQ_1(1\wedge \#)\wedge \varphi_{con}\wedge  \StrongNeg (1 \vee [\StrongNeg (2 \vee \varphi_{check})]). 
\]
Note that the first conjunct ensures that the consistent team contains all the $1$-traces, and we are done. 
\end{proof} 

By using Propositions~\ref{prop:CountableModelProperty} and~\ref{prop:UncountableModelPropertyTeamLTLNeg}, we easily deduce the following expressiveness result.

\begin{prop}\label{prop:FromInqLTLtoTeamLTL}   $\InqLTL(\StrongNeg)$ is strictly  less expressive than $\TeamLTL(\StrongNeg)$.
 \end{prop}
 \begin{proof} We first show that $\InqLTL(\StrongNeg)$ is  subsumed by $\TeamLTL(\StrongNeg)$.
 It suffices to 
  observe that $\varphi_1 \rightarrow \varphi_2 \equiv \AQ (\StrongNeg \varphi_1\StrongDis \varphi_2)$,
 $\varphi_1\StrongDis \varphi_2 \equiv \StrongNeg(\StrongNeg \varphi_1 \wedge \StrongNeg \varphi_2)$, $\AQ \varphi \equiv \StrongNeg \EQ \StrongNeg \varphi$,
 and $\EQ \varphi \equiv \top \vee \varphi$ (recall that $\vee$ is split disjunction in $\TeamLTL$). 
Hence, each  $\InqLTL(\StrongNeg)$ formula can be converted in linear time into an equivalent 
 $\TeamLTL(\StrongNeg)$ formula.

\noindent It remains to prove that there are $\TeamLTL(\StrongNeg)$ formulas which cannot be expressed in $\InqLTL(\StrongNeg)$. By Proposition~\ref{prop:UncountableModelPropertyTeamLTLNeg}, there exists satisfiable $\TeamLTL(\StrongNeg)$
formulas whose unique models are uncountable. 
On the other hand, by Proposition~\ref{prop:CountableModelProperty},  each satisfiable  $\InqLTL(\StrongNeg)$ formula has a countable model. Hence, the result follows.  
 \end{proof} 
 
 \paragraph{Expressiveness Comparison between $\InqLTL$ and $\TeamLTL$} 
While Proposition~\ref{prop:FromInqLTLtoTeamLTL} establishes the strict inclusion $\InqLTL(\StrongNeg) \subsetneq \TeamLTL(\StrongNeg)$, determining the exact relative expressive power between pure $\InqLTL$ and pure $\TeamLTL$ remains a fundamental open problem. 

Both formalisms satisfy downward closure, empty team satisfaction, and singleton equivalence (Proposition~\ref{prop:BasicPropertiesInqLTL} and~\cite{KrebsMV018}).
However, the two logics govern subteams through fundamentally orthogonal mechanisms:
\begin{itemize}
  \item In $\TeamLTL$, the core connective is the split disjunction $\vee$, which inherently embodies an \emph{existential decomposition} $\Lang = \Lang_1 \cup \Lang_2$ of the current team.
  \item In $\InqLTL$, split disjunction is replaced by inquisitive disjunction $\StrongDis$, while intuitionistic implication $\rightarrow$ natively provides an intrinsic mechanism
  for conditional universal quantification over subteams (or system refinements), constraining all subteams that satisfy a given antecedent.
\end{itemize}
Based on this foundational duality, we formulate the following conjectures regarding their expressiveness.

\begin{conj}\label{conj:IncomparabilityInqLTLTeamLTL}
$\InqLTL$ and $\TeamLTL$ are expressively incomparable:
\begin{enumerate}
  \item There exists an $\InqLTL$ formula (even in the left-positive fragment) that cannot be expressed in $\TeamLTL$.
  \item There exists a $\TeamLTL$ formula that cannot be expressed in $\InqLTL$.
\end{enumerate}
\end{conj}

Regarding item (1) of Conjecture~\ref{conj:IncomparabilityInqLTLTeamLTL}, natural candidates are properties requiring conditional universal quantification over operational subteams, such as the uniform response property $\Always(q \rightarrow \Future p)$ introduced in Section~\ref{sec:SpecificationExamples}. Since $\TeamLTL$ can only decompose teams existentially via $\vee$, it lacks the means to conditionally constrain all synchronous subteams that resolve a precondition without resorting to contradictory negation $\StrongNeg$. A similar intuition applies to generalized dependence atoms $\dep(\varphi_1, \dots, \varphi_n, \psi)$ over arbitrary $\LTL$ formulas, which are directly expressible in $\InqLTL$ (Section~\ref{sec:SpecificationExamples}) but are conjectured not to be encodable in pure $\TeamLTL$.

For item (2) of Conjecture~\ref{conj:IncomparabilityInqLTLTeamLTL}, we conjecture a substantially stronger separation: the existential splitting mechanism of $\TeamLTL$ cannot be subsumed even when allowing full contradictory negation $\StrongNeg$ in inquisitive logic.

\begin{conj}\label{conj:TeamLTLNotInqLTLNeg}
$\TeamLTL$ is not subsumed by $\InqLTL(\StrongNeg)$, i.e., there exists a $\TeamLTL$ formula that is not equivalent to any $\InqLTL(\StrongNeg)$ formula.
\end{conj}

A natural candidate to witness Conjecture~\ref{conj:TeamLTLNotInqLTLNeg} is the simple split formula
\[
\varphi_{\vee} \DefinedAs \Future p \vee \Future p
\]
Notice that semantic equivalence between team formulas requires agreement on \emph{arbitrary} (possibly uncountable) teams. 
To justify Conjecture~\ref{conj:TeamLTLNotInqLTLNeg}, we observe that each trace $w$ over the singleton alphabet $\AP = \{p\}$ encodes the subset $A_w$ of natural numbers defined by $k \in A_w \iff p \in w(k)$. 
Under this identification, an arbitrary team $\Lang$ satisfies $\Future p$ iff there exists a common index $n \in \Nat$ belonging to the intersection of all sets associated with the traces of $\Lang$, i.e., $n \in \bigcap_{w \in \Lang} A_w$.
Now, let $n_1, n_2 \in \Nat$ be two distinct timestamps with $n_1 \neq n_2$. We can define two uncountable families $S_1, S_2 \subseteq 2^{\Nat}$ of subsets of $\Nat$ such that:
\[
\bigcap_{A \in S_1} A = \{n_1\} \quad \text{and} \quad \bigcap_{A \in S_2} A = \{n_2\}
\]
Let $\Lang_1 = \{w \mid A_w \in S_1\}$ and $\Lang_2 = \{w \mid A_w \in S_2\}$ be the corresponding uncountable teams, and consider the team $\Lang = \Lang_1 \cup \Lang_2$.
Because $n_1 \neq n_2$, the global intersection $\bigcap_{C \in S_1 \cup S_2} C =\emptyset$ is empty, so that $\Lang$ fails to satisfy $\Future p$. 
However, $\Lang \models \Future p \vee \Future p$, as witnessed by the covering $\Lang = \Lang_1 \cup \Lang_2$.
Crucially, evaluating $\Future p \vee \Future p$ on such uncountable teams intrinsically requires an existential choice over uncountably many traces to isolate two sets of continuum size with disjoint uniform witnesses. 
In contrast, $\InqLTL(\StrongNeg)$ lacks split disjunction and can only quantify over subteams individually via $\AQ$ and $\EQ$, without the capability to non-deterministically partition an uncountable team into complementary components. This fundamental discrepancy strongly supports the conjecture that $\Future p \vee \Future p$ cannot be captured within $\InqLTL(\StrongNeg)$. 
\section{Undecidability of $\InqLTL(\StrongNeg)$}\label{sec:UndecidabilityResults}

In this section, we show that model checking and satisfiability of $\InqLTL(\StrongNeg)$ are highly undecidable since they can encode truth in second-order arithmetic.

Recall that second-order arithmetic (see e.g.~\cite{Rogers87}) is second-order predicate logic with equality over the signature \mbox{$(<,+,*,\in)$} evaluated over
the set $\Nat$ of natural numbers, where  $<$ is interpreted as the standard ordering over $\Nat$, $+$ and $*$ are interpreted as standard addition and multiplication in
$\Nat$, respectively, and $\in$ is the set membership operator. Note that first-order variables (denoted by the lower-case letters $x,y,z,\ldots$) 
range over natural numbers, while second-order variables (denoted by the upper-case letters $X,Y,Z,\ldots$) 
range over sets of natural numbers. W.l.o.g.~we assume that arithmetical formulas are in prenex normal form, i.e., consisting of a prefix of  existential ($\exists$) or universal ($\forall$) quantifiers, applied to  first-order or second-order variables, followed by a
Boolean combination of atomic formulas of the form $x<y$ or $x=y+z$ or $x=y*z$ or $x\in X$ for first-order variables $x,y,z$  and second-order variables $X$. 
 Truth in second-order arithmetic is the decision problem consisting of checking whether an arithmetical sentence (i.e., a formula with no free variables) is true over $\Nat$.

In the following, we fix an arithmetic sentence
\[
\Phi= Q_1 \nu_1\ldots  Q_k \nu_k.\, \Psi
\]
where $\Psi$ is quantifier-free and for each $1\leq i\leq k$, $Q_i\in \{\exists,\forall\}$  and $\nu_i$ is a first-order or second-order variable. We build an $\InqLTL(\StrongNeg)$ formula $\Enc(\Phi)$ which is satisfiable iff $\Phi$ is true over $\Nat$.
Moreover, at the end of the section, we show that $\InqLTL(\StrongNeg)$ model checking is at least as hard as $\InqLTL(\StrongNeg)$ satisfiability.

\paragraph{Encoding of natural numbers and arithmetic operations} Given an atomic proposition $p$,
each natural number $n\in\Nat$ can be encoded by the trace over $2^{\{p\}}$ where proposition
$p$ holds exactly at position $n$. Subsets of natural numbers 
can then be encoded by teams consisting of traces of the previous form. However, in the valuation of the quantifier prefix
$Q_1 \nu_1\ldots  Q_k \nu_k$ of the given arithmetic sentence $\Phi$, we need to distinguish the natural numbers (resp., the sets of natural numbers) which are assigned   
to distinct first-order variables (resp., distinct second-order variables). This justifies the following definition.
Let $\AP_{num} \DefinedAs \{\nu_1,\ldots,\nu_k, \#\}$. For each $1\leq i\leq k$, a \emph{$\nu_i$-trace} is a trace of the form 
\[
\{\nu_i\}^{n-1} \{\#,\nu_i\} \{\nu_i\}^\omega
\]
for some $n\in\Nat$ (i.e., $\nu_i$ holds at each position and $\#$ holds exactly at position $n$). The  previous trace encodes the natural number $n$.

Another important feature in the reduction is the encoding of addition and multiplication in $\InqLTL(\StrongNeg)$.
To this end, we use a coloured variant of the encoding considered in~\cite{DFrenkel025} for second-order $\HLTL$.
In this latter logic, there is explicit first-order quantification over traces and propositions are parameterized by first-order variables  so that one can distinguish the single traces that compose a team of traces. Since this is not directly possible in $\InqLTL(\StrongNeg)$,  we use different colors to distinguish the traces in the encoding. 
 
Let $\AP_{arith} \DefinedAs \{arg_1, arg_2,res, +,*, 0,1\}$. For all $c\in \{0,1\}$ and $op\in \{+,*\}$, an \emph{$op$-trace with colour $c$} is a trace $w$ 
over $2^{\{c,op,arg_1,arg_2,res\}}$ satisfying the following:
\begin{itemize}
  \item $c\in w(i)$ and $op\in w(i)$ for all $i\geq 0$;
  \item there are unique $n_1,n_2,n_3\in\Nat$ with $arg_1\in w(n_1)$, $arg_2 \in w(n_2)$, and $res \in w(n_3)$. We write $arg_1(w)$ (resp., $arg_2(w)$) to mean $n_1$
  (resp., $n_2$), and $res(w)$ to mean $n_3$.
\end{itemize} 

\noindent An \emph{$op$-trace} is an  $op$-trace with colour $c$ for some $c\in \{0,1\}$. An $op$-trace $w$ is \emph{\underline{well-formed}} if $res(w)=arg_1(w)+arg_2(w)$ when $op$ is $+$, and
   $res(w)=arg_1(w)*arg_2(w)$ otherwise  (i.e., $op$ is $*$). 
   
\noindent The  set $\AP$ of  atomic  propositions  used in the reduction is then defined as follows:
\[
\AP\DefinedAs \AP_{num}\cup \AP_{arith}.
\]   
\begin{figure}[t]
\centering
\begin{tikzpicture}[xscale=1.05, yscale=0.9, >=stealth]
  \tikzstyle{tstep}=[circle, fill=black!10, draw=black, inner sep=1.5pt]

  \node[anchor=east, font=\footnotesize] at (-0.3, 2.2) {$\nu_i$-trace ($n=3$):};
  \draw[->, thick, gray] (0,2.2) -- (8.5,2.2);
  \foreach \t in {0,1,2,3,4,5,6,7} {
    \node[tstep] (n\t) at (\t, 2.2) {};
    \node[below=2pt of n\t, font=\tiny, text=gray] {$t=\t$};
  }
  \node[right=2pt of n7, font=\scriptsize] {$\dots$};
  \node[above=2pt of n0, font=\scriptsize] {$\nu_i$};
  \node[above=2pt of n1, font=\scriptsize] {$\nu_i$};
  \node[above=2pt of n2, font=\scriptsize] {$\nu_i$};
  \node[above=2pt of n3, font=\scriptsize, text=blue!80!black] {$\{\nu_i, \#\}$};
  \node[above=2pt of n4, font=\scriptsize] {$\nu_i$};
  \node[above=2pt of n5, font=\scriptsize] {$\nu_i$};
  \node[above=2pt of n6, font=\scriptsize] {$\nu_i$};
  \node[above=2pt of n7, font=\scriptsize] {$\nu_i$};

  \draw[dashed, gray!40] (-0.5, 1.1) -- (8.5, 1.1);

  \node[anchor=east, font=\footnotesize] at (-0.3, 0) {well-formed $+$-trace ($c=0$):};
  \draw[->, thick, gray] (0,0) -- (8.5,0) node[right, font=\tiny, text=gray] {$t$};
  \foreach \t in {0,1,2,3,4,5,6,7} {
    \node[tstep] (o\t) at (\t, 0) {};
    \node[below=2pt of o\t, font=\tiny, text=gray] {$t=\t$};
  }
  \node[right=2pt of o7, font=\scriptsize] {$\dots$};

  \node[above=2pt of o0, font=\scriptsize, text=black!70] {$\{+, 0\}^\omega$};
  \node[above=2pt of o2, font=\scriptsize, fill=red!15, rounded corners=2pt] {$arg_1$};
  \node[above=2pt of o4, font=\scriptsize, fill=green!20, rounded corners=2pt] {$arg_2$};
  \node[above=2pt of o6, font=\scriptsize, fill=cyan!20, rounded corners=2pt] {$res$};

  \draw[<-, thick, red!70!black] (2, -0.65) -- (2, -1.15) node[below, font=\scriptsize] {$arg_1=2$};
  \draw[<-, thick, green!50!black] (4, -0.65) -- (4, -1.15) node[below, font=\scriptsize] {$arg_2=4$};
  \draw[<-, thick, cyan!70!black] (6, -0.65) -- (6, -1.15) node[below, font=\scriptsize] {$res=6$};
\end{tikzpicture}
\caption{Encoding of natural numbers via a $\nu_i$-trace (top) and a \textit{well-formed} $+$-trace with colour $c=0$ encoding $2 + 4 = 6$ (bottom).}
\label{fig:traces_encoding}
\end{figure}
 A trace $w$ is \emph{consistent} if either
 $w$ is a $\nu_i$-trace for some $1\leq i\leq k$, or $w$ is an $op$-trace for some $op\in \{+,*\}$ (note that we do \emph{not} require that the $op$-trace is well-formed). 
A team is \emph{consistent} if it contains only consistent traces.  
One can trivially construct an $\LTL$ formula $\psi_{con}$ characterizing the consistent traces. Hence,  $\varphi_{con}\DefinedAs\neg\neg\psi_{con}$
characterizes the consistent teams  under the $\InqLTL$ semantics.  Moreover, for each $1 \leq i\leq k$, let $\Lang_{all}^{\nu_i}$ be the team consisting of all $\nu_i$-traces. Evidently,
$\Lang_{all}^{\nu_i}$ is the unique model of the formula 
\[
\varphi_{all}^{\nu_i}:=\varphi_{con} \wedge \Always (\nu_i \wedge  \EQOne \#).
\]
Note that $\EQOne\#$ asserts the existence of a trace $w$ belonging to the given team
such that $\{w\}\models \#$ (i.e., $\#\in w(0)$). Thus, we obtain the following result.
  
\begin{prop}\label{prop:ConsistentTeams} One can build two $\InqLTL(\StrongNeg)$ formulas $\varphi_{con}$ and $\varphi_{all}^{\nu_i}$
such that $\varphi_{con}$ captures the consistent teams and $\varphi_{all}^{\nu_i}$ has a unique model which is $\Lang_{all}^{\nu_i}$.  
\end{prop}

Let $\Lang_{arith}$ be the team consisting of  all and only 
the well-formed $+$-traces and the well-formed $*$-traces. The following result will allow us to implement 
addition and multiplication in $\InqLTL(\StrongNeg)$.

\begin{prop}\label{prop:CapturingOperation} One can construct an $\InqLTL(\StrongNeg)$ formula $\varphi_{arith}$  such that
$\Lang\models \varphi_{arith}$ \emph{iff} $\Lang$ is a consistent team whose set of $+$-traces and $*$-traces is $\Lang_{arith}$. 
\end{prop}
 \begin{proof} 
Formula $\varphi_{arith}$ is defined as follows: 
\[
\displaystyle{\varphi_{arith}  \DefinedAs \varphi_{con} \wedge \bigwedge_{op\in \{+,*\}} (\varphi^{op}_{all} \wedge \varphi^{op}_{wf}) }
\]
where $\varphi_{con}$ is the formula of Proposition~\ref{prop:ConsistentTeams}  capturing the consistent teams.
Conjunct $\varphi^{op}_{all}$ ensures that for all natural numbers $n_1$ and $n_2$ and for each colour $c\in \{0,1\}$, there is an $op$-trace $w$ with colour $c$ whose 
arguments $arg_1(w)$ and $arg_2(w)$ are $n_1$ and $n_2$, respectively: 
\[
   \displaystyle{\varphi^{op}_{all} \DefinedAs \bigwedge_{c\in \{0,1\}}\bigwedge_{\ell\in \{1,2\}} \Always\EQ(op\wedge c\wedge arg_\ell \wedge \Always\EQOne arg_{3-\ell}).}
\]
Conjuncts $\varphi^{+}_{wf}$ and $\varphi^{*}_{wf}$ activate recursion by encoding the inductive definition of addition and multiplication.

\paragraph{Definition of $\varphi^{+}_{wf}$}  We use the  auxiliary 
formula  $\theta^+_{0,1}$ requiring that for each consistent team $\Lang$, $\Lang$ consists of one $+$-trace with colour 
$0$ and one $+$-trace with colour $1$:  
\[   
 \displaystyle{ \theta^+_{0,1} \DefinedAs + \wedge \bigwedge_{c\in \{0,1\}}(\EQOne c \wedge   (c \rightarrow \Card_{\leq 1})).}
\]
Then, the formula $\varphi^{+}_{wf}$ enforces the following requirements for each colour  $c\in\{0,1\}$.
\begin{itemize}
\item For each $+$-trace $w$ with colour $c$ such that  $arg_1(w)=arg_2(w)=0$, it holds that   $res(w)=0$. This is trivially expressible.  
\item For all $\ell \in \{1,2\}$, $+$-traces $w$ with colour $c$ and $+$-traces $w'$ with colour $1-c$ such that $arg_\ell(w)=arg_\ell(w')$ and 
     $arg_{3-\ell}(w')=arg_{3-\ell}(w)+1$, it holds that $res(w')=res(w)+1$. This can be expressed as  follows:
\[
\begin{array}{l}
  \displaystyle{\bigwedge_{c\in \{0,1\}}\bigwedge_{\ell\in \{1,2\}}}  ([\Future arg_\ell \wedge \theta^+_{0,1} \wedge  \psi(c,arg_{3-\ell}) ]  \rightarrow  \psi(c,res) )\vspace{0.1cm}\\  
  \psi(c,p)\DefinedAs \Future (\EQOne(c\wedge p) \wedge \Next \EQOne((1-c)\wedge p))
\end{array}
\]
\end{itemize}
Note that we use two distinct colours $c$ and $1-c$  in  order to ensure that for the two compared $+$-traces,  the one having the greatest argument $arg_{3-\ell}$ is  also the one having the
greatest result $res$. Figure~\ref{fig:inductive_addition} illustrates the encoding of the inductive definition of addition.

\begin{figure}[t]
\centering
\begin{tikzpicture}[xscale=1.1, yscale=0.95, >=stealth]
  \tikzstyle{tstep}=[circle, fill=black!10, draw=black, inner sep=1.4pt]

  \node[anchor=east, font=\footnotesize] at (-0.2, 1.8) {$w$ (colour $0$):};
  \draw[->, thick, gray] (0,1.8) -- (6.5,1.8);
  \foreach \t in {0,1,2,3,4,5} \node[tstep] (w\t) at (\t, 1.8) {};
  \node[above=2pt of w1, font=\scriptsize, fill=red!15, rounded corners=2pt] {$arg_1$};
  \node[above=2pt of w2, font=\scriptsize, fill=green!20, rounded corners=2pt] {$arg_2$};
  \node[above=2pt of w3, font=\scriptsize, fill=cyan!20, rounded corners=2pt] {$res$};

  \node[anchor=east, font=\footnotesize] at (-0.2, 0) {$w'$ (colour $1$):};
  \draw[->, thick, gray] (0,0) -- (6.5,0);
  \foreach \t in {0,1,2,3,4,5} {
    \node[tstep] (wp\t) at (\t, 0) {};
    \node[below=2pt of wp\t, font=\tiny, text=gray] {$t=\t$};
  }

  \node[below=14pt of wp1, font=\scriptsize, fill=red!15, rounded corners=2pt] {$arg_1$};
  \node[below=14pt of wp3, font=\scriptsize, fill=green!20, rounded corners=2pt] {$arg_2$};
  \node[below=14pt of wp4, font=\scriptsize, fill=cyan!20, rounded corners=2pt] {$res$};

  \draw[<->, dashed, thick, red!80!black] (w1) -- (wp1);

  \draw[->, thick, green!60!black] (w2.south) to[out=-40, in=140] (wp3.north);

  \draw[->, thick, cyan!80!black] (w3.south) to[out=-40, in=140] (wp4.north);

\end{tikzpicture}
\caption{The inductive step enforcing  $res(w') = res(w) + 1$ whenever 
$arg_1(w) = arg_1(w')$ and $arg_2(w') = arg_2(w) + 1$, synchronised across opposite colours via subteams satisfying $\theta^+_{0,1}$.}
\label{fig:inductive_addition}
\end{figure}

\paragraph{Definition of $\varphi^{*}_{wf}$} We now illustrate the definition of the conjunct
  $\varphi^{*}_{wf}$ of $\varphi_{arith}$ which encodes the inductive definition of multiplication based on the correct implementation of addition (this is ensured by 
  the conjunct $\varphi^{+}_{wf}$).  We  exploit the  auxiliary formula  $\theta^*_{0,1,+}$ requiring that for the given consistent team $\Lang$, $\Lang$ consists of one $*$-trace with color    $0$, one $*$-trace with color $1$, and one $+$-trace with arbitrary color. The definition of formula $\theta^*_{0,1,+}$ is similar to the definition of the auxiliary formula $\theta^+_{0,1}$ in the construction 
   of $\varphi^{+}_{wf}$, and we omit the details here.
   Then, the formula $\varphi^{*}_{wf}$ ensures that  for each color $c\in\{0,1\}$, the following two requirements hold.
\begin{itemize}
     \item For each $*$-trace $w$ with color $c$ such that $arg_1(w)=arg_2(w)=0$, it holds that $res(w)=0$. This can be trivially expressed.
     \item Let $\ell \in \{1,2\}$,  $w$ be a $*$-trace with color $c$,   $w'$ a   $*$-trace with color $1-c$, and $w_+$ a $+$-trace. If $arg_\ell(w)=arg_\ell(w')=arg_\ell(w_+)$,   
     $arg_{3-\ell}(w')=arg_{3-\ell}(w)+1$, and $arg_{3-\ell}(w_+)=res(w)$, then $res(w')=res(w_+)$. Note that by the conjunct 
     $\varphi^{+}_{wf}$, we can assume that $res(w_+)= arg_1(w_+)+arg_2(w_+)$.  Thus, the previous requirement ensures that $res(w')= res(w) + arg_\ell(w)$.
     The requirement can be expressed as follows:
   \[
    \begin{array}{l}
    \displaystyle{\bigwedge_{c\in \{0,1\}}\bigwedge_{\ell\in \{1,2\}}} \bigl([\Future arg_\ell \wedge \theta^*_{0,1,+} \wedge  \psi_1(c,\ell) \wedge \psi_2(c,\ell) ]    \,\rightarrow\, \psi_3(c) \bigr)\vspace{0.1cm}\\  
    \psi_1(c,\ell)\DefinedAs \Future (\EQOne(c\wedge * \wedge arg_{3-\ell}) \, \wedge  \,\Next \EQOne((1-c)\wedge * \wedge arg_{3-\ell}))\vspace{0.1cm}\\
    \psi_2(c,\ell)\DefinedAs \Future (\EQOne(c\wedge * \wedge res) \wedge  \EQOne(+ \wedge arg_{3-\ell}))\vspace{0.1cm}\\
    \psi_3(c)\DefinedAs \Future (\EQOne(+ \wedge res) \wedge  \EQOne(* \wedge (1-c) \wedge res)) 
    \end{array}
   \]
   Note that $\psi_1(c,\ell)$ requires that $arg_{3-\ell}(w')=arg_{3-\ell}(w)+1$, $\psi_2(c,\ell)$ requires that $arg_{3-\ell}(w_+)=res(w)$, and
   $\psi_3$ ensures that $res(w')=res(w_+)$.
\end{itemize}   
This concludes the proof of Proposition~\ref{prop:CapturingOperation}.
\end{proof}

Let $\Lang_{all}$ be the team defined as follows:
\[
  \Lang_{all}\DefinedAs \Lang_{arith}\cup \bigcup_{i=1}^{i=k}\Lang_{all}^{\nu_i}.
\]
Note that for each variable $\nu_i$, $\Lang_{all}$ contains all the $\nu_i$-traces. 
 By Propositions~\ref{prop:ConsistentTeams}--\ref{prop:CapturingOperation},  $\Lang_{all}$ is the unique model of the $\InqLTL(\StrongNeg)$ formula $\varphi_{all}$
\[
 \varphi_{all}\DefinedAs \varphi_{con}\wedge \varphi_{arith}\wedge \bigwedge_{i=1}^{i=k}\EQ \varphi_{all}^{\nu_i}.
 \]
 Hence, the following holds.
 
\begin{prop}\label{prop:CapturingWellFormedTeam} One can build an $\InqLTL(\StrongNeg)$ formula $\varphi_{all}$ whose unique model is $\Lang_{all}$. 
\end{prop}

\paragraph{Encoding of variable valuations} Let $g$ be a variable valuation over the set $\{\nu_1,\ldots,\nu_k\}$, i.e., a mapping assigning to each variable $\nu_i$ a natural number if
$\nu_i$ is a first-order variable, and a subset of natural numbers otherwise. We encode $g$ by the consistent team 
\[
\Lang_g \DefinedAs \Lang_{arith}\cup \Lang'_g,
\]
where $\Lang'_g $ does not contain $+$-traces and $*$-traces and  the following holds for each variable $\nu_i$:
\begin{itemize}
\item if $\nu_i$ is a first-order variable, then $\Lang'_g$ contains exactly one $\nu_i$-trace. Moreover, this trace encodes the natural number $g(\nu_i)$.
\item If instead $\nu_i$ is a second-order variable, then
 the set of natural numbers encoded by the $\nu_i$-traces which belong to $\Lang'_g$ is exactly $g(\nu_i)$.
\end{itemize} 
By exploiting the previous encoding, we now show how to express the evaluation of quantifier-free arithmetic formulas over $\{\nu_1,\ldots,\nu_k\}$ in  $\InqLTL(\StrongNeg)$.

\begin{prop}\label{prop:CapturingQuantifierFreeArith} Given a quantifier-free arithmetic formula $\Psi$ with variables in  $\{\nu_1,\ldots,\nu_k\}$, one can construct 
an $\InqLTL(\StrongNeg)$ formula $\Enc(\Psi)$ such that for each variable valuation $g$ it holds that 
$g$ satisfies $\Psi$ \emph{iff} $\Lang_g\models \Enc(\Psi)$.
\end{prop}
\begin{proof}  
For each nonempty set $P\subseteq \{\nu_1,\ldots,\nu_k,+,*\}$, we first build  an $\InqLTL(\StrongNeg)$ formula $\chi_P$ such that a consistent team $\Lang$ is a model 
of $\chi_P$ \emph{iff} $\Lang$ has cardinality $|P|$ and for each $t\in P$, there is exactly  one $t$-trace in $\Lang$:   
\[
\displaystyle{\chi_P \DefinedAs (\AQOne \bigvee_{t\in P}\,t) \wedge \bigwedge_{t\in P}(\EQOne t \wedge (t \rightarrow \Card_{\leq 1})).}
\] 
Fix a quantifier-free arithmetic formula $\Psi$ with variables in  $\{\nu_1,\ldots,\nu_k\}$. Since Boolean connectives can be expressed in $\InqLTL(\StrongNeg)$, w.l.o.g.~we can assume that 
$\Psi$ is an atomic formula. There are the following cases:
\begin{itemize}
\item $\Psi$ is of the form $x<y$: 
\[
\Enc(\Psi)\DefinedAs \EQ[\chi_{\{x,y\}} \wedge \Future (\EQOne (x\wedge \#) \wedge \Next \Future\EQOne(y\wedge \#) ) ].
\]
\item $\Psi$ is of the form $x=y+z$:   
\[
\Enc(\Psi)\DefinedAs \EQ \bigl(\chi_{\{x,y,z,+\}} \wedge \Future [\EQOne (x\wedge \#) \wedge  \EQOne res ] \wedge  
                           \Future [\EQOne (y\wedge \#) \wedge  \EQOne arg_1 ]\wedge \Future [\EQOne (z\wedge \#) \wedge  \EQOne arg_2 ] \bigr).
\]
\item $\Psi$ is of the form $x=y*z$: this case is similar to the previous one. 
\item $\Psi$ is of the form $x\in X$: 
\[
\Enc(\Psi)\DefinedAs \EQ[\chi_{\{x,X\}} \wedge \Future \#].
\]
\end{itemize}
Correctness of the construction easily follows  from the definitions of the team encodings  $\Lang_g$ of variable valuation  $g$.
\end{proof}

\begin{figure}[t]
\centering
\begin{tikzpicture}[xscale=1.1, yscale=0.9, >=stealth]
  \tikzstyle{teamnode}=[draw, rounded corners=3pt, thick, align=center, inner sep=5pt, font=\small]
  \tikzstyle{blocknode}=[draw, dashed, rounded corners=2pt, inner sep=3pt, font=\scriptsize]

  \node[teamnode, fill=gray!10] (all) at (0, 0) {
    $\Lang_{all} \;=\; \Lang_{arith} \;\cup\; \Lang_{all}^{\nu_1} \;\cup\; \dots \;\cup\; \Lang_{all}^{\nu_k}$
  };

  \draw[->, thick] (0, -0.6) -- (0, -1.4) 
    node[midway, right, font=\footnotesize] {$\theta_1 \colon Q_1$-choice on $\nu_1$-traces};

  \node[teamnode, fill=blue!5] (l1) at (0, -2.0) {
    $\Lang_1 \;=\; \Lang_{arith} \;\cup\; T_1 \;\cup\; \Lang_{all}^{\nu_2} \;\cup\; \dots \;\cup\; \Lang_{all}^{\nu_k}$\\[2pt]
    \scriptsize ($T_1 \subseteq \Lang_{all}^{\nu_1}$, singleton if $\nu_1$ is first-order)
  };

  \draw[->, thick, dashed] (0, -2.7) -- (0, -3.5)
    node[midway, right, font=\footnotesize] {inductive elimination of $\nu_2, \dots, \nu_{k-1}$};

  \node[teamnode, fill=green!10] (lg) at (0, -4.1) {
    $\Lang_g \;=\; \Lang_{arith} \;\cup\; T_1 \;\cup\; \dots \;\cup\; T_k$\\[2pt]
    \scriptsize ($T_i$ encodes valuation $g(\nu_i)$ for each $1 \le i \le k$)
  };

  \draw[->, thick] (0, -4.8) -- (0, -5.5);
  \node[draw, thick, fill=yellow!15, rounded corners=3pt, font=\small, inner sep=4pt] at (0, -5.9) {
    Quantifier-free arithmetical check: $\Lang_g \models \Enc(\Psi)$
  };

\end{tikzpicture}
\caption{Schematic progression of subteam selection simulating the quantifier prefix $Q_1 \nu_1 \dots Q_k \nu_k$, isolating a team $\Lang_g$ that encodes a variable valuation $g$.}
\label{fig:quantifier_elimination}
\end{figure}

\paragraph{Expressing truth of the arithmetic sentence $\Phi$ in $\InqLTL(\StrongNeg)$ }  For the given arithmetic sentence $\Phi =Q_1 \nu_1\ldots  Q_k \nu_k.\,\Psi $, where 
$\Psi$ is quantifier-free, the arithmetical quantifiers $Q_i\nu_i$ are emulated in $\InqLTL(\StrongNeg)$ as follows. We start with the consistent team 
$\Lang_{all}$ which is the unique model of the  $\InqLTL(\StrongNeg)$ formula $\varphi_{all}$ of Proposition~\ref{prop:CapturingWellFormedTeam}. Recall that 
$\Lang_{all}= \Lang_{arith}\cup \bigcup_{i=1}^{i=k}\Lang_{all}^{\nu_i}$.
 Then by exploiting the 
$\InqLTL(\StrongNeg)$ formulas $\varphi_{all}^{\nu_1},\ldots,\varphi_{all}^{\nu_k},\varphi_{arith}$ of Propositions~\ref{prop:ConsistentTeams} and~\ref{prop:CapturingOperation},
we can select, existentially or universally (depending on the polarity of $Q_1\in\{\exists,\forall\}$), a subteam $\Lang_1\subseteq \Lang_{all}$ of the form 
$\Lang_1= (\Lang_{all}\setminus \Lang_{all}^{\nu_1})\cup T_1$ where $T_1\subseteq \Lang_{all}^{\nu_1}$ and $T_1$ is a singleton if $\nu_1$ is a first-order variable. Then, we proceed with the team $\Lang_1$ and apply the previous procedure by selecting a subteam of the form
$(\Lang_1\setminus \Lang_{all}^{\nu_2})\cup T_2$ where  $T_2\subseteq \Lang_{all}^{\nu_2}$ and $T_2$ is a singleton if $\nu_2$ is a first-order variable, and so on.
At the end of this process, we obtain a subteam $\Lang_g$ of $\Lang_{all}$ encoding  a variable valuation $g$ on $\{\nu_1,\ldots,\nu_k\}$. 

Let $\Enc(\Psi)$ be the $\InqLTL(\StrongNeg)$ formula of Proposition~\ref{prop:CapturingQuantifierFreeArith} for the quantifier-free arithmetic formula $\Psi$. Moreover, let 
$\theta_{k+1},\ldots,\theta_{1}$ be the $\InqLTL(\StrongNeg)$ formulas  defined as follows: 
\[
\theta_{k+1}\DefinedAs \Enc(\Psi)
\]
and for each $i=k,\ldots,1$: 
\[
    \begin{array}{ll}
        \theta_{i}\DefinedAs &
        \begin{cases}
            \EQ \,\xi_i & \text{if $Q_i$ is $\exists$}\\
            \AQ \,\xi_i & \text{otherwise } 
        \end{cases}    \quad     \xi_i \DefinedAs \theta_{i+1}\wedge \varphi_{arith } \wedge select(\nu_i) \wedge   \displaystyle{\bigwedge_{\ell=i+1}^{\ell = k}}\EQ\, \varphi_{all}^{\nu_\ell}\vspace{0.2cm}\\
         select(\nu_i)\DefinedAs &
        \begin{cases}
             (\nu_i \rightarrow \Card_{\leq 1}) \wedge \EQ_1\, \nu_i  & \text{if $\nu_i$ is first-order}\\
            \top & \text{otherwise }
        \end{cases} 
    \end{array}   
\]
Figure~\ref{fig:quantifier_elimination} illustrates the progression of subteam selection in simulating arithmetic quantification in the prefix of $\Phi$. Finally, define 
\[
\Enc(\Phi)\DefinedAs \varphi_{all}\wedge \theta_1.
\]
 By Propositions~\ref{prop:ConsistentTeams}--\ref{prop:CapturingQuantifierFreeArith}, we obtain 
that $\Enc(\Phi)$ is satisfiable \emph{iff} $\Lang_{all}$ is a model of $\Enc(\Phi)$ iff $\Phi$ is true over $\Nat$. 
Now, let us consider the Kripke structure $\KS_{full}=\tpl{2^{\AP},2^{\AP},2^{\AP}\times 2^{\AP},\Lab}$, where $\Lab$ is the identity mapping. Since $\Lang(\KS_{full})$ is the set of all the 
traces (over $\AP$), an
$\InqLTL(\StrongNeg)$ formula $\theta$ is satisfiable \emph{iff} $\KS_{full}\models \EQ \theta$. Hence, $\InqLTL(\StrongNeg)$ satisfiability is reducible to 
$\InqLTL(\StrongNeg)$ model checking.  Thus, we obtain the following result.

\begin{thm}\label{theo:UndecidabilityStrongInqLTL} Satisfiability and model checking  of $\InqLTL(\StrongNeg)$ are undecidable. In particular, 
the truth in second-order arithmetic is reducible to satisfiability (resp., model checking) of $\InqLTL(\StrongNeg)$.   
\end{thm}

\begin{rem}
For ease of presentation, the proposed reduction uses a number of propositions which depends on the length $k$ of the quantifier
prefix. Indeed, recall that 
\[
\AP\DefinedAs \AP_{num}\cup \AP_{arith}\DefinedAs \{arg_1, arg_2,res, +,*, 0,1,\nu_1,\ldots,\nu_k, \#\}.
\]
A slightly  more involved reduction allows us to use a fixed and small set of propositions. The idea is to exploit just one proposition, say $p$, instead of the
$k$ propositions $\nu_1,\ldots,\nu_k$  for encoding the different types of traces associated with distinct quantifiers: in particular, for each $1\leq i\leq k$, a $\nu_i$-trace can be replaced with a trace over $2^{\{p,\#\}}$ whose projection over $\{p\}$ is the word  $(\{p\}^i\emptyset)^\omega$.  
\end{rem}

\section{Decidability results}\label{sec:DecidabilityResults}

In this section,  we show that for left-positive $\InqLTL$, model checking  
is decidable. Moreover, we prove that for each $k\geq 0$, model checking left-positive $\InqLTL_k$ formulas
is in  $k$-\EXPSPACE. The upper bounds are obtained 
in two steps. In the first step, we define an abstract semantics of $\InqLTL$ on Kripke structures,
which we call \emph{macro-path semantics}. In this setting, for a given Kripke structure $\KS$, $\InqLTL$ formulas are interpreted 
over infinite sequences of subsets of $\KS$-states (\emph{macro-paths}), which provide a word-encoding of sets (teams) of $\KS$-paths. Not all the teams
of $\KS$-paths can be represented by macro-paths.  However, we show that for  left-positive $\InqLTL$, the macro-path
semantics captures the teams semantics over Kripke structures. Then, in the second step, we provide an automata-theoretic approach
for solving the $\InqLTL$ model-checking problem under the macro-path semantics.

\subsection{Macro-path semantics for $\InqLTL$}\label{sec:MacroStates}

Fix a Kripke structure $\KS=\tpl{S,S_0,R,\Lab}$. For a set $\Pi$ of paths  of $\KS$, we denote 
by $\Lang_{\KS}(\Pi)$ the set of traces induced by the paths in $\Pi$. For an $\InqLTL$ formula
$\varphi$, we write $\Pi\models_{\KS} \varphi$ to mean that $\Lang_{\KS}(\Pi)\models \varphi$.

A \emph{macro-state} of $\KS$ is a (possibly empty)
set $S'$ of states of $\KS$, that is $S'\subseteq S$. Given two macro-states $S'$ and $S''$, we say that \emph{$S''$ is a successor
of $S'$} if  the following  two conditions hold:
\begin{itemize}
  \item for each $s'\in S'$, there is $s''\in R[s']\cap S''$,
  \item for each $s''\in S''$, there is $s'\in S'$ such that $s''\in R[s']$.
\end{itemize}
\noindent A \emph{macro-path} $\rho$ of $\KS$ is an infinite sequence of macro-states $\rho=S_1 S_2 \ldots$ such that 
$S_{i+1}$ is a successor of $S_i$ for each $i\geq 1$. A \emph{macro-path} $\rho$ encodes the set   $\Paths_{\KS}(\rho)$ of paths
defined as follows:
\[
\text{$\Paths_{\KS}(\rho)$ is the set of  $\KS$-paths $\pi$ such that $\pi(i)\in \rho(i)$ for each $i\geq 0$.}
\]
Given two macro-paths $\rho$ and $\rho'$, we write $\rho \sqsubseteq \rho'$ to mean that for each $i\geq 0$,
$\rho(i)\subseteq \rho'(i)$. Evidently, if $\rho \sqsubseteq \rho'$, then $\Paths_{\KS}(\rho)\subseteq \Paths_{\KS}(\rho')$.  A \emph{singleton} macro-path is a macro-path $\rho$
such that $\Paths_{\KS}(\rho)$ is a singleton. Note that for a singleton macro-path $\rho$, 
$\rho(i)$ is a singleton for each $i\geq 0$.

Not all the sets of paths can be encoded by macro-paths. Intuitively,  the macro-paths can only represent \emph{memoryless}
sets $\Pi$ of paths where for all $i\geq 0$ and $\pi,\pi'\in\Pi$ such that $\pi(i)=\pi'(i)$, the following holds for all $\nu\in S^\omega$: $\pi(0)\ldots \pi(i) \cdot \nu\in \Pi$ iff
$\pi'(0)\ldots \pi'(i) \cdot \nu\in \Pi$.   As an example, assume that $S=\{s_0,s_1\}$ and $(s_i,s_j)\in R$ for all $i,j\in \{0,1\}$.
Then, there is no macro-path $\rho$ such that $\Paths_{\KS}(\rho)=\{s_0^\omega,s_1^\omega\}$. Indeed, the unique macro-path encoding a set of paths 
which contains both the paths $s_0^\omega$ and $s_1^\omega$ is  $S^\omega$, but $\Paths_{\KS}(S^\omega)$ is the set of all the paths.  

\begin{figure}[t]
\centering
\begin{tikzpicture}[
    >=Stealth,
    state/.style={circle, draw=black, thick, fill=white, inner sep=0pt, minimum size=6.5mm, font=\small},
    macro/.style={draw=black!70, dashed, rounded corners=6pt, fill=blue!6, inner sep=5pt},
    path1/.style={->, thick, color=blue!80!black},
    path2/.style={->, thick, color=orange!85!black},
    spurious/.style={->, thick, dashed, color=gray!70}
]

\node[font=\bfseries] at (1.8, 2.2) {(a) Team of paths $\Pi = \{s_0^\omega, s_1^\omega\}$};

\foreach \t in {0,1,2,3} {
    \node[state] (s0_\t) at (\t*1.2, 1.1) {$s_0$};
    \node[state] (s1_\t) at (\t*1.2, -0.1) {$s_1$};
    \node[font=\footnotesize, text=gray] at (\t*1.2, -1.5) {$t=\t$};
}
\node[font=\large] at (4.2, 1.1) {$\dots$};
\node[font=\large] at (4.2, -0.1) {$\dots$};

\foreach \t [evaluate=\t as \next using int(\t+1)] in {0,1,2} {
    \draw[path1] (s0_\t) -- (s0_\next);
    \draw[path2] (s1_\t) -- (s1_\next);
}
\draw[path1] (s0_3) -- (3.9, 1.1);
\draw[path2] (s1_3) -- (3.9, -0.1);

\node[font=\footnotesize, color=blue!80!black, anchor=west] at (4.5, 1.1) {$\pi_1 = s_0^\omega$};
\node[font=\footnotesize, color=orange!85!black, anchor=west] at (4.5, -0.1) {$\pi_2 = s_1^\omega$};

\begin{scope}[xshift=7.4cm, yshift=-0.5cm]
\node[font=\bfseries] at (1.8, 2.7) {(b) Macro-path $\MP(\Pi) = \{s_0, s_1\}^\omega$};

\foreach \t in {0,1,2,3} {
    \node[state] (m0_\t) at (\t*1.2, 1.1) {$s_0$};
    \node[state] (m1_\t) at (\t*1.2, -0.1) {$s_1$};
    \node[font=\footnotesize, text=gray] at (\t*1.2, -1.0) {$t=\t$};
    
    \begin{scope}[on background layer]
        \node[macro, fit=(m0_\t) (m1_\t), label=above:{\scriptsize $\{s_0,s_1\}$}] (box_\t) {};
    \end{scope}
}
\node[font=\large] at (4.2, 0.5) {$\dots$};

\foreach \t [evaluate=\t as \next using int(\t+1)] in {0,1,2} {
    \draw[path1] (m0_\t) -- (m0_\next);
    \draw[path2] (m1_\t) -- (m1_\next);
    \draw[spurious] (m0_\t) -- (m1_\next);
    \draw[spurious] (m1_\t) -- (m0_\next);
}
\draw[path1] (m0_3) -- (3.9, 1.1);
\draw[path2] (m1_3) -- (3.9, -0.1);
\draw[spurious] (m0_3) -- (3.9, 0.7);
\draw[spurious] (m1_3) -- (3.9, 0.3);
\end{scope}

\end{tikzpicture}
\caption{Comparison between (a) the original set of paths $\Pi = \{s_0^\omega, s_1^\omega\}$ and (b) its abstraction via the macro-path $\MP(\Pi) = \{s_0,s_1\}^\omega$. Dashed gray arrows represent spurious paths in $\Paths_{\KS}(\MP(\Pi)) \setminus \Pi$ resulting from the memoryless abstraction.}
\label{fig:MacroPathAbstraction}
\end{figure}

However, each set $\Pi$ of paths can be abstracted away by the macro-path, denoted by $\MP(\Pi)$, whose $i$-th  macro state is the collection of states associated with the $i$-th position
of the paths in $\Pi$. Formally, for each $i\geq 0$: 
\[
\MP(\Pi)(i)\DefinedAs \{\pi(i)\mid \pi\in\ \Pi\}.
\]
Note that $\Paths_{\KS}(\MP(\Pi))\supseteq \Pi$ and, in general, $\Pi\neq \Paths_{\KS}(\MP(\Pi))$. With reference to the previous example, let $\Pi = \{s_0^\omega,s_1^\omega\}$.
We have that $\MP(\Pi)= \{s_0,s_1\}^\omega$  and $\Paths_{\KS}(\MP(\Pi))$ is the set of all the paths of $\KS$. Hence, $\Pi\subset\Paths_{\KS}(\MP(\Pi))$.  
In Figure~\ref{fig:MacroPathAbstraction}, we give a graphical comparison between the set of paths $\Pi$ and its macro-path abstraction $\MP(\Pi)$.

\paragraph{Macro-path semantics} We now provide a semantics of $\InqLTL$ interpreted over macro-paths of the given Kripke structure    $\KS=\tpl{S,S_0,R,\Lab}$.
For a macro-path $\rho$ and an $\InqLTL$ formula $\varphi$,
the satisfaction relation $\rho\models_{\KS} \varphi$ is inductively defined as follows (we omit the semantics of
temporal modalities which is defined as for $\LTL$ but we replace traces $w$ with macro-paths $\rho$): 
\[
\begin{array}{ll}
\rho \models_{\KS}  \bot  &  \iff \Paths_{\KS}(\rho)= \emptyset \\   
     \rho \models_{\KS}  p  &  \iff \text{for each $s\in \rho(0)$,\,} p\in \Lab(s)\\
     \rho \models_{\KS}  \varphi_1\StrongDis\varphi_2  &  \iff \rho \models_{\KS}  \varphi_1 \text{ or }  \rho \models_{\KS}  \varphi_2 \\
    \rho \models_{\KS}  \varphi_1\wedge\varphi_2  &  \iff  \rho \models_{\KS}  \varphi_1 \text{ and }  \rho \models_{\KS}  \varphi_2\\
     \rho \models_{\KS}  \varphi_1 \rightarrow \varphi_2  &  \Leftrightarrow \text{for each macro-path $\rho' \sqsubseteq \rho$, } \rho' \models_{\KS}  \varphi_1 \text{ implies }  \rho' \models_{\KS}  \varphi_2
\details{
     \rho \models_{\KS}  \Next\varphi &  \Leftrightarrow \rho_{\geq 1}\models_{\KS}  \varphi\\
     \rho \models_{\KS}  \varphi_1\Until \varphi_2  &
  \Leftrightarrow  \text{for some $i\geq 0$}: \rho_{\geq i}
  \models_{\KS}  \varphi_2
  \text{ and }  \rho_{\geq k} \models_{\KS}   \varphi_1 \text{ for all }0\leq k<i \\
  \rho \models_{\KS}  \varphi_1\Release \varphi_2  &
  \Leftrightarrow  \text{for each $i\geq 0$}: \rho_{\geq i}
  \models_{\KS}  \varphi_2
  \text{ or }  \rho_{\geq k} \models_{\KS}   \varphi_2 \text{ for some }0\leq k< i }
\end{array}
\]
Note that $\Paths_{\KS}(\rho)= \emptyset$ iff $\rho(i)=\emptyset$ for each $i\geq 0$. The macro-path semantics is downward closed, that is for all macro-paths $\rho,\rho'$ such that $\rho \sqsubseteq \rho'$, $\rho'\models_{\KS}\varphi$ implies $\rho\models_{\KS}\varphi$.

\paragraph{Equivalence of macro-path semantics and team semantics for left-positive $\InqLTL$.} 
Recall that the positive fragment of $\InqLTL$ is defined as follows.   
\[
\varphi ::=  \bot \mid  p \mid  \neg p   \mid  \varphi \StrongDis \varphi  \mid  \varphi \wedge \varphi   \mid \Next \varphi     \mid \varphi \Until \varphi \mid \varphi \Release \varphi
\]
By construction for each set of paths $\Pi$ of the given Kripke structure $\KS$, it holds that $\Pi\models_{\KS} p$ iff $\MP(\Pi)\models_{\KS} p$. Moreover,  
 $\Pi\models_{\KS} \neg p$ iff $\MP(\Pi)\models_{\KS} \neg p$. Additionally, for each $i\geq 0$, $\MP(\Pi_{\geq i})=(\MP(\Pi))_{\geq i}$. 
 Thus, by a straightforward induction on the structure of the given positive $\InqLTL$ formula, we 
 obtain the following result.
 
\begin{prop}\label{prop:MacroStateSemanticsPositiveInqLTL} Let $\KS$ be a Kripke structure and $\varphi$ be a positive $\InqLTL$ formula $\varphi$. Then, for each set $\Pi$ of $\KS$-paths,
  $\Pi\models_{\KS}\varphi$ \emph{iff} $\MP(\Pi)\models_{\KS}\varphi$ .
\end{prop}
 
We now show that for the left-positive fragment of $\InqLTL$, the team semantics over Kripke structures and the macro-path semantics are equivalent.
Intuitively, while macro-paths can only represent memoryless sets of paths and lose the history of individual executions, formulas in the left-positive fragment cannot enforce 
memory-dependent constraints on paths.

\begin{prop}\label{prop:MacroStateSemanticsLeftPositiveInqLTL} Let $\KS$ be a Kripke structure and $\varphi$ be a left-positive $\InqLTL$ formula $\varphi$. Then, for each macro-path $\rho$ of $\KS$,
  $\rho\models_{\KS}\varphi$ \emph{iff} $\Paths_{\KS}(\rho)\models_{\KS}\varphi$.
\end{prop}
\begin{proof} The proof is by induction on the structure of the left-positive formula $\varphi$. 
The cases  where $\varphi =p$  and $\varphi=\bot$ easily follow from the macro-path semantics.
The cases where the root modality is a temporal modality or a connective in $\{\StrongDis,\wedge\}$ directly follow from the induction hypothesis and the fact that  $(\Paths_{\KS}(\rho))_{\geq i}=\Paths_{\KS}(\rho_{\geq i})$.
For the remaining cases, where the root modality of $\varphi$ is $\neg$ or $\rightarrow$, we proceed as follows:  
\begin{itemize}
  \item $\varphi=\neg\psi$, where $\psi$ is an arbitrary $\InqLTL$ formula: by downward closure of $\InqLTL$ formulas, we have that $\Paths_{\KS}(\rho)\models_{\KS}\neg\psi$ iff for each path
   $\pi\in  \Paths_{\KS}(\rho)$,
  $\{\pi\}\not\models_{\KS} \psi$. Moreover, since the macro-path semantics is downward closed, we have that $\rho\models_{\KS} \neg \psi$ iff for each \emph{singleton} macro-path $\rho'$ with $\rho'\sqsubseteq \rho$,  $\rho'\not\models_{\KS} \psi$ iff (by the macro-path semantics)  for each $\pi\in \Paths_{\KS}(\rho)$,  $\{\pi\}\not\models_{\KS} \psi$. Hence, the result follows.    
  \item $\varphi= \psi_1 \rightarrow \psi_2$, where $\psi_1$ is a positive $\InqLTL$ formula: first assume that $\Paths_{\KS}(\rho)\models_{\KS}\psi_1 \rightarrow \psi_2$. Let $\rho'$ be a macro-path such that $\rho'\sqsubseteq \rho$ and 
  $\rho'\models_{\KS} \psi_1$. We need to show that $\rho'\models_{\KS} \psi_2$. By the induction hypothesis, $\Paths_{\KS}(\rho')\models_{\KS} \psi_1$. Since $\Paths_{\KS}(\rho')\subseteq \Paths_{\KS}(\rho)$ and
  $\Paths_{\KS}(\rho)\models_{\KS}\psi_1 \rightarrow \psi_2$, it follows that  $\Paths_{\KS}(\rho')\models_{\KS} \psi_2$. Thus, by the induction hypothesis, we obtain that $\rho'\models_{\KS} \psi_2$.
  
  For the converse direction, 
  let  $\rho\models_{\KS}\psi_1 \rightarrow \psi_2$ and $\Pi$ be a set of $\KS$-paths such that $\Pi\subseteq \Paths_{\KS}(\rho)$ and $\Pi\models_{\KS} \psi_1$. We need to 
  show that $\Pi\models_{\KS} \psi_2$. We note that $\MP(\Pi) \sqsubseteq \rho$. Moreover, since $\psi_1$ is a positive $\InqLTL$ formula, by Proposition~\ref{prop:MacroStateSemanticsPositiveInqLTL}, it follows that
  $\MP(\Pi)\models_{\KS} \psi_1$. Thus, being $\MP(\Pi) \sqsubseteq \rho$ and $\rho\models_{\KS}\psi_1 \rightarrow \psi_2$, we have that $\MP(\Pi)\models_{\KS} \psi_2$ and by the induction hypothesis,
   $\Paths_{\KS}(\MP(\Pi))\models_{\KS} \psi_2$. Since $\Pi \subseteq \Paths_{\KS}(\MP(\Pi))$ and $\psi_2$ is downward closed, we conclude that $\Pi\models_{\KS} \psi_2$, and we are done. \qedhere
\end{itemize}
\end{proof} 

Given a Kripke structure $\KS=\tpl{S,S_0,R,\Lab}$, the
 \emph{initial macro-path} of $\KS$ is the macro-path $\rho_0$ starting at $S_0$ of the form  $\rho_0=S_0,S_1,\ldots$ where 
\[  
S_{i+1}\DefinedAs\{s'\in S \mid s'\in R[s] \text{ for some } s\in S_i\} \text{ for each }  i\geq 0.
\]
We crucially observe that for the initial macro-path $\rho_0$ of $\KS$,  $\Paths_{\KS}(\rho_0)$ is the set of initial paths of $\KS$.
Hence, by Proposition~\ref{prop:MacroStateSemanticsLeftPositiveInqLTL}, we obtain the following result.

\begin{cor}\label{cor:InitialMacroPath} Given a Kripke structure $\KS$ with initial macro-path $\rho_0$ and a left-positive $\InqLTL$ formula $\varphi$, $\rho_0\models_{\KS} \varphi$ iff $\Lang(\KS)\models \varphi$.
\end{cor}

\subsection{Model checking of left-positive $\InqLTL$}

In this section, we provide an asymptotically optimal automata-theoretic approach for checking whether the initial macro-path of a finite Kripke structure $\KS$ satisfies 
an $\InqLTL$ formula $\varphi$ under the macro-path semantics. In particular, 
we show how to construct  a \emph{hesitant alternating word automaton} (\HAA)~\cite{KupfermanVW00} $\Au_{\KS,\varphi}$   accepting the set of macro-paths of $\KS$  satisfying $\varphi$. As a consequence, the considered problem is reduced to the membership problem $\rho_0\in\Lang(\Au_{\KS,\varphi})$, where $\rho_0$ is the initial macro-path of $\KS$. The latter problem can be reduced to nonemptiness of one-letter $\HAA$
which in turn can be efficiently solved in polylogarithmic space in the number of states~\cite{KupfermanVW00}. Thus, by Corollary~\ref{cor:InitialMacroPath}, for each $k\geq 0$, we obtain 
that model checking the left-positive fragment of $\InqLTL_k$ is decidable and precisely in $k$-\EXPSPACE. A matching lower bound is proved in Section~\ref{sec:LowerBounds}.

\paragraph{Syntax and semantics of $\HAA$~\cite{KupfermanVW00}} 
 An  $\HAA$ is a tuple $\Au=\tpl{\Sigma,Q,q_0,\delta,\Fam}$, where
$\Sigma$ is a finite input alphabet, $Q$ is a finite set of states, $q_0\in Q$ is the initial state,
  $\delta: Q\times \Sigma \rightarrow \Bool^+(Q)$ is the transition function, with $\Bool^+(Q)$ being the set of positive Boolean formulas over $Q$  (we also allow the formulas $\True$ and
$\False$), and the acceptance condition $\Fam$ is encoded as an ordered set $\Fam=\{(Q_1,F_1,t_1),\ldots,(Q_h,F_h,t_h)\}$ of \emph{strata}, where $Q_i\subseteq Q$, $F_i\subseteq Q_i$, and $t_i\in \{\bu,\co,\tr\}$. 
Each stratum   $(Q_i,F_i,t_i)$ is classified either as transient ($t_i= \tr$) or   \emph{B\"{u}chi} ($t_i=\bu$)  or \emph{coB\"{u}chi} ($t_i= \co$).
Moreover, we require that the components $Q_1,\ldots,Q_h$ form a partition of $Q$ and moves from states in $Q_i$ lead to states in components $Q_j$ so that $j\geq i$ (\emph{partial-order requirement}):
formally, for each $(q,\sigma)\in Q_i\times \Sigma$, $ \delta(q,\sigma)$ contains only states in components $Q_j$ with $j\geq i$.
Additionally,  for each component $Q_i$ and $(q,\sigma)\in Q_i\times \Sigma$, the following holds (\emph{hesitant requirement}):
 \begin{itemize}
   \item if $Q_i$ is transient, then 
         $\delta(q,\sigma)$ does not contain states in $Q_i$;
   \item if $Q_i$ is B\"{u}chi, then 
          each conjunct in the disjunctive normal form of $\delta(q,\sigma)$ contains at most one state in $Q_i$; 
   \item if $Q_i$ is coB\"{u}chi, then 
           each disjunct in the conjunctive normal form of $\delta(q,\sigma)$ contains at most one state in $Q_i$. 
 \end{itemize} 
 Intuitively, when  $\Au$ is in state $q$, reading the input
symbol $\sigma\in\Sigma$,  then $\Au$ chooses a set of states  $\{q_1,\ldots,q_k\}$ satisfying $\delta(q,\sigma)$ and
splits in $k$ copies such that the $i$-th copy moves  to the next input symbol  in state $q_i$.  
Formally, a run over an infinite word $w\in \Sigma^\omega$ is  a $Q\times \Nat$-labeled tree $T_r$ 
such that the root is labeled by $(q_0,0)$ 
and for each $T_r$-node $x$ with label $(q,i)\in Q\times \Nat$ (describing a copy of $\Au$ in state $q$ which reads $w(i)$),
there is a (possibly empty) set
$H=\{q_1 ,\ldots,q_k\}\subseteq  Q$ satisfying $\delta(q,w(i))$ such
that $x$ has $k$ children $x_1,\ldots, x_k$, and for
 $\ell\in [1,k]$, $x_\ell$ has label $(q_\ell,i+1)$.
 
 The hesitant and partial-order requirements ensure  that  every infinite path  $\pi$ of the run  gets trapped in some B\"{u}chi or coB\"{u}chi component.  
 Then, the run $T_r$ is accepting if for every infinite path $\pi$, denoting with $Q_i$ the B\"{u}chi/coB\"{u}chi component in which $\pi$ gets trapped, $\pi$ satisfies the
B\"{u}chi/coB\"{u}chi acceptance condition $F_i$ associated with $Q_i$: formally, $\pi$ visits infinitely (resp., finitely) many times nodes labeled by states in $F_i$ if $t_i=\bu$ (resp., $t_i =\co$).
 We denote by $\Lang(\Au)$ the set of inputs $w\in\Sigma^\omega$ such that there is an accepting run over $w$.
  The dual $\widetilde{\Au}$ of $\Au$ is the $\HAA$ obtained from $\Au$ by dualizing the transition function and  by converting each B\"{u}chi (resp., coB\"{u}chi) stratum
 into a coB\"{u}chi (resp., B\"{u}chi) stratum.  The \emph{depth} of $\Au$ is the number of $\Au$-components.
 A $1$-letter $\HAA$ is an $\HAA$ over a singleton alphabet.
 It is known~\cite{KupfermanVW00} that nonemptiness of $1$-letter $\HAA$ can be solved efficiently. In particular,
  we  exploit the following known results.

\begin{prop}\label{prop:HAAProperties}\emph{\cite{KupfermanVW00}} Given an  $\HAA$ $\Au$, the dual $\widetilde{\Au}$ of $\Au$ is an $\HAA$ accepting the complement of $\Lang(\Au)$. Moreover,
nonemptiness of $1$-letter $\HAA$ with $n$ states and depth $k$ can be solved in space $O(k\cdot \log^2 n)$.
 \end{prop}
 
\begin{rem}\label{rem:HAA-optimality}
Resorting to hesitant alternating automata (\HAA), rather than general alternating parity automata, is essential to achieve an asymptotically optimal upper bound.
Indeed, unlike $\HAA$, nonemptiness of general alternating parity word automata over a singleton alphabet corresponds to solving parity games, for which no polylogarithmic-space algorithm in the number of states is known; treating parity games without such space bounds would induce an exponential blowup when searching the state space on-the-fly, leading to a suboptimal $(k+1)$-\EXPSPACE\ upper bound for model checking the left-positive fragment of $\InqLTL_k$. By contrast, the restricted transition and acceptance structure of $\HAA$ allows the 
$1$-letter nonemptiness problem to be decided in space polylogarithmic in the number of states (Proposition~\ref{prop:HAAProperties}), keeping the overall model-checking procedure strictly within $k$-\EXPSPACE.
\end{rem}

We also consider standard B\"{u}chi nondeterministic word automata (B\"{u}chi $\NWA$)   which are tuples
 $\Nu=\tpl{\Sigma,Q,Q_0,\delta,F}$, where
$\Sigma$ and $Q$ are defined as for $\HAA$,  $Q_0\subseteq Q$
is the nonempty set of initial states,
$\delta:Q\times \Sigma\rightarrow 2^Q$ is
a transition function, and $F\subseteq Q$  (\emph{B\"{u}chi condition}).
 A run of $\Nu$ over an input $w\in \Sigma^{\omega}$ is an 
 infinite sequence of states $q_0q_1\ldots$, where $q_0\in Q_0$ and $q_{i+1}\in \delta(q_i,w(i))$ for each $i\geq 0$.
The run is accepting if for infinitely many $i\geq 0$, $q_i\in F$. The language $\Lang(\Nu)$ accepted by $\Nu$ is the set of infinite words 
$w$ over $\Sigma$ such that there is an accepting run of $\Nu$ over $w$.
Note that a B\"{u}chi $\NWA$ can be trivially converted in linear time into an $\HAA$ with just one   B\"{u}chi stratum.

\paragraph{Translation into $\HAA$} For a finite Kripke structure $\KS$ and an $\InqLTL$ formula $\varphi$, we denote by $\MP(\KS)$ the set of macro-paths of $\KS$ and by $\MP(\KS,\varphi)$ 
the set of macro-paths $\rho$ of $\KS$ such that $\rho \models_{\KS} \varphi$.
 In the following, for the given finite Kripke structure $\KS$, we consider  B\"{u}chi $\NWA$ and 
 $\HAA$ over the alphabet $2^S$, where $S$ is the set of $\KS$-states.  The following result is straightforward.

\begin{prop}\label{prop:AutomataForMacroPaths} Let $\KS$ be a finite Kripke structure with set of states $S$ and $\Au$ be an  $\HAA$  over $2^S$ with $n$ states and 
depth $k$. Then, one can construct in time $O(n+ 2^{|S|})$ an $\HAA$ $\Au'$ with depth $k+1$ such that $\Lang(\Au')=\Lang(\Au)\cap \MP(\KS)$.  
\end{prop}

By~\cite{DaxK08,SanchezS12}, given an $\HAA$ with $n$ states, one can construct an equivalent B\"{u}chi $\NWA$  in time
$2^{O(n\cdot \log n)}$. Moreover, a B\"{u}chi $\NWA$ corresponds to an $\HAA$ with just one   B\"{u}chi stratum. Thus, since 
B\"{u}chi $\NWA$ are closed under projection and intersection, by Proposition~\ref{prop:HAAProperties}, we easily obtain the following result
which  allows handling intuitionistic implication under the macro-path semantics.

\begin{prop}\label{prop:AutomataForInqImplication}Let $\KS$ be a finite Kripke structure with set of states $S$  and for each $i=1,2$, let $\varphi_i$ be an  $\InqLTL$ formula and 
$\Au_i$ be an $\HAA$ with $n_i$ states  accepting $\MP(\KS,\varphi_i)$. Then, one can construct in time $2^{O(n)}$, where $n= n_1 \log n_1 + n_2 \log n_2 + |S|$,
an   $\HAA$ $\Au$ with  depth $O(1)$ such that $\Lang(\Au)=\MP(\KS,\varphi_1 \rightarrow \varphi_2)$. 
\end{prop} 
\begin{proof} Let $\Au'_2$ be the dual of $\Au_2$. By Proposition~\ref{prop:HAAProperties}, $\Lang(\Au'_2)\cap \MP(\KS)$
is the set of macro-paths $\rho$ of $\KS$ such that $\rho \not\models_{\KS} \varphi_2$. Moreover, by~\cite{DaxK08,SanchezS12}, for each $i=1,2$,  
one can construct    a B\"{u}chi $\NWA$  $\Nu_i=\tpl{2^S,Q_i,Q_{i}^0,\delta_i,F_i}$  with $2^{O(n_i\log n_i)}$ states such that 
$\Lang(\Nu_1)=\Lang(\Au_1)=\MP(\KS,\varphi_1)$ and $\Lang(\Nu_2)= \Lang(\Au'_2)$.
We first construct a B\"{u}chi $\NWA$ $\Nu$ with $O(|Q_1|\cdot |Q_2|)$ states
which accepts a macro-path  $\rho$ of $\KS$ iff   there exists a nonempty macro-path $\rho' \sqsubseteq \rho$ so that $\rho'\models_{\KS} \varphi_1$ and $\rho'\not\models_{\KS} \varphi_2$.
Intuitively, given an input macro-path $\rho$, $\Nu$ guesses a nonempty macro-path $\rho' \sqsubseteq \rho$ and checks that there is an accepting run of $\Nu_1$ over $\rho'$ and an accepting run of $\Nu_2$ 
over $\rho'$. Formally, $\Nu= \tpl{2^S,Q,\{\top\},\delta,F}$   where:
\begin{itemize}
  \item $Q\DefinedAs (2^S\times Q_1\times Q_2 \times \{1,2\})\cup \{\top\}$.
\item $\delta(\top,S')$ consists of the states $(T,q_1,q_2,1)$ such that $T$ is a nonempty subset of $S'$ and for each $i=1,2$, $q_i\in \delta_i(q_i^0,T)$ for some 
$q_i^0\in Q_{i}^0$ .
  \item $\delta((T,q_1,q_2,\ell),S')$ consists of the states $(T',q'_1,q'_2,\ell')$ such that the macro-state $T'$ is a nonempty successor of $T$,
   $T'\subseteq S'$, $q'_1\in \delta_1(q_1,T')$, $q'_2\in \delta_2(q_2,T')$ and the following holds:
\begin{itemize}
  \item case $\ell=1$: $\ell'=2$ if $q_1\in F_1$, and $\ell'=1$ otherwise;
    \item case $\ell=2$: $\ell'=1$ if $q_2\in F_2$, and $\ell'=2$ otherwise.
\end{itemize}
\item $F\DefinedAs \{(T,q_1,q_2,2)\in Q\mid q_2\in F_2 \}$.
\end{itemize} 
\noindent  By hypothesis, correctness of the  construction easily follows. Since a B\"{u}chi $\NWA$ can be trivially converted into an  $\HAA$ with depth $1$, by  Proposition~\ref{prop:HAAProperties}, the dual $\Au$ of 
$\Nu$ is an  $\HAA$ such that  $\Lang(\Au)\cap \MP(\KS)=\MP(\KS,\varphi_1 \rightarrow \varphi_2)$. Thus, by Proposition~\ref{prop:AutomataForMacroPaths}, the result follows. 
\end{proof}

Intuitionistic negation can be managed by a generalization of the standard automata-theoretic approach for $\LTL$.
For an $\InqLTL$ formula $\varphi$, we denote by $\Lang(\varphi)$ the set of traces satisfying $\varphi$ under the standard $\LTL$ semantics.

\begin{prop} 
\label{prop:AutomataForInqNegation} Given  a finite Kripke structure   $\KS$  with set of states $S$ and an $\InqLTL$ formula $\varphi$, one can build  
in time    $2^{O(|\varphi|+|S|)}$ an  $\HAA$ $\Au$ with depth $O(1)$ such that $\Lang(\Au)=\MP(\KS,\neg \varphi)$.
\end{prop}
\begin{proof} Let $\KS=\tpl{S,S_0,R,\Lab}$. By downward closure of $\neg\varphi$ under the macro-path semantics, we have that for each macro-path $\rho$ of $\KS$, $\rho\models \neg\varphi$ iff for each $\pi\in\Paths_{\KS}(\rho)$, $\Lab(\pi)\models_{\LTL} \neg\varphi$. By~\cite{VardiW94}, one can construct 
a B\"{u}chi $\NWA$ $\Nu=\tpl{2^{\AP},Q,Q_0,\delta,F}$  accepting $\Lang(\varphi)$ with $2^{O(|\varphi|)}$ states. 
We first construct a B\"{u}chi $\NWA$ $\Nu'$ over $2^S$ which accepts a macro-path $\rho$ of $\KS$ iff there is $\pi\in \Paths_{\KS}(\rho)$ such that $\Lab(\pi)\in\Lang(\Nu)$ (i.e., $\Lab(\pi)\models_{\LTL} \varphi$). 
Intuitively, given an input macro-path $\rho$, the B\"{u}chi $\NWA$ $\Nu'$  guesses a path $\pi\in \Paths_{\KS}(\rho)$ and simulates the behaviour of $\Nu$ over $\Lab(\pi)$.
Formally, $\Nu'=\tpl{2^{S},Q \times (S\cup \{\top\}),Q_0\times \{\top\},\delta',F\times S}$, where for each $S'\subseteq S$:
\[
\begin{array}{ll}
  \delta((q,\top),S') & \DefinedAs \{(q',s')\mid s'\in S' \text{ and }
    q'\in\delta(q,\Lab(s'))\}.\\
   \delta((q,s),S') & \DefinedAs \{(q',s')\mid s'\in S'\cap R[s] \text{ and }  
    q'\in\delta(q,\Lab(s'))\}.
\end{array}
\]
\noindent Since a B\"{u}chi $\NWA$ can be trivially converted in linear-time into an  $\HAA$ with depth $1$, by Proposition~\ref{prop:HAAProperties}, the dual $\Au'$ of $\Nu'$ accepts a macro-path $\rho$ of $\KS$ iff $\rho \models_{\KS}\neg\varphi$.  Hence, $\Lang(\Au')\cap \MP(\KS)=\MP(\KS,\neg \varphi)$ and by Proposition~\ref{prop:AutomataForMacroPaths}, the result follows.      
\end{proof}  

By exploiting Propositions~\ref{prop:AutomataForMacroPaths}--\ref{prop:AutomataForInqNegation}, we deduce the following result.

\begin{prop}\label{prop:FromFormulasToHAA}Let $k\geq 0$, $\KS$ be a finite Kripke structure with set of states $S$, and $\varphi$ be an  $\InqLTL_k$ formula. Then, one can construct in time 
$\Tower_2(k+1,|S|+|\varphi|)$ an  $\HAA$  with depth $O(|\varphi|)$ accepting $\MP(\KS,\varphi)$. 
\end{prop} 
\begin{proof}
The proof is by induction on $k\geq 0$. Let $FS(\varphi)$ be the set of subformulas $\psi$ of $\varphi$ such that some occurrence of $\psi$
is not preceded by the connectives in $\{\neg ,\rightarrow\}$ in the syntax tree of $\varphi$. Moreover, let $H_\neg$ be the set of formulas in $FS(\varphi)$ of the form
$\neg \theta$, and $H_{\rightarrow}$ the set of formulas in $FS(\varphi)$ of the form
$  \theta_1\rightarrow \theta_2$. Note that $H_{\rightarrow}=\emptyset$ if $k=0$.  By Proposition~\ref{prop:AutomataForInqNegation}, for each $\psi\in H_\neg$, one can construct 
in time    $2^{O(|S|+|\psi| )}$ an  $\HAA$ $\Au_{\psi}$  
accepting $\MP(\KS,\psi)$.  Moreover, if $k>0$, then by the induction hypothesis and Proposition~\ref{prop:AutomataForInqImplication}, for each $\psi\in H_\rightarrow$, one can construct 
in time  $\Tower_2(k+1,|S|+ |\psi|)$   an $\HAA$ $\Au_{\psi}$ with depth $O(|\psi|)$ accepting $\MP(\KS,\psi)$. Then, by an easy generalization of the standard linear-time translation of $\LTL$ formulas into B\"{u}chi alternating word automata and by using the   $\HAA$ $\Au_\psi$ with $\psi\in H_\neg\cup H_\rightarrow$, we show how to construct
 in time $\Tower_2(k+1,O(|S|+|\varphi|))$ an $\HAA$ $\Au_\varphi$ with depth $O(|\varphi|)$ such that $\Lang(\Au_\varphi)\cap \MP(\KS) = \MP(\KS,\varphi)$. Hence, by 
 Proposition~\ref{prop:AutomataForMacroPaths}, the result follows.
 Intuitively, given an input macro-path of $\KS$, each copy of  $\Au_\varphi$ keeps track of the current subformula in $FS(\varphi)$ which needs to be evaluated. The evaluation simulates the macro-path semantics of $\InqLTL$, but when the current subformula $\psi$ is in $ H_\neg\cup H_\rightarrow$, then the current copy of $\Au_\varphi$ activates a copy of $\Au_\psi$ in the initial state.
Formally, for each $\psi\in H_\neg\cup H_\rightarrow$, let $\Au_\psi=\tpl{2^S,Q_\psi,q_\psi,\delta_\psi,  \Fam_\psi}$. W.l.o.g., we assume that the state sets of the $\HAA$ $\Au_\psi$ are pairwise distinct. Then, $\Au_\varphi=\tpl{2^S,Q,q_0,\delta, \Fam}$, where:
 \begin{itemize}
   \item $Q \DefinedAs   FS(\varphi)\cup{\bigcup_{\psi\in H_\neg\cup H_\rightarrow}Q_\psi}$ and $q_0=\varphi$;
   \item The transition function $\delta$ is defined as follows: $\delta(q,\sigma) = \delta_\psi(q,\sigma)$ if $q\in Q_\psi$ for some $\psi\in  H_\neg\cup H_\rightarrow$.
   If instead  $ q\in FS(\varphi)$, then $\delta(q,\sigma)$ is inductively defined as follows: 
   \begin{itemize}
   \item $\delta(\bot,\sigma) = \True$ if $\sigma =\emptyset$, and $\delta(\bot,\sigma) = \False$ otherwise;
     \item $\delta(p,\sigma) = \True$ if $p\in \Lab(s)$ for each $s\in\sigma$, and $\delta(p,\sigma) = \False$ otherwise; 
     \item  $\delta(\phi_1\StrongDis \phi_2,\sigma)=\delta(\phi_1,\sigma)\vee\delta(\phi_2,\sigma)$;
        \item $\delta(\phi_1\wedge \phi_2,\sigma)=\delta(\phi_1,\sigma)\wedge\delta(\phi_2,\sigma)$;
     \item $\delta(\Next\phi,\sigma)=\phi$;
      \item $\delta(\phi_1\Until \phi_2,\sigma)=\delta(\phi_2,\sigma)\vee (\delta(\phi_1,\sigma)\wedge  \phi_1\Until \phi_2)$;
       \item $\delta(\phi_1\Release \phi_2,\sigma)=\delta(\phi_2,\sigma)\wedge (\delta(\phi_1,\sigma)\vee  \phi_1\Release \phi_2)$;
      \item for each $\psi\in H_\neg\cup H_\rightarrow$, $\delta(\psi,\sigma)=\delta(q_\psi, \sigma)$.
   \end{itemize}
   \item $\Fam  = \displaystyle{\bigcup_{\psi\in H_\neg\cup H_\rightarrow}\Fam_\psi}\cup \bigcup_{\phi\in FS(\varphi)}\{\Com_\phi\}$, where for each
   $\phi\in FS(\varphi)$, the stratum $\Com_\phi$ is defined as: 
   \begin{itemize}
     \item if $\phi$ has the form $\psi_1\Until \psi_2$, then $\Com_\phi$ is the coB\"{u}chi stratum $(\{\phi\}, \{\phi\},\co)$;
          \item if $\phi$ has the form $\psi_1\Release \psi_2$, then $\Com_\phi$ is the B\"{u}chi stratum  $(\{\phi\}, \{\phi\},\bu)$;
     \item otherwise, $\Com_\phi$ is the transient stratum $(\{\phi\}, \emptyset,\tr)$.\qedhere
   \end{itemize}
 \end{itemize} 
 \end{proof}

\paragraph{Model-checking the left-positive fragment of $\InqLTL$.} 
Let $k\geq 0$,  $\KS=\tpl{S,S_0,R,\Lab}$ be a finite Kripke structure with initial macro-path $\rho_0$, and $\varphi$ be a left-positive $\InqLTL_k$
formula. By  Corollary~\ref{cor:InitialMacroPath} and Proposition~\ref{prop:FromFormulasToHAA}, $\Lang(\KS)\models \varphi$ iff $\rho_0\in \Lang(\Au_\varphi)$, where $\Au_\varphi=\tpl{2^S,Q,q_0,\delta, \Fam}$ is the $\HAA$ of Proposition~\ref{prop:FromFormulasToHAA} accepting $\MP(\KS,\varphi)$. We construct a $1$-letter $\HAA$ $\Au'_\varphi$  which simulates the behaviour of $\Au_{\varphi}$ over $\rho_0$ and accepts  
iff $\Au_\varphi$ accepts $\rho_0$. Formally, $\Au'_\varphi=\tpl{\{1\},Q\times 2^S,(q_0,S_0),\delta', \Fam'}$, where:
\begin{itemize}
\item for all $(q,T) \in  Q\times 2^S$, $\delta'((q,T),1)$ is obtained from $\delta(q,T)$ by replacing each state $q'$ occurring in $\delta(q,T)$ with $(q',T')$,
where $T'\DefinedAs  \{s'\in S\mid s'\in R[s] \text{ for some }s\in T\}$;
\item $\Fam'$ is obtained from $\Fam$ by replacing each stratum $(Q',F',t)\in \Fam$ with $(Q'\times 2^S,F'\times 2^S,t)$.
\end{itemize}
By Proposition~\ref{prop:FromFormulasToHAA}, $\Au'_\varphi$ has depth $O(|\varphi|)$ and size  $\Tower_2(k+1,|S|+ |\varphi|)$.
Hence, by Proposition~\ref{prop:HAAProperties}, we obtain the following result, where for the lower bounds, we provide a detailed proof in Section~\ref{sec:LowerBounds}. Note that $\InqLTL_0$ coincides with  left-positive $\InqLTL_0$, allows an unrestricted use of intuitionistic negation, and subsumes the split-disjunction-free fragment
of $\TeamLTL$ for which model checking is known to be in \PSPACE~\cite{KrebsMV018}.

\begin{thm}\label{theo:ComplexityMCLeftPositiveInqLTL} For each $k\geq 0$, 
model checking of left-positive  $\InqLTL_k$ is   $k$-\EXPSPACE-complete. In particular, model checking of
$\InqLTL_0$ is \PSPACE-complete.
\end{thm}
 
\section{Lower bounds for model checking left-positive InqLTL}\label{sec:LowerBounds}

In this section, we establish the following result. 
 
\begin{thm}\label{theo:lowerBounds} For each $k\geq 0$, the model checking problem for left-positive 
$\InqLTL_k$ is  $k$-\EXPSPACE-hard.
\end{thm}

Given $k\geq 0$, Theorem~\ref{theo:lowerBounds} for left-positive 
$\InqLTL_k$  is proved by a polynomial-time
reduction from a domino-tiling problem for grids with  rows of length
$\Tower_c(k,n^d)$~\cite{Boas97} for some integer constants $d\geq 1$ and $c>1$, where $n$ is an input
parameter. In the following, for ease  of presentation,  we assume that $c=2$ and $d=1$. 

 Formally, an instance $\Instance$ of the considered domino-tiling problem is a
tuple $\Instance =\tpl{C,\Delta,n,d_{in}}$, where $C$ is a
finite set of colors, $\Delta \subseteq C^{4}$ is a set of tuples
$\tpl{c_{down},c_{left},c_{up},c_{right}}$ of four colors, called
\emph{domino-types}, $n>0$ is a natural number encoded in
\emph{unary}, and $d_{in}$ is the initial domino-type.  Given
$k\geq 0$, a \emph{$k$-grid of $\Instance$} is a mapping
$f:\Nat\times [0,\Tower_2(k,n)-1] \rightarrow \Delta$. Intuitively, a $k$-grid  is a grid consisting of an infinite number of rows, where
each row  consists of $\Tower_2(k,n)$ cells, and each cell contains a
domino type.  A \emph{\mbox{$k$-tiling} of $\Instance$} is a $k$-grid
$f$ satisfying the following additional constraints:\vspace{0.2cm}
 \begin{description}
 \item [Initialization]
$f(0,0)=d_{in}$.
     \item [Row adjacency]  two adjacent cells in a row
       have the same color on the shared edge: for all
       $(i,j)\in \Nat\times [0,\Tower_2(k,n)-2]$,
      $ [f(i,j)]_{right}=[f(i,j+1)]_{left}$.
  \item [Column adjacency]  two adjacent cells in a column have the same color on the shared edge: for all $(i,j)\in \Nat\times [0,\Tower_2(k,n)-1]$,
   $[f(i,j)]_{up}=[f(i+1,j)]_{down}$.
\end{description} \vspace{0.1cm}
 
\noindent Given $k\geq 0$, the  problem of checking the existence  of a $k$-tiling for $\Instance$ is $k$-\EXPSPACE-complete~\cite{Boas97}.
In the following, we show that one can build, in time polynomial in the size of $\Instance$, a finite Kripke structure 
$\KS_{\Instance,k}$ and a left-positive $\InqLTL_k$ formula $\varphi_{\Instance,k}$ such that 
$\Lang(\KS_{\Instance,k})\models \varphi_{\Instance,k}$ iff there is \emph{no} $k$-tiling of $\Instance$. Hence, since $k$-\EXPSPACE\ and its complement coincide, Theorem~\ref{theo:lowerBounds} directly follows.

\paragraph{Trace encoding of $k$-grids} Fix $k\geq 0$. In the following, we assume that $k\geq 1$ (the proof
of Theorem~\ref{theo:lowerBounds} for the case $k=0$ being simpler). 
We define a suitable encoding of the \mbox{$k$-grids} by using the set
$\AP$ of atomic propositions defined as follows:
\[
\begin{array}{ll}
\AP \DefinedAs & \AP_{main}\cup \{\col\} \vspace{0.1cm}\\
  \AP_{main}\DefinedAs & \Delta \cup \{0,1\}\cup  
  \{\Symb,\Symb_1,\ldots,\Symb_{k-1}\}
\end{array}  
\]
The propositions in $\AP_{main}$ are used to encode the \mbox{$k$-grids}, while the special 
proposition $\col$ is used to mark exactly one position along a trace.  
Essentially, the \emph{unmarked} trace code of a \mbox{$k$-grid} $f$ is obtained by
concatenating the codes of the rows of $f$ starting from the first row. The code of
a row is in turn obtained by concatenating the codes of the row's cells starting from the first cell.

In the encoding of a cell of a \mbox{$k$-grid}, we keep track of
the content of the cell together with a suitable encoding of the cell
number which is a natural number in $[0,\Tower_2(k,n)-1]$. Thus, for all
$1\leq h\leq k$, we define the notions of \emph{\mbox{$h$-block}} and
\emph{well-formed \mbox{$h$-block}}. Essentially, for $1\leq h<k$, well-formed
$h$-blocks are finite traces over $2^{\{0,1,\Symb_1,\ldots,\Symb_{h}\}}$
which encode integers in $[0,\Tower_2(h,n)-1]$, while well-formed
\mbox{$k$-blocks} are finite traces over $2^{\AP_{main}\setminus \{\Symb\}}$  which encode the cells of
\mbox{$k$-grids}.  In particular, for $h>1$, a
well-formed $h$-block encoding a natural number $m\in
[0,\Tower_2(h,n)-1]$ is a sequence of $\Tower_2(h-1,n)$ $(h-1)$-blocks,
where the $i$-th $(h-1)$-block encodes both the value and
(recursively) the position of the $i$-th bit in the binary
representation of $m$. Formally, the set of (well-formed) $h$-blocks
is defined by induction on $h$ as follows:

\paragraph{Case $h=1$} A  $1$-block $\bl$ is a finite trace of the form 
 $\bl=\{\Symb_{1},\tau\} \{bit_1\}\ldots \{bit_j\}$ for some $j\geq 1$
 such that $bit_1,\ldots,bit_j\in\{0,1\}$ and $\tau\in \{0,1\}$ if
 $1<k$, and $\tau\in \Delta$ otherwise.  The
 \emph{content} of $\bl$ is $\tau$. The $1$-block $\bl$ is \emph{well-formed} if $j=n$. In this case,  the \emph{index} of
 $\bl$ is the natural number in $[0,\Tower_2(1,n)-1]$ (recall that
 $\Tower_2(1,n)=2^n$) whose binary code is $bit_1\ldots bit_n$.\footnote{We assume that the first bit in the binary encoding of a natural number is the least significant one.}    

\paragraph{Case $1<h\leq k$} An \emph{$h$-block} is
a finite trace $\bl$ having the form $\{\Symb_{h},\tau\}\, \bl_0
\ldots \bl_j$ for some $j\geq 0$ such that $\bl_0,\ldots,\bl_j$ are
$(h-1)$-blocks, and $\tau\in \{0,1\}$ if $h<k$, and $\tau\in \Delta$ otherwise.  
 The \emph{content} of $\bl$ is $\tau$.  The $h$-block $\bl$ is
\emph{well-formed} if additionally, the following holds:
$j=\Tower_2(h-1,n)-1$ and for all $ 0\leq i\leq j$, $\bl_i$ is
well-formed and has index $i$. If $\bl$ is well-formed, then
its \emph{index} is the natural number in $[0,\Tower_2(h,n)-1]$
whose binary code is given by $bit_0,\ldots,bit_j$, where $bit_i$
is the content of the $(h-1)$-sub-block $\bl_i$ for all $0\leq i\leq j$. Figure~\ref{CellCode}
illustrates the encoding of a cell for $k=2$ (well-formed $k$-block). \vspace{0.2cm}

\begin{figure*}[!tb]
\centering
\begin{tikzpicture}[
    scale=0.97,
    >=Stealth,
    headStyle/.style={fill=blue!8, draw=blue!70!black, rounded corners=2pt, inner sep=2.5pt},
    blockTag/.style={fill=white, inner sep=1.5pt, font=\footnotesize},
    indexTag/.style={fill=white, inner sep=1.5pt, font=\footnotesize, text=blue!80!black}
]

\node[headStyle] (L0) at (0.0,-0.3) {\footnotesize $\{\Symb_2,d\}$};

\coordinate [label=center:{\footnotesize $\{\Symb_1,bit_0\}\{bit_{0,1}\},\ldots,\{bit_{0,n}\}$}] (L1) at (2.8,-0.3);
\coordinate [label=center:{\footnotesize $\dots\dots$}] (L4) at (5.7,-0.3); 
\coordinate [label=center:{\footnotesize $\{\Symb_1,bit_{2^n-1}\}\{bit_{2^n-1,1}\},\ldots,\{bit_{2^n-1,n}\}$}] (L6) at (9.3,-0.3);

\draw[<->, semithick, color=black!85] (0.7,0.4) -- (4.9,0.4) 
    node[midway, above, blockTag] {well-formed $1$-block};

\draw[<->, semithick, color=black!85] (6.5,0.4) -- (12.1,0.4) 
    node[midway, above, blockTag] {well-formed $1$-block};

\coordinate [label=center:{\footnotesize $bit_0$}] (Lb0) at (1.6,-1.1);
\coordinate [label=center:{\footnotesize $bit_{2^n-1}$}] (Lbn) at (7.6,-1.1);
\coordinate [label=center:{\footnotesize $\dots\dots\ldots$}] (Dots) at (4.7,-1.1); 

\draw[<->, semithick, color=blue!75!black] (1.2,-1.35) -- (8.2,-1.35) 
    node[midway, below, indexTag] {binary encoding of cell index};

\end{tikzpicture}
\caption{Encoding of a cell of a $k$-grid for $k=2$}
\label{CellCode}
\end{figure*}
 A \emph{$k$-row} is a finite trace of the form $w_r=\{\Symb\} \bl_0\ldots \bl_j $ such that $j\geq 0$ and 
$\bl_0,\ldots, \bl_j$ are $k$-blocks.    The $k$-row $w_r$ is \emph{well-formed} if additionally,
 $j=\Tower_2(k,n)-1$ and for all $0\leq i\leq j$, $\bl_i$ is well-formed
 and has index $i$. 
 
 A \emph{$k$-grid code} (resp., \emph{well-formed $k$-grid code}) is an infinite concatenation  of 
 $k$-rows (resp., well-formed $k$-rows).   A
 $k$-grid code is \emph{initialized} if the first $k$-block of the
 first $k$-row has content $d_{in}$.  Note that while $k$-grid codes encode
 grids of $\Instance$ having rows of arbitrary length, \emph{well-formed} $k$-grid codes encode the $k$-grids of
 $\Instance$.  In particular, there is exactly one well-formed
 $k$-grid code associated with a given $k$-grid of $\Instance$. 
 
 It is worth noting that the special proposition $\col$ is not used in 
 the definition of (well-formed) $k$-grid codes. It is not difficult to construct an $\LTL$ 
 formula $\theta_k$ over $\AP_{main}$ of size polynomial in the size of $\Instance$ which characterizes the traces
 which are $k$-grid codes. The construction of $\theta_k$ is tedious, and we omit the details here.
 
 \begin{prop}\label{prop:LTLforPseudoGrids} One can build in time polynomial in the size of
 $\Instance$ an $\LTL$ formula $\theta_k$ over $\AP_{main}$ such that $\Lang(\theta_k)$ is the set of initialized $k$-grid codes. 
 \end{prop} 
 
 \paragraph{Team encoding of marked $k$-grid codes} For a trace $w$, we denote by $w_{main}$ the projection of $w$ over $\AP_{main}$.
 For a set of traces (team) $\Lang$, $\Lang_{main}$ is the team obtained from $\Lang$ by replacing each trace $w$ in $\Lang$ with $w_{main}$. 
 We say that a team $\Lang$ is \emph{consistent} if the following two conditions are fulfilled:
 \begin{itemize}
   \item $\Lang_{main}$ is a singleton and the unique trace $w_k$ in $\Lang_{main}$ is an initialized $k$-grid code. We say that 
   $w_k$ is the $k$-grid code associated with $\Lang_{main}$. 
   \item For each trace $w\in \Lang$, there is at most one position $i$ such that $\col\in w(i)$.
 \end{itemize} 
 
 The finite Kripke structure $\KS_{\Instance,k}$ used in the reduction simply ensures that for each trace $w$,  $w\in \Lang(\KS_{\Instance,k})$ if and only if the following 
 two conditions hold:
 \begin{itemize}
   \item $w$ is  \emph{$3$-sparse}, i.e., $|w(i)\cap \AP|\leq 3$  for each $i\geq 0$, and
   \item  there is at most one position $i$ of $w$ which is marked by proposition $\col$ (i.e., $\col\in w(i)$).
 \end{itemize} 

Notice that in the trace encoding of $k$-grids, no position contains more than two propositions from $\AP_{main}$ (for instance, the marker $\{\Symb_h, \tau\}$ at the beginning of an $h$-block). 
Restricting the labeling of $\KS_{\Instance,k}$ to subsets of $\AP=\AP_{main}\cup\{\col\}$ of cardinality at most three keeps the state space polynomial in $|\AP|$ rather than exponential in $|\AP|$.

 \begin{prop}[Construction of $\KS_{\Instance,k}$]\label{prop:LowerBoundKS} One can build, in time polynomial in the size of $\Instance$, a finite Kripke structure $\KS_{\Instance,k}$
such that $\Lang(\KS_{\Instance,k})$ is the set of all the \emph{$3$-sparse} traces $w$   so that $|\{i\in\Nat\mid \col\in w(i)\}|\leq 1$.
 \end{prop}
 \begin{proof} Let $P$ be the set of subsets $\sigma$ of $\AP$ such that $|\sigma|\leq 3$.   Then, the finite Kripke structure  $\KS_{\Instance,k}$ is given by $\KS_{\Instance,k}=\tpl{S,S_0,R,\Lab}$, where:
 \begin{itemize}
   \item $S\DefinedAs P\times \{0,1\}$. Intuitively, the additional flag $f$ in a state $(\sigma,f)\in P\times \{0,1\}$ is used for keeping track whether the marker $\col$ has been already generated.
   \item  $S_0$ is the set of states $(\sigma,f)$ such that $f=1$ iff $\col\in\sigma$.
   \item $((\sigma,f),(\sigma',f'))\in R$ iff either (i) $f'=f$ and $\col\notin\sigma'$, or (ii) $f=0$, $f'=1$, and $\col\in \sigma'$.
   \item $\Lab((\sigma,f))=\sigma$ for all $(\sigma,f)\in S$. 
 \end{itemize}
 Correctness of the construction easily follows. 
 \end{proof}
 
 \begin{rem} It is worth noting  that each consistent team is a subset of  $\Lang(\KS_{\Instance,k})$.
 \end{rem}
 
\paragraph{Definition of the formula $\varphi_{\Instance,k}$} For the fixed $k\geq 1$, the difficult part of the reduction concerns the polynomial-time construction 
 of the left-positive $\InqLTL_k$ formula $\varphi_{\Instance,k}$  ensuring that $\Lang(\KS_{\Instance,k})\models \varphi_{\Instance,k}$ iff there is \emph{no} consistent
 team whose associated initialized $k$-grid code is well-formed and satisfies the row and column adjacency requirements. 
   
 For the definition of $\varphi_{\Instance,k}$, we use some auxiliary formulas. For each $h\in [1,k]$, we use the notations $p_{\leq h}$ and $\Symb_{\geq h}$ for denoting the following 
 propositional formulas:
 \[
 \begin{array}{ll}
 p_{\leq h} &\DefinedAs 0 \StrongDis 1 \StrongDis \Symb_1 \StrongDis \ldots \StrongDis \Symb_h\vspace{0.1cm}\\
 \Symb_{\geq h} &\DefinedAs \Symb \StrongDis  \Symb_h \StrongDis \Symb_{h+1} \StrongDis \ldots \StrongDis \Symb_k
 \end{array}
 \]
For each $h\in [1,k]$, we now illustrate how to check in polynomial time whether for two  well-formed $h$-blocks $\bl$ and $\bla$   of a consistent team $\Lang$, their indices are \emph{not} equal.   At this end, we use the following notion. 
Let $\Lang$ be a consistent team and $i\geq 0$. We say that $\Lang_{\geq i}$ is \emph{$h$-marked} iff the set of $\col$-positions in $\Lang_{\geq i}$ (i.e., the positions $\ell$ such that for some trace 
$w\in\Lang_{\geq i}$, $\col\in w(\ell)$) exactly corresponds to an $h$-block of $\Lang_{\geq i}$.  

\begin{figure*}[!tb]
\centering
\begin{tikzpicture}[
    scale=0.95,
    >=Stealth,
    headStyle/.style={fill=blue!8, draw=blue!70!black, rounded corners=2pt, inner sep=2.5pt},
    markerStyle/.style={fill=orange!15, draw=orange!80!black, rounded corners=2pt, inner sep=2.5pt},
    subBlockStyle/.style={draw=black!70, rounded corners=2pt, inner sep=2.5pt, fill=white},
    tagStyle/.style={fill=white, inner sep=2pt, font=\footnotesize}
]

\coordinate [label=center:{\footnotesize \textbf{Well-formed $h$-block $\bl$}}] (Blk_l_title) at (-4.7, 1.3);
\coordinate [label=center:{\scriptsize (sub-block $\sbl$ with index $idx$ and content $c$)}] (Sbl_desc) at (-2.8, 0.75);

\node[headStyle] (L_head) at (-6.8, 0.0) {\footnotesize $\{\Symb_h, \tau\}$};
\coordinate [label=center:{\footnotesize $\sbl_0$}] (L_start) at (-5.6, 0.0);
\coordinate [label=center:{\footnotesize $\dots$}] (L_dots1) at (-4.5, 0.0);
\node[subBlockStyle] (Sbl) at (-2.8, 0.0) {\footnotesize $\{\Symb_{h-1}, \mathbf{c}\} \dots$};
\coordinate [label=center:{\footnotesize $\dots$}] (L_dots2) at (-1.3, 0.0);
\coordinate [label=center:{\footnotesize $\sbl_j$}] (L_end) at (-0.4, 0.0);
\coordinate [label=center:{\tiny ($j = \Tower_2(h-1,n)-1$)}] (L_end_idx) at (-0.4, -0.50);

\coordinate [label=center:{\footnotesize $\dots\dots\dots$}] (Trace_sep) at (0.7, 0.0);

\coordinate [label=center:{\footnotesize \textbf{Marked well-formed $h$-block $\bla$}}] (Blk_la_title) at (5.1, 1.3);
\coordinate [label=center:{\scriptsize (sub-block $\sbla$ marked via $\MaxCol_{h-1}$)}] (Sbla_desc) at (5.3, 0.75);

\node[headStyle] (La_head) at (1.8, 0.0) {\footnotesize $\{\Symb_h, \tau'\}$};
\coordinate [label=center:{\footnotesize $\sbla_0$}] (La_start) at (3.0, 0.0);
\coordinate [label=center:{\footnotesize $\dots$}] (La_dots1) at (3.9, 0.0);
\node[markerStyle] (Sbla) at (5.3, 0.0) {\footnotesize $\{\Symb_{h-1}, \mathbf{c'}, \col\} \dots$};
\coordinate [label=center:{\footnotesize $\dots$}] (La_dots2) at (6.8, 0.0);
\coordinate [label=center:{\footnotesize $\sbla_j$}] (La_end) at (7.6, 0.0);
\coordinate [label=center:{\tiny ($j = \Tower_2(h-1,n)-1$)}] (La_end_idx) at (7.0, -0.50);

\draw[<->, thick, color=blue!80!black, dashed] (Sbl.south) to [bend right=32] 
    node[midway, below=2mm, tagStyle, align=center] {
        \footnotesize \textbf{$h$-inequality check via $\phi_{h,\neq}$:}\\[-0.5mm]
        \footnotesize Same sub-block index $idx$ (checked via $\psi_{h-1,\neq}$) $\implies \mathbf{c \neq c'}$
    } (Sbla.south);

\end{tikzpicture}
\caption{Checking index inequality between well-formed $h$-blocks $\bl$ and $\bla$ via formula $\psi_{h,\neq}$ (Proposition~\ref{prop:CheckingInequalIndex}). Sub-block $\sbla$ of the $\col$-marked block $\bla$ is isolated by $\MaxCol_{h-1}$; if its index matches that of $\sbl$, their contents $c$ and $c'$ must differ.}
\label{fig:CheckingInequalIndex}
\end{figure*}

\begin{prop}\label{prop:CheckingInequalIndex} For each $h\in [1,k]$, one can construct in time polynomial in the size of $\Instance$ a left-positive $\InqLTL_{h-2}$ formula $\psi_{h,\neq}$ (where 
$\InqLTL_{-1}$ is for $\InqLTL_{0}$) such that for each consistent team $\Lang$ and $\Symb_h$-position $i$ of $\Lang$ so that $\Lang_{\geq i}$ is $h$-marked, the following holds. Let $\bl$ be the $h$-block
starting at position $i$ and $\bla$ the  marked $h$-block of $\Lang_{\geq i}$. If $\bl$  precedes $\bla$ and $\bl$ and $\bla$ are well-formed, then:
\[
\Lang_{\geq i}\models \psi_{h,\neq} \Leftrightarrow \text{ $\bl$ and $\bla$ do \emph{not} have the same index.}
\]
\end{prop}
\begin{proof} We assume that $h>1$ (the case where $h=1$ is straightforward).  
 Formula $\psi_{h,\neq}$ requires that  there exists an $(h-1)$-sub-block $\sbl$
of $\bl$ such that for each $(h-1)$-sub-block $\sbla$ of $\bla$, whenever $\sbl$ and $\sbla$ have the same index, then $\sbl$ and $\sbla$ have distinct content (\emph{$h$-inequality requirement}).
 
 We construct   $\psi_{h,\neq}$ by induction on $h\geq 2$. The $\InqLTL_0$ formula $\psi_{2,\neq}$ is defined as follows.  
 \[
 \begin{array}{ll}
 \psi_{2,\neq} &  \DefinedAs  \Next (  p_{\leq 1}    \,\Until \,  ( \Symb_1 \wedge \phi_{2,\neq}) )\vspace{0.3cm}\\
 \phi_{2,\neq} & \DefinedAs \neg\neg \Bigl[(\Always \neg \col) \StrongDis  \Future(\col \wedge \neg \Symb_1)\, \StrongDis \,
   \displaystyle{\bigvee_{i\in [1,n]}\bigvee_{b\in \{0,1\}}} \bigl(\Next^i b\wedge \Future(\col\wedge \Next^i \neg b)\bigr) \, \StrongDis \\
    & \phantom{\DefinedAs \neg\neg \Bigl[} \displaystyle{\bigvee_{c\in \{0,1\}}} \bigl(c\wedge \Future(\col\wedge \neg c)\bigr) \Bigr]  
 \end{array}
 \]
Note that $\psi_{2,\neq}$ checks that the subformula $\phi_{2,\neq}$ holds at the starting $\Symb_1$-position of some $1$-sub-block    $\sbl$ of $\bl$. For such a sub-block $\sbl$, 
$\phi_{2,\neq}$ requires that for each trace $\pi\in \Lang$ such that $\col$ marks the starting position of a $(h-1)$-block $\sbla$ (since $\bla$ is the unique $\col$-marked $h$-block of $\Lang_{\geq i}$, $\sbla$ is necessarily a
sub-block of $\bla$), then whenever $\sbl$ and $\sbla$  have the same index, then $\sbl$ and $\sbla$ have distinct content.
 Now assume that $h>2$. In this case, we exploit the following auxiliary \emph{positive} $\InqLTL$ formula $\MaxCol_{h-1}$:
 \[
\MaxCol_{h-1} \DefinedAs (\neg\col) \,\Until \, (\Symb_{h-1}\wedge \Next [p_{\leq h-2}\,\Until \, (\Symb_{\geq h-1}\wedge\Always \neg\col)])
 \]
For the given consistent team $\Lang$ such that $\Lang_{\geq i}$ is $h$-marked, the previous formula is satisfied at the starting position $\ell$ of an $(h-1)$-sub-block of $\bl$   by all and only the subteams $\Lang'$ of $\Lang_{\geq \ell}$ such that $\Lang'$ is contained in some $(h-1)$-marked subteam $\Lang_{h-1}$ of $\Lang_{\geq \ell}$. 
Hence, the $(h-1)$-marked subteams $\Lang_{h-1}$ of $\Lang_{\geq \ell}$ are the maximal subteams of $\Lang_{\geq \ell}$  which satisfy $\MaxCol_{h-1}$.
Moreover, note that the marked $(h-1)$-block of $\Lang_{h-1}$ is necessarily a sub-block of  the marked  $h$-block $\bla$ of $\Lang_{\geq i}$.
Thus,  the  $\InqLTL_{h-2}$ formula $\psi_{h,\neq}$ exploits the formula $\psi_{h-1,\neq}$ and is defined as: 
 \[
 \begin{array}{ll}
 \psi_{h,\neq }  & \hspace{-0.2cm} \DefinedAs  \Next (  p_{\leq h-1}    \,\Until \,  ( \Symb_{h-1} \wedge \phi_{h,\neq}) )\vspace{0.3cm}\\
 \phi_{h,\neq}  & \hspace{-0.2cm}\DefinedAs \MaxCol_{h-1} \longrightarrow  
   \Bigl[ \psi_{h-1,\neq} \StrongDis \neg\neg \Bigl\{ (\Always \neg \col) \StrongDis  \Future(\col \wedge \neg \Symb_{h-1}) \,\StrongDis \,   
    \displaystyle{\bigvee_{c\in \{0,1\}}} \bigl(c\wedge \Future(\col\wedge \neg c\bigr) \Bigr\}\Bigr]  
 \end{array}
 \]
Formula  $\psi_{h,\neq}$ checks that the subformula $\phi_{h,\neq}$ holds at the starting $\Symb_{h-1}$-position $\ell$ of some $1$-sub-block    $\sbl$ of $\bl$. 
Since each subteam of $\Lang_{\geq \ell}$ satisfying $\MaxCol_{h-1}$ is contained in some $(h-1)$-marked subteam of $\Lang_{\geq \ell}$  and $\InqLTL$ formulas are downward closed,
by the induction hypothesis, when asserted at the starting position of $\sbl$, 
$\phi_{h,\neq}$ requires that for each $(h-1)$-sub-block $\sbla$ of $\bla$, \emph{either} $\sbl$ and $\sbla$ have distinct index, \emph{or} $\sbl$ and $\sbla$ have distinct content.
Figure~\ref{fig:CheckingInequalIndex} illustrates the $h$-inequality check performed by formula $\psi_{h,\neq}$.
This concludes the proof of Proposition~\ref{prop:CheckingInequalIndex}.
\end{proof}

Next, for each $h\in [1,k]$, we show how to check in polynomial time whether for two adjacent well-formed $h$-blocks $\bl$ and $\bla$ of a consistent team $\Lang$, their indices are \emph{not} consecutive, i.e., the index of $\bla$ is \emph{not} the increment of the index of $\bl$. At this end, we exploit \emph{fully-marked} consistent  teams  which are consistent teams $\Lang$ such that 
for each position $i$, there is a trace $w$ of $\Lang$ so that $\col$ holds exactly at position $i$.

\begin{prop}\label{prop:CheckingBadIncrement} For each $h\in [1,k]$, one can construct in time polynomial in the size of $\Instance$ a left-positive $\InqLTL_{h-1}$ formula $\widetilde{\psi}_{h,inc}$ such that for each fully-marked consistent team $\Lang$ and adjacent well-formed $h$-blocks $\bl$ and $\bla$ along $\Lang$, the following holds, where $i$ is the starting position of $\bl$:
\[
\Lang_{\geq i}\models \widetilde{\psi}_{h,inc} \Leftrightarrow \text{the indices of $\bl$ and $\bla$ are \emph{not} consecutive.}
\]
\end{prop}
\begin{proof} We assume that $h>1$ (the case where $h=1$ is straightforward). We observe that the indices of the two adjacent 
well-formed $h$-blocks $\bl$ and $\bla$ along $\Lang$ are \emph{not} consecutive iff \emph{either} 
the index of $\bl$ is maximal (i.e., each $(h-1)$-sub-block of $\bl$ has content $1$), \emph{or} denoted 
by $\sbl_0$ the first $(h-1)$-sub-block of $\bl$ with content $0$, one of the following three conditions holds:
 
\begin{enumerate}
  \item[(1)] there exists an $(h-1)$-sub-block $\sbl$ of $\bl$ strictly preceding 
  $\sbl_0$ such that the $(h-1)$-sub-block of $\bla$ having the same index as $\sbl$ has content $1$;
  \item[(2)] the $(h-1)$-sub-block of $\bla$ having the same index as $\sbl_0$ has content $0$; 
  \item[(3)] there exists an $(h-1)$-sub-block $\sbl$ of $\bl$ strictly following 
  $\sbl_0$ such that for the $(h-1)$-sub-block $\sbla$ of $\bla$ having the same index as $\sbl$, $\sbl$ and $\sbla$ have distinct content.
\end{enumerate}
 
\noindent Thus, $\widetilde{\psi}_{h,inc}\DefinedAs \psi_{last} \StrongDis\widetilde{\psi}_{h,inc,1} \StrongDis \widetilde{\psi}_{h,inc,2}  \StrongDis \widetilde{\psi}_{h,inc,3}$,
where $\psi_{last}$ requires that $\bl$ has maximal index and $\widetilde{\psi}_{h,inc,1}$ (resp., $\widetilde{\psi}_{h,inc,2}$, resp., $\widetilde{\psi}_{h,inc,3}$)
expresses the previous requirement~(1) (resp., requirement~(2), resp., requirement~(3)). We focus on the definition of formulas $\psi_{last}$ and $\widetilde{\psi}_{h,inc,1}$ (formulas $\widetilde{\psi}_{h,inc,2}$ and $\widetilde{\psi}_{h,inc,3}$ are similar to $\widetilde{\psi}_{h,inc,1}$).
\[
\psi_{last} \DefinedAs  \Next\Bigl( (p_{\leq h-1}\wedge (\neg \Symb_{h-1}\StrongDis 1))  \,\Until\, \Symb_{h}\Bigr) 
\]
For the construction of $\widetilde{\psi}_{h,inc,1}$, we exploit the formula $\psi_{h-1,\neq}$ of Proposition~\ref{prop:CheckingInequalIndex} and the following \emph{positive} 
 $\InqLTL$ formula $\MaxCol'_{h-1}$:
 \[
 \begin{array}{ll}
 \MaxCol'_{h-1} & \DefinedAs   (p_{\leq h-1}\wedge \neg\col) \,\Until \,  ( \Symb_{h} \wedge \neg\col \wedge \Next\xi_{h-1}) \vspace{0.1cm}\\
 \xi_{h-1} & \DefinedAs    (p_{\leq h-1}\wedge \neg\col) \,\Until \,   
    \Bigl( \Symb_{h-1} \wedge  \Next [p_{\leq h-2} \,\Until \, (\Symb_{\geq h-1}\wedge\Always\neg \col)]  \Bigr)
 \end{array}
 \]
 \begin{figure*}[!tb]
\centering
\begin{tikzpicture}[
    scale=0.95,
    >=Stealth,
    headStyle/.style={fill=blue!8, draw=blue!70!black, rounded corners=2pt, inner sep=2.5pt},
    markerStyle/.style={fill=orange!15, draw=orange!80!black, rounded corners=2pt, inner sep=2.5pt},
    subBlockStyle/.style={draw=black!70, rounded corners=2pt, inner sep=2.5pt, fill=white},
    tagStyle/.style={fill=white, inner sep=2pt, font=\footnotesize}
]

\coordinate [label=center:{\footnotesize \textbf{Well-formed $h$-block $\bl$}}] (Blk_l_title) at (-4.7, 1.4);
\coordinate [label=center:{\scriptsize (sub-block $\sbl$ with content $\mathbf{1}$)}] (Sbl_desc) at (-4.1, 0.8);

\node[headStyle] (L_head) at (-6.8, 0.0) {\footnotesize $\{\Symb_h, \tau\}$};
\coordinate [label=center:{\footnotesize $\dots$}] (L_dots0) at (-5.7, 0.0);
\node[subBlockStyle] (Sbl) at (-4.1, 0.0) {\footnotesize $\{\Symb_{h-1}, \mathbf{1}\} \dots$};
\coordinate [label=center:{\footnotesize $\dots$}] (L_dots1) at (-2.7, 0.0);
\node[subBlockStyle] (Sbl0) at (-1.4, 0.0) {\footnotesize $\{\Symb_{h-1}, \mathbf{0}\}\dots$};
\coordinate [label=center:{\footnotesize $\dots$}] (L_dots2) at (-0.1, 0.0);
\coordinate [label=center:{\footnotesize $\sbl_j$}] (L_end) at (0.6, 0.0);

\coordinate [label=center:{\footnotesize \textbf{Adjacent well-formed $h$-block $\bla$}}] (Blk_la_title) at (5.7, 1.4);
\coordinate [label=center:{\scriptsize (sub-block $\sbla$ marked via $\MaxCol'_{h-1}$)}] (Sbla_desc) at (4.7, 0.8);

\node[headStyle] (La_head) at (2.0, 0.0) {\footnotesize $\{\Symb_h, \tau'\}$};
\coordinate [label=center:{\footnotesize $\dots$}] (La_dots0) at (3.1, 0.0);
\node[markerStyle] (Sbla) at (4.7, 0.0) {\footnotesize $\{\Symb_{h-1}, \mathbf{1}, \col\} \dots$};
\coordinate [label=center:{\footnotesize $\dots$}] (La_dots1) at (6.3, 0.0);
\coordinate [label=center:{\footnotesize $\sbla_j$}] (La_end) at (7.3, 0.0);
\coordinate [label=center:{\tiny ($j = \Tower_2(h-1,n)-1$)}] (La_end_idx) at (7.0, -0.5);

\draw[<->, thick, color=blue!80!black, dashed] (Sbl.south) to [bend right=32] 
    node[midway, below=2mm, tagStyle, align=center] {
        \footnotesize \textbf{Bad increment check (Condition 1) via $\phi_{h,1}$:}\\[-0.5mm]
        \footnotesize Same sub-block index (checked via $\psi_{h-1,\neq}$) $\implies$ bit content remains $\mathbf{1}$ instead of $\mathbf{0}$
    } (Sbla.south);

\end{tikzpicture}
\caption{Detecting a bad increment between adjacent $h$-blocks $\bl$ and $\bla$ via $\widetilde{\psi}_{h,inc,1}$ (Proposition~\ref{prop:CheckingBadIncrement}). Sub-block $\sbl$ (with content $1$) precedes  the first sub-block with content $0$; formula $\phi_{h,1}$ uses $\MaxCol'_{h-1}$ to locate the sub-block $\sbla$ in $\bla$ with the same index and detects a violation if its content is still $1$.}
\label{fig:CheckingBadIncrement}
\end{figure*}
For the given fully-marked consistent team $\Lang$, the previous formula is satisfied at the starting position $j$ of an $(h-1)$-sub-block of $\bl$   by all and only the subteams $\Lang'$ of $\Lang_{\geq j}$ such that $\Lang'$ is contained in some $(h-1)$-marked subteam $\Lang_{h-1}$ of $\Lang_{\geq j}$ and the  marked $(h-1)$-block of $\Lang_{h-1}$ is  a sub-block of  $\bla$.
Thus,     $\widetilde{\psi}_{h,inc,1}$ exploits the formula $\psi_{h-1,\neq}$ of Proposition~\ref{prop:CheckingInequalIndex} and is defined as: 
 \[
 \begin{array}{ll}
 \widetilde{\psi}_{h,inc,1} & \hspace{-0.2cm}\DefinedAs  \Next \Bigl( ( p_{\leq h-1} \wedge (\neg\Symb_{h-1}\StrongDis 1))   \,\Until \,   
     ( \Symb_{h-1} \wedge 1 \wedge  \phi_{h,1}) \Bigr)\vspace{0.2cm}\\
 \phi_{h,1} & \hspace{-0.2cm} \DefinedAs   \MaxCol'_{h-1} \longrightarrow \Bigl[ \psi_{h-1,\neq}\, \StrongDis \, \neg\neg \Bigl\{ (\Always \neg \col) \StrongDis  \Future(\col \wedge \neg \Symb_{h-1}) \StrongDis   \Future(\col\wedge 1) \Bigr\}\Bigr]  
 \end{array}
 \]
Formula $\widetilde{\psi}_{h,inc,1}$ checks that the subformula $\phi_{h,1}$ holds at the starting $\Symb_{h-1}$-position $j$ of some $(h-1)$-sub-block    $\sbl$ of $\bl$ strictly preceding the
first $(h-1)$-sub-block of $\bl$, if any, having content $0$.  Recall that 
  each subteam of $\Lang_{\geq j}$ satisfying $\MaxCol'_{h-1}$ is contained in some $(h-1)$-marked subteam  $\Lang_{h-1}$ of $\Lang_{\geq i}$ whose marked $(h-1)$-block  is  a sub-block of  $\bla$. Thus, since $\InqLTL$ formulas are downward closed,
by Proposition~\ref{prop:CheckingInequalIndex}, when asserted at the starting position of $\sbl$, 
$\phi_{h,1}$ requires that for each $(h-1)$-sub-block $\sbla$ of $\bla$, \emph{either} $\sbl$ and $\sbla$ have distinct index, \emph{or} $\sbl$ has content $1$.
The intuition behind formula $\widetilde{\psi}_{h,inc,1}$ is illustrated in Figure~\ref{fig:CheckingBadIncrement}.
This concludes the proof of Proposition~\ref{prop:CheckingBadIncrement}.
\end{proof}    

By exploiting Propositions~\ref{prop:CheckingInequalIndex}--\ref{prop:CheckingBadIncrement}, we now establish 
the core result in the proposed reduction  which together with Proposition~\ref{prop:LowerBoundKS}
concludes the proof of Theorem~\ref{theo:lowerBounds}.
 
\begin{prop}[Construction of $\varphi_{\Instance,k}$]\label{prop:LowerBoundFormula} Let $\KS_{\Instance,k}$ be the Kripke structure of Proposition~\ref{prop:LowerBoundKS}. One can construct, in time polynomial in the size of $\Instance$, a left-positive $\InqLTL_k$ formula $\varphi_{\Instance,k}$ such that $\Lang(\KS_{\Instance,k})\models \varphi_{\Instance,k}$ iff there is \emph{no} $k$-tiling of $\Instance$.
 \end{prop}
 \begin{proof} Here, we assume that $k>1$ (the case $k=1$ is simpler). We construct $\varphi_{\Instance,k}$ in such a way that $\Lang(\KS_{\Instance,k})\models \varphi_{\Instance,k}$ iff every fully-marked consistent team does \emph{not} encode  a
 $k$-tiling of $\Instance$. Recall that $\Lang(\KS_{\Instance,k})$ is the set of all the \emph{$3$-sparse} traces $w$  such that proposition $\col$ holds at most at one position of $w$. By Proposition~\ref{prop:LTLforPseudoGrids},
 one can construct in polynomial time an $\LTL$ formula $\theta_k$ over $\AP\setminus \{\col\}$ which captures the initialized $k$-grid codes. Note that
 $\theta_k$ can be seen as a positive $\InqLTL$ formula. Hence, by Proposition~\ref{prop:LowerBoundKS},  the following positive $\InqLTL$ formula $\varphi_{con}$ is satisfied by a \emph{nonempty} subteam $\Lang$ of
  $\Lang(\KS_{\Instance,k})$ iff $\Lang$ is consistent (note that each consistent team is a subset of
 $\Lang(\KS_{\Instance,k})$):
 \[
 \varphi_{con}\DefinedAs \theta_k \wedge \displaystyle{\bigwedge_{p\in\AP\setminus \{\col\}}} (p\StrongDis \neg p)
 \]
 Then, the left-positive $\InqLTL_k$ formula $\varphi_{\Instance,k}$ is defined as follows:
 \[
 \varphi_{\Instance,k}\DefinedAs \varphi_{con} \longrightarrow ((\Future \neg \col) \StrongDis \widetilde{\varphi}_{1} \StrongDis \ldots \StrongDis \widetilde{\varphi}_{k+1} \StrongDis  
 \widetilde{\varphi}_{row} \StrongDis \widetilde{\varphi}_{col})
 \]
Note that the disjunct $(\Future \neg \col)$ in the definition of $ \varphi_{\Instance,k}$  is satisfied by a consistent team $\Lang$ iff $\Lang$ is \emph{not} fully-marked.
Now, let us define the disjuncts  $\widetilde{\varphi}_{1}, \ldots,\widetilde{\varphi}_{k+1}$ in $\varphi_{\Instance,k}$. They enforce the following requirements:
\begin{figure*}[!tb]
\centering
\begin{tikzpicture}[
    scale=0.95,
    >=Stealth,
    headStyle/.style={fill=blue!8, draw=blue!70!black, rounded corners=2pt, inner sep=2.5pt},
    markerStyle/.style={fill=orange!15, draw=orange!80!black, rounded corners=2pt, inner sep=2.5pt},
    tagStyle/.style={fill=white, inner sep=2pt, font=\footnotesize}
]

\coordinate [label=left:{\footnotesize \textbf{Row $m$:}}] (R1_lbl) at (-6.2, 1.2);
\coordinate [label=center:{\footnotesize $\{\Symb\} \dots$}] (R1_start) at (-4.7, 1.2);
\coordinate [label=center:{\footnotesize $\dots$}] (R1_dots1) at (-3.2, 1.2);

\node[headStyle] (Cell_m) at (-0.5, 1.2) {\footnotesize $\{\Symb_k, \mathbf{d}\} \bl_0 \dots \bl_j$};
\coordinate [label=center:{\scriptsize (cell $\bl$ with domino $d$)}] (Cell_m_desc) at (-0.5, 1.75);

\coordinate [label=center:{\footnotesize $\dots$}] (R1_dots2) at (2.2, 1.2);
\coordinate [label=center:{\footnotesize $\bl_{\Tower_2(k,n)-1}$}] (R1_end) at (4.0, 1.2);

\coordinate [label=center:{\footnotesize $\dots\dots\dots$}] (Trace_sep) at (-0.5, 0.0);

\coordinate [label=left:{\footnotesize \textbf{Row $m+1$:}}] (R2_lbl) at (-6.2, -1.2);
\coordinate [label=center:{\footnotesize $\{\Symb\} \dots$}] (R2_start) at (-4.7, -1.2);
\coordinate [label=center:{\footnotesize $\dots$}] (R2_dots1) at (-3.2, -1.2);

\node[markerStyle] (Cell_m1) at (-0.5, -1.2) {\footnotesize $\{\Symb_k, \mathbf{d'}, \col\} \bla_0 \dots \bla_j$};
\coordinate [label=center:{\scriptsize (cell $\bla$ with domino $d'$ marked via $\MaxCol'_k$)}] (Cell_m1_desc) at (-0.5, -1.75);

\coordinate [label=center:{\footnotesize $\dots$}] (R2_dots2) at (2.2, -1.2);
\coordinate [label=center:{\footnotesize $\bla_{\Tower_2(k,n)-1}$}] (R2_end) at (4.0, -1.2);

\draw[<->, dashed, thick, color=blue!80!black] (Cell_m.south) -- (Cell_m1.north)
    node[midway, right=8mm, tagStyle, align=left] {
        \footnotesize \textbf{Column adjacency check via $\widetilde{\xi}_{col}$:}\\[-0.5mm]
        \footnotesize $\bullet$ Same cell index (enforced via $\psi_{k,\neq}$)\\[-0.5mm]
        \footnotesize $\bullet$ Color mismatch: $[d]_{up} \neq [d']_{down}$  
    };

\end{tikzpicture}
\caption{Checking column adjacency between two consecutive rows via formula $\widetilde{\varphi}_{col}$ (Proposition~\ref{prop:LowerBoundFormula}). The auxiliary formula $\MaxCol'_k$ isolates a cell $\bla$ in the adjacent row; formula $\widetilde{\xi}_{col}$ verifies that whenever $\bl$ and $\bla$ share the same index, their vertical colors are incompatible ($[d]_{up} \neq [d']_{down}$).}
\label{fig:ColumnAdjacencyCheck}
\end{figure*}
\begin{itemize}
  \item for each fully-marked consistent team $\Lang$, $\Lang\models \widetilde{\varphi}_{1}$ iff some $1$-block of $\Lang$ is not well-formed;
  \item for each $h\in [2,k]$ and fully-marked consistent team $\Lang$ such that all the $(h-1)$-blocks of $\Lang$ are well-formed, 
  $\Lang\models \widetilde{\varphi}_{h}$ iff some $h$-block of $\Lang$ is not well-formed;
  \item for each fully-marked consistent team $\Lang$ such that all the $k$-blocks of $\Lang$ are well-formed, 
  $\Lang\models \widetilde{\varphi}_{k+1}$ iff some $k$-row of $\Lang$ is not well-formed.
\end{itemize}
The definition of the disjunct $\widetilde{\varphi}_{1}$, which is a positive $\InqLTL$ formula, is straightforward, while 
for each $h\in [2,k+1]$, the disjunct $\widetilde{\varphi}_{h}$ is a left-positive $\InqLTL_{h-2}$ formula which uses as a subformula 
the left-positive $\InqLTL_{h-2}$ formula $\widetilde{\psi}_{h-1,inc}$ of Proposition~\ref{prop:CheckingBadIncrement}. We focus on the construction  of 
$\widetilde{\varphi}_{k+1}$ (the definitions of  $\widetilde{\varphi}_{2},\ldots,\widetilde{\varphi}_{k}$ are similar). 
 \[
 \begin{array}{ll}
 \widetilde{\varphi}_{k+1} & \DefinedAs  \widetilde{\varphi}_{k,init} \StrongDis \widetilde{\varphi}_{k,last} \StrongDis \widetilde{\varphi}_{k,inc}\vspace{0.2cm}\\
 \widetilde{\varphi}_{k,init} & \DefinedAs \Future(\Symb \wedge \Next^2 [p_{\leq k-1}\,\Until \, (\Symb_{k-1}\wedge 1)])\vspace{0.2cm}\\
  \widetilde{\varphi}_{k,last} & \DefinedAs \Future(\Symb_k \wedge \Next[p_{\leq k-1}\,\Until \,\Symb ] \, \wedge \,
     \Next [p_{\leq k-1}\,\Until \, (\Symb_{k-1}\wedge 0)])\vspace{0.2cm}\\
  \widetilde{\varphi}_{k,inc} & \DefinedAs \Future(\Symb_k \wedge \Next[p_{\leq k-1}\,\Until \,\Symb_k] \wedge \widetilde{\psi}_{k,inc})
 \end{array} 
 \]
 For each fully-marked consistent team $\Lang$ such that all the $k$-blocks of $\Lang$ are well-formed,
 the conjunct $\widetilde{\varphi}_{k,init}$ requires that there is some $k$-row whose first $k$-block has index distinct from $0$,
 while the conjunct $\widetilde{\varphi}_{k,last}$ requires that there is some $k$-row whose last $k$-block $\bl$ has an  index which is not maximal (i.e., some 
 $(k-1)$-sub-block of $\bl$ has content $0$). Note that $\widetilde{\varphi}_{k,init}$ and $\widetilde{\varphi}_{k,last}$ are positive $\InqLTL$ formulas.
 The last conjunct $\widetilde{\varphi}_{k,inc}$ is a left-positive  $\InqLTL_{k-1}$ formula which exploits the left-positive $\InqLTL_{k-1}$ formula $\widetilde{\psi}_{k-1,inc}$ of Proposition~\ref{prop:CheckingBadIncrement}, and requires that there are two adjacent  $k$-blocks along a $k$-row of $\Lang$ whose indices are not consecutive.
 
 Finally, the disjuncts  $\widetilde{\varphi}_{row}$ and $\widetilde{\varphi}_{col}$ in the definition of $\varphi_{\Instance,k}$ enforce the following requirements
 for each fully-marked consistent team $\Lang$ whose $k$-rows are well-formed: 
 \begin{itemize}
  \item  $\Lang\models \widetilde{\varphi}_{row}$ iff the well-formed $k$-grid encoded by $\Lang$ does not satisfy the row adjacency requirement of $k$-tilings.
  \item  $\Lang\models \widetilde{\varphi}_{col}$ iff the well-formed $k$-grid encoded by $\Lang$ does not satisfy the column adjacency requirement of $k$-tilings.
\end{itemize} 
We focus on the definition of $\widetilde{\varphi}_{col}$ (the construction of $\widetilde{\varphi}_{row}$, which is a positive $\InqLTL$ formula, is straightforward).
Let $\psi_{k,\neq}$ be the left-positive $\InqLTL_{k-2}$ formula of Proposition~\ref{prop:CheckingInequalIndex}. Moreover, by proceeding as in the proof of  
Proposition~\ref{prop:CheckingBadIncrement}, we can define a positive $\InqLTL$ formula $\MaxCol_k'$ such that the following holds:
  \begin{itemize}
  \item  for each fully-marked consistent team $\Lang$, starting position $i$ of a $k$-block $\bl$, and subteam $\Lang'$ of $\Lang_{\geq i}$,
   $\Lang'\models \MaxCol_k'$ iff $\Lang'$ is contained in some $k$-marked subteam $\Lang_{k}$ of $\Lang_{\geq i}$ and the  marked $k$-block of $ \Lang_{k}$ belongs to the $k$-row adjacent to the $k$-row of 
   $\bl$.
\end{itemize}
Intuitively, $\MaxCol_k'$ allows to mark a $k$-block belonging to the  $k$-row adjacent to the $k$-row of 
   $\bl$. Then, $\widetilde{\varphi}_{col}$ is an $\InqLTL_{k-1}$ formula defined as follows, where $Bad$ denotes the set of
   pairs $(d,d')\in \Delta\times \Delta$ such that $[d]_{up}\neq [d']_{down}$:
 \[
 \begin{array}{ll}
 \widetilde{\varphi}_{col} & \DefinedAs  \displaystyle{\bigvee_{(d,d')\in Bad }}\Future \Bigl( \Symb_k \wedge d   \wedge [ \MaxCol'_{k} \rightarrow \widetilde{\xi}_{col} ] \Bigr)\vspace{0.2cm}\\
 \widetilde{\xi}_{col} & \DefinedAs  
      \psi_{k,\neq} \, \StrongDis \,
      \neg\neg \Bigl( (\Always \neg \col) \StrongDis  \Future(\col \wedge \neg \Symb_{k}) \StrongDis   \Future(\col\wedge d') \Bigr)  
 \end{array}
 \]
 Essentially, $\widetilde{\varphi}_{col}$ asserts that there are $(d,d')\in Bad$ and a $k$-block $\bl$ with content $d$ such that the $k$-block having the same index as $\bl$ and belonging to the row adjacent 
 to the $\bl$-row has content $d'$. The intuition behind formula $\widetilde{\varphi}_{col}$ is illustrated in Figure~\ref{fig:ColumnAdjacencyCheck}. This concludes the proof of Proposition~\ref{prop:LowerBoundFormula}.
\end{proof}

\section{Conclusion}\label{sec:Conclusions}

We have introduced $\InqLTL$, a team semantics for $\LTL$ inspired by inquisitive logic. 
The logic replaces the split disjunction of $\TeamLTL$ with inquisitive disjunction and intuitionistic implication. 
  We have shown that, when enhanced with strong negation $\StrongNeg$, the logic has the countable model property, is highly undecidable, and is strictly less expressive than
  $\TeamLTL(\StrongNeg)$. 
We then have identified a  fragment of $\InqLTL$, called left-positive $\InqLTL$, with a decidable model checking problem,  which does not allow for nesting of implication in the left side of an implication
(with the exception of applying intuitionistic negation to atomic propositions). We have illustrated how left-positive $\InqLTL$  can capture meaningful classes of hyperproperties
such as  information-flow security properties. 
It is worth noting that the decidability status of model checking $\TeamLTL$ and  fragments of $\TeamLTL(\StrongNeg)$ is a challenging open question. The known positive results~\cite{KrebsMV018,VirtemaHFK021} have been achieved by imposing drastic syntactical or semantical restrictions. This paper shows, for the first time, a hyper logic, namely left-positive $\InqLTL$, with unrestricted use of temporal modalities and conditional universal second-order quantification over system refinements with a decidable model-checking problem.
Our results were obtained with non-trivial proof techniques. The abstraction technique introduced for left-positive $\InqLTL$ is a notable contribution in its own right, with promising potential for generalization to other logics such as $\TeamLTL$ and unrestricted $\InqLTL$.
 
Several intriguing directions remain open for future research.
First, our foundational comparison between $\InqLTL$ and $\TeamLTL$ highlights their dual treatment of subteams, namely conditional universal quantification 
over refinements via intuitionistic implication versus existential decomposition via split disjunction. A major open goal is to formally settle our conjectures regarding their relative incomparability, as well as the conjecture that split formulas such as $\Future p \vee \Future p$ strictly separate $\TeamLTL$ from $\InqLTL(\StrongNeg)$.
Second, a crucial open problem is whether the model-checking problem for unrestricted $\InqLTL$ remains decidable, or whether nested implications in the antecedent cross the decidability boundary.
Moreover, it would be interesting to analyze the exact complexity of model checking within more constrained sub-fragments of left-positive $\InqLTL$ (such as those restricted to unary temporal operators), as well as to investigate inquisitive team semantics for asynchronous hyperproperties where traces of a team progress with different speed. 
Finally, a natural direction is to develop branching-time and strategic counterparts of $\InqLTL$. Existing intuitionistic branching-time and strategic logics, such as $\ICTL$ and $\IATL$~\cite{bozzelli2025intuitionistic,DBLP:conf/eumas/Capone25}, provide relevant semantic and algorithmic starting points for combining temporal branching or strategic quantification with constructive notions of information growth~\cite{bozzelli2025intuitionistic,DBLP:conf/eumas/Capone25}.





\bibliographystyle{alphaurl}
\bibliography{biblio}

@article{AbramskyV09,
  author       = {S.~Abramsky and
                  J.A. V{\"{a}}{\"{a}}n{\"{a}}nen},
  title        = {From {IF} to {BI}. A tale of dependence and separation},
  journal      = {Synth.},
  volume       = {167},
  number       = {2},
  pages        = {207--230},
  year         = {2009},
    doi        = {10.1007/S11229-008-9415-6}
}

@article{BalbianiBDF20,
  author       = {P.~Balbiani and
                  J.~Boudou and
                  M.~Di{\'{e}}guez and
                  D.F.~Duque},
  title        = {Intuitionistic Linear Temporal Logics},
  journal      = {{ACM} Trans. Comput. Log.},
  volume       = {21},
  number       = {2},
  pages        = {14:1--14:32},
  year         = {2020},
  doi          = {10.1145/3365833}
}

@inproceedings{BaumeisterCBFS21,
  author    = {J.~Baumeister and
               N.~Coenen and
               B.~Bonakdarpour and
               B.~Finkbeiner and
               C.~S{\'{a}}nchez},
  title     = {A {T}emporal {L}ogic for {A}synchronous {H}yperproperties},
  booktitle = {Proc. 33rd {CAV}},
  series    = {LNCS 12759},
  volume    = {12759},
  pages     = {694--717},
  publisher = {Springer},
  year      = {2021},
  doi       = {10.1007/978-3-030-81685-8\_33},
}

@article{BittnerBCGTV22,
  author       = {B.~Bittner and
                  M.~Bozzano and
                  A.~Cimatti and
                  M.~Gario and
                  S.~Tonetta and
                  V.~Voz{\'{a}}rov{\'{a}}},
  title        = {Diagnosability of fair transition systems},
  journal      = {Artif. Intell.},
  volume       = {309},
  pages        = {103725},
  year         = {2022},
  doi          = {10.1016/J.ARTINT.2022.103725}
}

@incollection{Boas97,
  author    = {P.~{van Emde Boas}},
  title     = {The {C}onvenience of {T}ilings},
  booktitle = {Complexity, Logic, and Recursion Theory},
  series    = {Lecture Notes in Pure and Applied Mathematics},
  volume    = {187},
  pages     = {331--363},
  publisher = {Marcel Dekker Inc},
  year      = {1997}
}

@inproceedings{BozzelliMP15,
  author    = {L.~Bozzelli and
               B.~Maubert and
               S.~Pinchinat},
  title     = {Unifying {H}yper and {E}pistemic {T}emporal {L}ogics},
  booktitle = {Proc. 18th FoSSaCS},
  series    = {LNCS 9034},
  pages     = {167--182},
  publisher = {Springer},
  year      = {2015},
  doi       = {10.1007/978-3-662-46678-0\_11}
}

@inproceedings{BozzelliPS21,
  author    = {L.~Bozzelli and
               A.~Peron and
               C.~S{\'{a}}nchez},
  title     = {{A}synchronous {E}xtensions of {HyperLTL}},
  booktitle = {Proc. 36th {LICS}},
  pages     = {1--13},
  publisher = {{IEEE}},
  year      = {2021},
  doi       = {10.1109/LICS52264.2021.9470583}
}

@inproceedings{bozzelli2025intuitionistic,
  author       = {L.~Bozzelli and
                  A.~Capone and
                  D.~Catta and
                  A.~Murano},
  title        = {{An Intuitionistic Version of Alternating-Time Temporal Logic}},
  booktitle    = {Proc. 22nd {KR}},
  year         = {2025},
  doi          = {10.24963/KR.2025/19}
}

@inproceedings{DBLP:conf/eumas/Capone25,
  author       = {L.~Bozzelli and
                  A.~Capone and
                  D.~Catta and
                  V.~Malvone and
                  A.~Murano},
  title        = {An Intuitionistic Version of Computation Tree Logic},
  booktitle    = {Proc. 22nd {EUMAS}},
  series       = {LNCS},
  volume       = {16258},
  pages        = {195--211},
  publisher    = {Springer},
  year         = {2025},
  doi          = {10.1007/978-3-032-22817-8\_12} 
}

@inproceedings{BozzelliLMM26,
  author       = {L.~Bozzelli and
                  T.~Litak and
                  M.~Mittelmann and
                  A.~Murano}, 
  title        = {{Inquisitive Team Semantics of {LTL}}},
  booktitle    = {Proc. 29th {FoSSaCS}},
  series       = {LNCS},
  volume       = {16503},
  pages        = {110--132},
  publisher    = {Springer},
  year         = {2026}, 
  doi          = {10.1007/978-3-032-22730-0\_6}
}

@BOOK{ChagrovZ97:ml,
  AUTHOR = {A.~Chagrov and M.~Zakharyaschev},
  LOCATION = {Oxford},
  NUMBER = {35},
  PUBLISHER = {Clarendon Press},
  SERIES = {Oxford Logic Guides},
  TITLE = {Modal Logic},
  YEAR = {1997},
}

@mastersthesis{ciardelli2009inquisitive,
  title={Inquisitive semantics and intermediate logics},
  author={I.~Ciardelli},
 school       = {Universiteit van Amsterdam},
  year={2009}
}

@article{CiardelliR11,
  author       = {I.~Ciardelli and
                  F.~Roelofsen},
  title        = {{Inquisitive Logic}},
  journal      = {J. Philos. Log.},
  volume       = {40},
  number       = {1},
  pages        = {55--94},
  year         = {2011},
  doi          = {10.1007/S10992-010-9142-6}
}

@article{CiardelliR15,
  author  = {I.~Ciardelli and F.~Roelofsen},
  title   = {Inquisitive Dynamic Epistemic Logic},
  journal = {Synthese},
  volume  = {192},
  number  = {6},
  pages   = {1643--1687},
  year    = {2015},
  doi     = {10.1007/S11229-014-0404-7}
}

@PhdThesis{Ciardelli2016a,
  author    = {I.~Ciardelli},
  title     = {Questions in Logic},
  school    = {University of Amsterdam},
  year      = {2016} 
}

@article{Ciardelli16,
  author       = {I.~Ciardelli},
  title        = {Propositional inquisitive logic: a survey},
  journal      = {Comput. Sci. J. Moldova},
  volume       = {24},
  number       = {3},
  pages        = {295--311},
  year         = {2016}
}

@incollection{Ciardelli16dep,
  author       = {I.~Ciardelli},
  title        = {{Dependency as Question Entailment}},
  booktitle    = {{Dependence Logic, Theory and Applications}},
  pages        = {129--181},
  publisher    = {Springer},
  year         = {2016},
  doi          = {10.1007/978-3-319-31803-5\_8}
}

@inproceedings{CiardelliO17, 
  author       = {I.~Ciardelli and
                  M.~Otto},
  title        = {{Bisimulation in Inquisitive Modal Logic}},
  booktitle    = {Proc. 16 th {TARK}},
  series       = {{EPTCS}},
  volume       = {251},
  pages        = {151--166},
  year         = {2017},
  doi          = {10.4204/EPTCS.251.11} 
}

@article{CiardelliO22, 
  author       = {I.~Ciardelli and
                  M.~Otto},
  title        = {{Inquisitive Bisimulation}}, 
  volume={86}, 
  doi={10.1017/jsl.2020.77}, 
  number={1}, 
  journal={The Journal of Symbolic Logic}, 
  pages={77--109},
  year = {2022},
  }

@book{Ciardelli2022,
  author    = {I.~Ciardelli},
  title     = {Inquisitive {L}ogic: {C}onsequence and {I}nference in the {R}ealm of {Q}uestions},
  series    = {Trends in Logic},
  volume    = {59},
  publisher = {Springer},
  year      = {2022},
  doi       = {10.1007/978-3-031-09706-5}
}

@article{CiardelliG22,
  author  = {I.~Ciardelli and G.~Grilletti},
  title   = {Coherence in inquisitive first-order logic},
  journal = {Ann. Pure Appl. Logic},
  volume  = {173},
  number  = {9},
  pages   = {103--155},
  year    = {2022},
  doi     = {10.1016/J.APAL.2022.103155}
}

@inproceedings{Ciardelli22,
  author    = {I.~Ciardelli},
  title     = {{Describing Neighborhoods in Inquisitive Modal Logic}},
  booktitle = {Proc. 14th {AiML}},
  pages     = {217--236},
  publisher = {College Publications},
  year      = {2022},
  url       = {http://www.aiml.net/volumes/volume14/16-Ciardelli.pdf}
}

@article{ClarksonS10,
  author    = {M.R.~Clarkson and
               F.B.~Schneider},
  title     = {Hyperproperties},
  journal   = {{Journal of Computer Security}},
  volume    = {18},
  number    = {6},
  year      = {2010},
  pages     = {1157-1210},
  doi       = {10.3233/JCS-2009-0393}
}

@inproceedings{ClarksonFKMRS14,
  author    = {M.R.~Clarkson and
               B.~Finkbeiner and
               M.~Koleini and
               K.K.~Micinski and
               M.N.~Rabe and
               C.~S{\'a}nchez},
  title     = {Temporal {L}ogics for {H}yperproperties},
  booktitle = {Proc. 3rd {POST}},
  year      = {2014},
  pages     = {265-284},
  publisher = {Springer},
  series    = {LNCS 8414},
   doi       = {10.1007/978-3-642-54792-8\_15}
}

@inproceedings{CoenenFHH19,
  author    = {N.~Coenen and
               B.~Finkbeiner and
               C.~Hahn and
               J.~Hofmann},
  title     = {{The Hierarchy of Hyperlogics}},
  booktitle = {Proc. 34th {LICS}},
  pages     = {1--13},
  publisher = {{IEEE}},
  year      = {2019},
  doi       = {10.1109/LICS.2019.8785713}
}

@inproceedings{DaxK08,
  author       = {C.~Dax and
                  F.~Klaedtke},
  title        = {{Alternation Elimination by Complementation (Extended Abstract)}},
  booktitle    = {Proc. 15th  {LPAR}},
  series       = {LNCS 5330},
  pages        = {214--229},
  publisher    = {Springer},
  year         = {2008},
  doi          = {10.1007/978-3-540-89439-1\_16}
}

@article{EmersonH86,
  author    = {E.A.~Emerson and
               J.Y.~Halpern},
  title     = {"{S}ometimes" and "{N}ot {N}ever" revisited: on branching
               versus linear time temporal logic},
  journal   = {J. ACM},
  volume    = {33},
  number    = {1},
  year      = {1986},
  pages     = {151-178},
  doi       = {10.1145/4904.4999}
}

@incollection{FerreiraO12:full,
  author    = {G.~Ferreira and P.~Oliva},
  title     = {{On the Relation Between Various Negative Translations}},
  booktitle = {Logic, Construction, Computation},
  series    = {Mathematical Logic Series},
  volume    = {3},
  pages     = {227--258},
  publisher = {Ontos-Verlag},
  year      = {2012} 
}

@inproceedings{FinkbeinerRS15,
  author    = {B.~Finkbeiner and
               M.N.~Rabe and
               C.~S{\'{a}}nchez},
  title     = {Algorithms for {M}odel {C}hecking {H}yper{LTL} and {H}yper{CTL}*},
  booktitle = {Proc. 27th {CAV}  Part
               {I}},
  series    = {LNCS 9206},
  volume    = {9206},
  pages     = {30--48},
  publisher = {Springer},
  year      = {2015},
  doi       = {10.1007/978-3-319-21690-4\_3}
}

@article{FischerL79,
  author    = {M.J.~Fischer and
               R.E.~Ladner},
  title     = {{P}ropositional {D}ynamic {L}ogic of {R}egular {P}rograms},
  journal   = {J. Comput. Syst. Sci.},
  volume    = {18},
  number    = {2},
  pages     = {194--211},
  year      = {1979},
  doi       = {10.1016/0022-0000(79)90046-1}
}

@inproceedings{DFrenkel025,
  author       = {H.~Frenkel and
                  M.~Zimmermann}, 
  title        = {The Complexity of Second-Order {H}yper{LTL}},
  booktitle    = {Proc. 33rd {CSL}},
  series       = {LIPIcs},
  volume       = {326},
  pages        = {10:1--10:23},
  publisher    = {Schloss Dagstuhl - Leibniz-Zentrum f{\"{u}}r Informatik},
  year         = {2025}, 
  doi          = {10.4230/LIPICS.CSL.2025.10}
}

@inproceedings{goguen1982security,
  title={Security {P}olicies and {S}ecurity {M}odels},
  author={J.A.~Goguen   and J.~Meseguer},
  booktitle={{IEEE} Symposium on Security and Privacy},
  pages     = {11--20},
  publisher = {{IEEE} Computer Society},
  year={1982},
  doi       = {10.1109/SP.1982.10014}
}

@article{grilletti2019disjunction,
  title={{Disjunction and existence properties in inquisitive first-order logic}},
  author={G.~Grilletti},
  journal={Studia Logica},
  volume={107},
  number={6},
  pages={1199--1234},
  year={2019},
  publisher={Springer}
}

@article{Grilletti2021,
  author  = {G.~Grilletti},
  title   = {{Completeness for the Classical Antecedent Fragment of Inquisitive First-Order Logic}},
  journal = {J. Log. Lang. Inf.},
  volume  = {30},
  number  = {4},
  pages   = {725--751},
  year    = {2021},
  doi     = {10.1007/S10849-021-09341-Y}
}

@inproceedings{GutsfeldMO20,
  author    = {J.Oliver.~Gutsfeld and
               M.~M{\"{u}}ller{-}Olm and
               C.~Ohrem},
  title     = {Propositional Dynamic Logic for Hyperproperties},
  booktitle = {Proc. 31st {CONCUR}},
  series    = {LIPIcs 171},
  pages     = {50:1--50:22},
  publisher = {Schloss Dagstuhl - Leibniz-Zentrum f{\"{u}}r Informatik},
  year      = {2020},
  doi       = {10.4230/LIPIcs.CONCUR.2020.50}
}

@inproceedings{GutsfeldMOV22,
  author       = {J.Oliver.~Gutsfeld and
                  A.~Meier and
                  C.~Ohrem and
                  J.~Virtema},
  title        = {Temporal {T}eam {S}emantics {R}evisited},
  booktitle    = {Proc. 37th {LICS}},
  pages        = {44:1--44:13},
  publisher    = {{ACM}},
  year         = {2022},
  doi          = {10.1145/3531130.3533360}
}

@inproceedings{HalpernV86,
  author       = {J.Y.~Halpern and M.Y.~Vardi},
  title        = {The {C}omplexity of {R}easoning about {K}nowledge and {T}ime: {E}xtended {A}bstract},
  booktitle    = {Proc. 18th   {STOC}},
  pages        = {304--315},
  publisher    = {{ACM}},
  year         = {1986},
  doi          = {10.1145/12130.12161}
}

@article{HalpernO08,
  author    = {J.Y.~Halpern and
               K.R.~O'Neill},
  title     = {{Secrecy in Multiagent Systems}},
  journal   = {ACM Trans. Inf. Syst. Secur.},
  volume    = {12},
  number    = {1},
  pages     = {5:1--5:47},
  year      = {2008},
  doi          = {10.1145/1410234.1410239}
}

@article{KontinenSV25,
  author       = {J.~Kontinen and
                  M.~Sandstr{\"{o}}m and
                  J.~Virtema},
  title        = {{Set semantics for asynchronous TeamLTL: Expressivity and complexity}},
  journal      = {Inf. Comput.},
  volume       = {304},
  pages        = {105299},
  year         = {2025},
  doi          = {10.1016/J.IC.2025.105299}
}

@inproceedings{KrebsMV018,
  author    = {A.~Krebs and
               A.~Meier and
               J.~Virtema and
               M.~Zimmermann},
  title     = {Team {S}emantics for the {S}pecification and {V}erification of {H}yperproperties},
  booktitle = {Proc. 43rd {MFCS}},
  series    = {LIPIcs 117},
  pages     = {10:1--10:16},
  publisher = {Schloss Dagstuhl - Leibniz-Zentrum f{\"{u}}r Informatik},
  year      = {2018},
  doi       = {10.4230/LIPIcs.MFCS.2018.10}
}

@article{KupfermanVW00,
    author = "O.~Kupferman and M.Y.~Vardi and P.~Wolper",
    title = "An {A}utomata-{T}heoretic {A}pproach to {B}ranching-{T}ime {M}odel {C}hecking",
    journal = "J. ACM",
    volume = "47",
    number = "2",
    pages = "312--360",
    year = "2000",
    doi       = {10.1145/333979.333987}
}

@inproceedings{litakpr17,
  author    = {T.~Litak and M.~Polzer and U.~Rabenstein},
  title     = {{Negative Translations and Normal Modality}},
  booktitle = {Proc. 2nd {FSCD}},
  series    = {LIPIcs},
  volume    = {84},
  pages     = {27:1--27:18},
  publisher = {Schloss Dagstuhl--Leibniz-Zentrum f{\"u}r Informatik},
  year      = {2017},
  doi       = {10.4230/LIPICS.FSCD.2017.27}
}

@inproceedings{LitakS25,
  author       = {T.~Litak and
                  K.~Sano}, 
  title        = {Bounded Inquisitive Logics: Sequent Calculi and Schematic Validity},
  booktitle    = {Proc. 34th {TABLEAUX}},
  series       = {LNCS},
  volume       = {15980},
  pages        = {39--58},
  publisher    = {Springer},
  year         = {2025},
  doi          = {10.1007/978-3-032-06085-3\_3} 
}

@article{Luck20,
  author    = {M.~L{\"{u}}ck},
  title     = {On the complexity of linear temporal logic with team semantics},
  journal   = {Theor. Comput. Sci.},
  volume    = {837},
  pages     = {1--25},
  year      = {2020},
  doi       = {10.1016/j.tcs.2020.04.019}
}

@article{MaricP24,
  author       = {S.~Mari{\'c} and
                  T.~Perkov},
  title        = {{Decidability of Inquisitive Modal Logic via Filtrations}},
  journal      = {Stud Logica},
  volume       = {113},
  number       = {5},
  pages        = {1437--1455},
  year         = {2025},
  doi          = {10.1007/S11225-024-10134-0}
}

@article{McLean96,
  author    = {J.~McLean},
  title     = {A {G}eneral {T}heory of {C}omposition for a {C}lass of "{P}ossibilistic'' {P}roperties},
  journal   = {{IEEE} Trans. Software Eng.},
  volume    = {22},
  number    = {1},
  pages     = {53--67},
  year      = {1996},
  doi       = {10.1109/32.481534}
}

@article{MeissnerO22,
      title={A First-Order Framework for Inquisitive Modal Logic}, 
      author={Silke Mei{\ss}ner and Martin Otto},
      year={2022},
      eprint={1906.04981},
      doi = {10.1017/S175502032100037X},
      journal={The Review of Symbolic Logic},
      volume={15},
      number={2}
}

@article{Nygren23,
  author  = {K.~Nygren},
  title   = {{Free Choice in Modal Inquisitive Logic}},
  journal = {J. Philos. Log.},
  volume  = {52},
  number  = {2},
  pages   = {347--391},
  year    = {2023},
  doi     = {10.1007/S10992-022-09674-4} 
}

@inproceedings{Pnueli77,
  author    = {A.~Pnueli},
  title     = {The {T}emporal {L}ogic of {P}rograms},
  booktitle = {Proc. 18th FOCS},
  year      = {1977},
  pages     = {46-57},
  publisher = {IEEE Computer Society},
  doi       = {10.1109/SFCS.1977.32}
}

@phdthesis{Rabe2016,
  author       = {M.N.~Rabe},
  title        = {A temporal logic approach to information-flow control},
  school       = {Saarland University},
  year         = {2016}
}

@book{Rogers87,
 author       = "H.~Rogers",
 year         = "1987",
 title        = "Theory of {R}ecursive {F}unctions and {E}ffective {C}omputability",
 publisher    = "MIT Press",
address   = "Cambridge,MA, US",
}

@inproceedings{SabelfeldS05,
  author       = {A.~Sabelfeld and
                  D.~Sands},
  title        = {{Dimensions and Principles of Declassification}},
  booktitle    = {Proc. 18th  {CSFW}},
  pages        = {255--269},
  publisher    = {{IEEE} Computer Society},
  year         = {2005},
  doi          = {10.1109/CSFW.2005.15}
}

@article{SampathSLST95,
  author       = {M.~Sampath and
                  R.~Sengupta and
                  S.~Lafortune and
                  K.~Sinnamohideen and
                  D.~Teneketzis},
  title        = {Diagnosability of discrete-event systems},
  journal      = {{IEEE} Trans. Autom. Control.},
  volume       = {40},
  number       = {9},
  pages        = {1555--1575},
  year         = {1995},
  doi          = {10.1109/9.412626}
}

@inproceedings{SanchezS12,
  author       = {C.~S{\'{a}}nchez and
                  J.~Samborski{-}Forlese},
  title        = {{Efficient Regular Linear Temporal Logic Using Dualization and Stratification}},
  booktitle    = {Proc. 19th {TIME}},
  pages        = {13--20},
  publisher    = {{IEEE} Computer Society},
  year         = {2012}, 
  doi          = {10.1109/TIME.2012.25} 
}

@article{SistlaC85,
  author    = {A.P.~Sistla and
               E.M.~Clarke},
  title     = {The {C}omplexity of {P}ropositional {L}inear {T}emporal {L}ogics},
  journal   = {J. {ACM}},
  volume    = {32},
  number    = {3},
  pages     = {733--749},
  year      = {1985},
  doi       = {10.1145/3828.3837}
}

@article{SistlaVW87,
  author    = {A.P.~Sistla and
               M.Y.~Vardi and
               P.~Wolper},
  title     = {The {C}omplementation {P}roblem for {B}{\"u}chi {A}utomata with
               {A}pplications to {T}emporal {L}ogic},
  journal   = {Theoretical Computer Science},
  volume    = {49},
  year      = {1987},
  pages     = {217-237},
  doi       = {10.1016/0304-3975(87)90008-9}
}

@book{SorensenU06:book,
  author    = {M.~H.~S{\o}rensen and P.~Urzyczyn},
  title     = {Lectures on the {Curry}-{Howard} {I}somorphism},
  series    = {Stud. Logic Found. Math.},
  volume    = {149},
  publisher = {Elsevier},
  year      = {2006},
  doi       = {10.1016/s0049-237x(06)80004-1}
}

@book{TroelstraVD,
  author    = {A.~S.~Troelstra and D.~{van Dalen}},
  title     = {Constructivism in Mathematics: An Introduction},
  series    = {Stud. Logic Found. Math.},
  volume    = {121},
  publisher = {Elsevier},
  year      = {1988}
}

@book{Vaananen07,
  author       = {J.A.~V{\"{a}}{\"{a}}n{\"{a}}nen},
  title        = {{Dependence Logic - {A} New Approach to Independence Friendly Logic}},
  series       = {London Mathematical Society student texts},
  volume       = {70},
  publisher    = {Cambridge University Press},
  year         = {2007}
}

@article{VardiW94,
  author    = {M. Y. Vardi and
               P. Wolper},
  title     = {Reasoning About Infinite Computations},
  journal   = {Inf. Comput.},
  volume    = {115},
  number    = {1},
  pages     = {1--37},
  year      = {1994},
  doi       = {10.1006/inco.1994.1092}
}

@inproceedings{VirtemaHFK021,
  author    = {J.~Virtema and
               J.~Hofmann and
               B.~Finkbeiner and
               J.~Kontinen and
               F.~Yang},
  title     = {Linear-{T}ime {T}emporal {L}ogic with {T}eam {S}emantics: {E}xpressivity and {C}omplexity},
  booktitle = {Proc. 41st {IARCS}  {FSTTCS}},
  series    = {LIPIcs 213},
  pages     = {52:1--52:17},
  publisher = {Schloss Dagstuhl - Leibniz-Zentrum f{\"{u}}r Informatik},
  year      = {2021},
  doi       = {10.4230/LIPIcs.FSTTCS.2021.52}
}

@inproceedings{ZdancewicM03,
  author    = {S.~Zdancewic and
               A.C.~Myers},
  title     = {Observational {D}eterminism for {C}oncurrent {P}rogram {S}ecurity},
  booktitle = {Proc. 16th {IEEE} CSFW-16},
  pages     = {29-43},
  publisher = {{IEEE} Computer Society},
  year      = {2003},
  doi       = {10.1109/CSFW.2003.1212703}
}

\end{document}